\documentclass[%
 reprint,
nofootinbib,
amsmath,
amssymb,
aps,
prd,
floatfix,
twocolumn
]{revtex4-2}

\usepackage[dvipsnames]{xcolor}
\definecolor{customcolor}{rgb}{0.369, 0.118, 0.267}
\usepackage{graphicx}
\usepackage{dcolumn}
\usepackage{bm}
\usepackage{booktabs}
\usepackage{colortbl}
\usepackage{braket} 
\usepackage{amsmath}
\usepackage{float}
\usepackage{multirow}
\usepackage{hyperref}
\usepackage[compat=1.1.0]{tikz-feynman}
\usepackage[compat=1.1.0]{tikz-feynhand}
\hypersetup{
    colorlinks,
    linkcolor={red!60!black},
    citecolor={RoyalBlue},
    urlcolor={red!60!black}
}
\usepackage{tikz}
\usepackage[compat=1.1.0]{tikz-feynman}
\usepackage[compat=1.1.0]{tikz-feynhand}
\usepackage[export]{adjustbox}
\usepackage{color, colortbl}

\newcommand{\be}{\begin{eqnarray}}
\newcommand{\ee}{\end{eqnarray}}
\newcommand{\beq}{\begin{equation}}
\newcommand{\eeq}{\end{equation}}

\newcommand{\sig}{\sigma}
\newcommand{\lzm}{\left(}
\newcommand{\dzm}{\right)}
\newcommand{\lzs}{\left[}
\newcommand{\dzs}{\right]}
\newcommand{\lzv}{\left\{}
\newcommand{\dzv}{\right\}}
\newcommand{\lzu}{\left|}
\newcommand{\dzu}{\right|}
\newcommand{\cL}{\mathcal{L}}
\newcommand{\cO}{\mathcal{O}}

\newcommand{\cW}{{\mathcal W}}
\newcommand{\cB}{{\mathcal B}}

\newcommand{\cI}{{\mathcal I}}
\newcommand{\cC}{{\mathcal C}}
\newcommand{\cS}{{\mathcal S}}

\newcommand{\gev}{\mathrm{GeV}}
\newcommand{\tev}{\mathrm{TeV}}

\newcommand{\im}{{\mathrm{Im}} \,}

\newcommand{\eminus}{\vcenter{\hbox{\scalebox{0.6}[1]{$ - $}}}}	

\newcommand{\sscript}[1]{{\scriptscriptstyle \mathrm{#1}}}

\definecolor{wildstrawberry}{rgb}{1.0, 0.26, 0.64}

\definecolor{puke}{rgb}{0.7, 0.7, 0.4}

\begin{document}

\title{Leptonic Flavor from Modular $A_4$: UV Mediators and SMEFT Realizations}

\author{Adri\'an Moreno-S\'anchez}
\email{adri@ugr.es}
\affiliation{Departamento de F\'isica Te\'orica y del Cosmos, Universidad de Granada, Campus de Fuentenueva, E–18071 Granada, Spain}
\author{Ajdin Palavri\'c}
\email{ajdin.palavric@unibas.ch}
\affiliation{Department of Physics, University of Basel, Klingelbergstrasse 82, CH-4056 Basel, Switzerland}

\begin{abstract} 
The absence of direct evidence for new physics at current collider energies motivates the study of indirect effects within the framework of the Standard Model Effective Field Theory (SMEFT). In this work, we investigate the role of the $A_4$ discrete flavor symmetry in constraining the UV dynamics, giving rise to dimension-6 SMEFT operators involving leptons. We consider renormalizable interactions between SM fields and heavy mediators, classified according to their $A_4$ transformation properties in the modular framework, where Yukawa structures are encoded in modular forms. Restricting our analysis to UV interactions with at most a single modular form insertion, we study the resulting classification of $A_4$ UV mediator irreps and explore their implications across a range of observables, including both lepton-flavor-conserving and lepton-flavor-violating processes. A detailed phenomenological analysis is carried out, incorporating both tree-level and one-loop matching contributions for scalar and fermionic mediators, as well as leading-log RGE effects in case of vectors.
\end{abstract}

\maketitle

\section{Introduction}
The persistent absence of direct evidence for new physics (NP) at current collider energies strongly suggests that potential NP mediators are likely associated with a scale well above the electroweak (EW) regime. In this context, it becomes imperative to establish a systematic and theoretically sound approach to parameterize potential deviations from Standard Model (SM) predictions. The Standard Model Effective Field Theory (SMEFT)~\cite{Buchmuller:1985jz, Grzadkowski:2010es, Brivio:2017vri, Isidori:2023pyp, Giudice:2007fh, Henning:2014wua} addresses this need by extending the Standard Model with higher-dimensional operators organized in a systematic expansion suppressed by powers of the NP scale. These operators, built exclusively from SM fields and constrained by gauge and Poincar\'e symmetry, are weighted by Wilson coefficients (WCs) that encapsulate the influence of heavy degrees of freedom in a model-independent manner. As such, SMEFT provides a versatile and rigorous platform for exploring a wide spectrum of NP effects indirectly.

One of the central structural features of the SMEFT Lagrangian is that, at any fixed canonical dimension, the number of independent higher-dimensional operators remains finite. This property has enabled systematic classification efforts, leading to major advances in operator counting techniques~\cite{Henning:2015alf, Fonseca:2019yya} and the construction of complete bases up to high mass dimensions~\cite{Grzadkowski:2010es, Lehman:2014jma, Li:2023cwy, Ren:2022tvi, Li:2020xlh, Li:2020gnx, Murphy:2020rsh, Liao:2016hru, Liao:2020jmn, Harlander:2023psl, Harlander:2023ozs}. Nonetheless, as the dimension increases, the number of independent operators exhibits a steep combinatorial growth. While this is partly due to the increasing number of invariant structures permitted by the SM gauge and Lorentz symmetries, the dominant contribution arises from the replication of fermionic fields across three generations. To illustrate, the number of baryon-number-conserving dimension-six operators is 59 in a single-flavor setup, but grows to 2499 when all three SM flavors are considered.

In the SM, the fermion kinetic terms respect a large global flavor symmetry, $U(3)^5$, which independently acts on the three generations of each of the five distinct fermionic gauge representations: $q$, $u$, $d$, $\ell$, and $e$. This symmetry is broken explicitly by the Yukawa interactions, which introduce non-universal and non-diagonal couplings between generations. As a result, the full $U(3)^5$ symmetry is reduced to $U(1)_B \times U(1)^3_L$, corresponding to the conservation of baryon number and three individual lepton numbers. Nonetheless, the structure of the Yukawa matrices, manifested in the pronounced hierarchies of fermion masses and the specific texture of the CKM and PMNS matrices, strongly hints at the presence of an approximate, possibly spontaneously broken, flavor symmetry. The notion that such a symmetry underlies the observed flavor patterns forms the basis for many extensions of the SM, where flavor symmetries, either continuous or discrete, are introduced to organize and explain the fermion mass and mixing hierarchies.

The remarkable sensitivity of modern flavor experiments has transformed the flavor sector into one of the most powerful probes of NP, setting stringent bounds on deviations from the SM predictions~\cite{EuropeanStrategyforParticlePhysicsPreparatoryGroup:2019qin}. In light of these constraints, any NP scenario operating near the TeV scale, a regime amenable to current or upcoming collider searches, must exhibit a high degree of alignment with SM flavor symmetries. This phenomenological imperative dovetails with structural considerations within the SMEFT, where the rapid growth in the number of independent operators is largely attributable to flavor multiplicity. In particular, what appears as a dramatic increase in operator count at higher dimensions is, to a large extent, the result of numerous allowed flavor assignments applied to a limited set of underlying operator structures. By incorporating approximate flavor symmetries into the SMEFT framework, one not only achieves compliance with experimental constraints but also uncovers organizing principles that reduce redundancy and expose symmetry-driven correlations among Wilson coefficients~\cite{Greljo:2022cah, Faroughy:2020ina, Bruggisser:2022rhb, Machado:2022ozb}.

\begin{table*}[t]
    \centering
\scalebox{0.85}{
\begin{tabular}{c@{\hspace{1.0cm}}c@{\hspace{1.0cm}}c}
\toprule
\textbf{UV Field} 
&{$\bm{-\cL_{\sscript{UV}}^{(4)}\supset}$}
&{$\bm{\cL_{\sscript{SMEFT}}\supset}$}
\\
\midrule
$\cS_1\sim(\bm1,\bm1)_{1}$
&$[y_{\mathcal S_1}]_{rij}\mathcal S_{1r}^\dag \bar\ell_{i}i\sig_2\ell_{j}^c$
&\vspace{+0.05cm}$\frac{1}{M^2_{\cS_1}}[y_{\cS_1}]_{rjl}^*[y_{\cS_1}]_{rik}[\cO_{\ell\ell}]_{ijkl}$
\\
$\cS_2\sim(\bm1,\bm1)_{2}$
&$[y_{\mathcal S_2}]_{rij}\mathcal S_{2r}^\dag \bar e_{i}e_{j}^c$
&\vspace{+0.05cm}$\frac{1}{2M^2_{\cS_2}}[y_{\mathcal S_2}]_{rki}[y_{\mathcal S_2}]_{rlj}^*[\cO_{ee}]_{ijkl}$
\\
$\varphi\sim(\bm1,\bm2)_{\frac{1}{2}}$
&$[y_\varphi]_{rij}\varphi_r\bar\ell_ie_j$
&\vspace{+0.05cm}$-\frac{1}{2M^2_\varphi}[y_\varphi]_{rli}^*[y_\varphi]_{rkj}[\cO_{\ell e}]_{ijkl}$
\\
$\Xi_1\sim(\bm1,\bm3)_{1}$
&$[y_{\Xi_1}]_{rij}\Xi_{1r}^{a\dag}\bar\ell_{i}\sig^ai\sig_2\ell_{j}^c$
&\vspace{+0.05cm}$\frac{1}{M^2_{\Xi_1}}[y_{\Xi_1}]_{rki}[y_{\Xi_1}]_{rlj}^*[\cO_{\ell\ell}]_{ijkl}$
\\
\midrule
\midrule
$N\sim(\bm1,\bm1)_0$
&$[\lambda_N]_{ri}\bar N_{R,r}\tilde\phi^\dag \ell_i$
&\vspace{+0.05cm}$\frac{1}{4M_N^2}[\lambda_N]_{ri}^*[\lambda_N]_{rj}\lzs [\cO_{\phi\ell}^{(1)}]_{ij}-[\cO_{\phi\ell}^{(3)}]_{ij} \dzs$
\\
$E\sim(\bm1,\bm1)_{-1}$
&$[\lambda_E]_{ri}\bar E_{R,r}\phi^\dag\ell_{i}$
&\vspace{+0.05cm}$-\frac{1}{4M_E^2}[\lambda_E]_{ri}^*[\lambda_E]_{rj}\lzs [\cO_{\phi\ell}^{(1)}]_{ij}+[\cO_{\phi\ell}^{(3)}]_{ij} \dzs$
\\
$\Delta_1\sim(\bm1,\bm2)_{-\frac{1}{2}}$
&$[\lambda_{\Delta_1}]_{ri}\bar\Delta_{1L,r}\phi e_i$
&\vspace{+0.05cm}$\frac{1}{2M^2_{\Delta_1}}[\lambda_{\Delta_1}]_{ri}^*[\lambda_{\Delta_1}]_{rj}[\cO_{\phi e}]_{ij}$
\\
$\Delta_3\sim(\bm1,\bm2)_{-\frac{3}{2}}$
&$[\lambda_{\Delta_3}]_{ri}\bar\Delta_{3L,r}\tilde\phi e_{i}$
&\vspace{+0.05cm}$-\frac{1}{2M^2_{\Delta_3}}[\lambda_{\Delta_3}]_{ri}^*[\lambda_{\Delta_3}]_{rj}[\cO_{\phi e}]_{ij}$
\\
$\Sigma\sim(\bm1,\bm3)_{0}$
&$\frac{1}{2}[\lambda_\Sigma]_{ri}\bar\Sigma^a_{R,r}\tilde\phi^\dag\sig^a\ell_{i}$
&\vspace{+0.05cm}$\frac{1}{16M^2_{\Sigma}}[\lambda_\Sigma]_{ri}^*[\lambda_{\Sigma}]_{rj}\lzs 3[\cO_{\phi\ell}^{(1)}]_{ij}+[\cO_{\phi\ell}^{(3)}]_{ij} \dzs$
\\
$\Sigma_1\sim(\bm1,\bm3)_{-1}$
&$\frac{1}{2}[\lambda_{\Sigma_1}]_{ri}\bar\Sigma^a_{1R,r}\phi^\dag\sig^a\ell_{i}$
&\vspace{+0.05cm}$\frac{1}{16M^2_{\Sigma_1}}[\lambda_{\Sigma_1}]_{ri}^*[\lambda_{{\Sigma_1}}]_{rj}\lzs [\cO_{\phi\ell}^{(3)}]_{ij}-3[\cO_{\phi\ell}^{(1)}]_{ij} \dzs$
\\
\midrule
\midrule
$\cB\sim(\bm1,\bm1)_0$
&$[g_\cB^\ell]_{rij}\cB_r^\mu\bar\ell_{i}\gamma_\mu\ell_{j}+[g_\cB^e]_{rij}\cB_r^\mu\bar e_{i}\gamma_\mu e_{j}$
&\vspace{+0.05cm}$-\frac{1}{2M^2_{\cB}}[g_\cB^\ell]_{rkl}[g_\cB^\ell]_{rij}[\cO_{\ell\ell}]_{ijkl}-\frac{1}{2M^2_\cB}[g_\cB^e]_{rkl}[g_\cB^e]_{rij}[\cO_{ee}]_{ijkl}
-\frac{1}{M^2_\cB}[g_\cB^e]_{rkl}[g_\cB^\ell]_{rij}[\cO_{\ell e}]_{ijkl}$
\\
$\cW\sim(\bm1,\bm3)_0$
&$\frac{1}{2}[g_\cW^\ell]_{rij}\cW^{\mu\,a}_r\bar\ell_i\sigma^a\gamma_\mu\ell_j$
&\vspace{+0.05cm}$-\frac{1}{4M^2_\cW}[g_\cW^\ell]_{rkj}[g_\cW^\ell]_{ril}[\cO_{\ell\ell}]_{ijkl}+\frac{1}{8M^2_\cW}[g_\cW^\ell]_{rkl}[g_\cW^\ell]_{rij}[\cO_{\ell\ell}]_{ijkl}$
\\
$\cL_3\sim(\bm1,\bm2)_{-\frac{3}{2}}$
&$[g_{\cL_3}]_{rij}\cL_{3r}^{\mu\dag}\bar e_{i}^c\gamma_\mu \ell_{j}$
&\vspace{+0.05cm}$\frac{1}{M^2_{\cL_3}}[g_{\cL_3}]_{rki}^*[g_{\cL_3}]_{rlj}[\cO_{\ell e}]_{ijkl}$
\\
\bottomrule
\end{tabular}
}	
    \caption{Overview of the UV mediators analyzed in this work. In the first column we list the labels for different UV mediators along with their irreps under the SM gauge group. In the second column we include the interaction terms of the UV fields with the SM leptons and in the third column we list the tree-level SMEFT matching relations to the dimension-6 operators. We adopt the standard basis for dimension-6 SMEFT operators, as defined in Ref.~\cite{Grzadkowski:2010es}. $[y_{S}]_{rij}$, $[\lambda_F]_{ri}$ and $[g_V]_{rij}$ denote the coupling tensors for each UV scalar, fermion and vector mediator, respectively, which are also referred to as \textit{flavor tensors} in the forthcoming discussions. For more details on them see Sec.~\ref{sec:UV_flavor_invs}.}
    \label{tab:Intro_Table_Mediators_SMEFT}
\end{table*}

To complement the SMEFT description and pursue a more complete understanding of the possible UV origins of effective operators, one can take an additional step in the bottom-up direction by introducing explicit heavy mediators. The central aim in this context is to identify which classes of renormalizable NP fields, whether scalar, fermionic, or vector in nature, can couple to the Standard Model (SM) fields and, when integrated out at tree level, generate specific dimension-6 operators in the SMEFT. This procedure not only elucidates the structure of the effective theory but also allows for a systematic connection to potential UV completions without committing to a specific high-scale model. A full catalogue of such mediator representations and their associated tree-level matching relations has been provided in Ref.~\cite{deBlas:2017xtg}, offering a valuable resource for model-building and phenomenological analyses.

Building on the role of flavor symmetries within the SMEFT framework, a compelling extension involves imposing the flavor symmetry structure on the UV mediators responsible for generating effective operators. In this approach, the transformation properties of the mediators under the flavor group dictate which renormalizable couplings to SM fields are allowed, and consequently, which SMEFT operators can arise through the matching procedure. This symmetry-guided restriction refines the mapping between UV dynamics and low-energy observables, offering a more predictive framework for constructing models that are both phenomenologically viable and consistent with flavor data.\footnote{A systematic analysis of this kind, with $U(3)^5$ flavor symmetry assumption, was carried out in Ref.~\cite{Greljo:2023adz}, with the important RGE effects further analyzed in Ref.~\cite{Greljo:2023bdy}, revealing additional constraints beyond tree level.}

Restricting our attention to the lepton sector, discrete flavor symmetries have long played a central role in organizing its observed structure, offering symmetry-based explanations for the patterns of neutrino masses and mixing angles. Among the various non-Abelian discrete groups considered in the literature, $A_4$ has received particular attention due to its minimality and natural embedding of tribimaximal-like mixing patterns~\cite{Morisi:2009sc, Ferreira:2013oga, Ding:2019zxk, Feruglio:2008ht, Holthausen:2012wz, Abbas:2020qzc, Morisi:2013eca, DeLaVega:2018bkp, Peinado:2010iu, Pramanick:2015qga, Altarelli:2009kr, Pramanick:2019qpg, Carrolo:2022oyg, Ding:2021eva, King:2011ab, Ding:2024fsf, Kadosh:2010rm, Barman:2023idm, Borah:2018nvu, Boucenna:2011tj, Memenga:2013vc, Morisi:2007ft, Nomura:2019jxj, Zhang:2019ngf, Asaka:2019vev, Kumar:2023moh, Okada:2019mjf, Gogoi:2022jwf, Singh:2023jke, Kobayashi:2018scp, Criado:2018thu, Feruglio:2017spp, Altarelli:2010gt, CentellesChulia:2023osj, Kumar:2024zfb, Chauhan:2023faf, Borah:2017dmk, Karmakar:2014dva, Feruglio:2019ybq, Petcov:2018snn, King:2013eh,Feruglio:2024ytl,Feruglio:2023uof}. At the same time, other discrete groups such as $A_5$~\cite{Ding:2011cm, Novichkov:2018nkm, DiIura:2016ols, DiIura:2015kfa, Cooper:2012bd, Ding:2019xna, Li:2015jxa, Turner:2015mwa, Puyam:2023div, Gehrlein:2014wda, Hernandez:2012ra, deMedeirosVarzielas:2022ihu, Ballett:2015wia, Feruglio:2011qq, Everett:2008et, Gehrlein:2015dza, Gehrlein:2015dxa, Chen:2010ty} and $S_4$~\cite{Kobayashi:2019mna, Ishimori:2009ns, Thapa:2023fxu, Ishimori:2011mt, Hagedorn:2006ug, Bazzocchi:2012st, Zhang:2021olk, Ishimori:2010fs, Izawa:2023qay, Bazzocchi:2008ej, Wang:2019ovr, Nomura:2021ewm, Penedo:2018nmg, Novichkov:2018ovf, King:2021fhl, Ding:2019gof, Okada:2019lzv, Behera:2025tpj,Criado:2019tzk,Belfkir:2024uvj} have also been extensively explored and remain viable candidates for capturing the full structure of lepton flavor. These frameworks have informed a wide array of models aiming to reproduce viable neutrino textures and mixing matrices.\footnote{Apart from these discrete symmetries, a range of other viable options relevant for model building can be found in Ref.~\cite{Ishimori:2010au}.} Within this broad context, Ref.~\cite{Palavric:2024gvu} investigated the interplay between exact non-Abelian discrete symmetries and UV, systematically analyzing how different group-theoretic assignments constrain the structure of renormalizable interactions and shape the resulting SMEFT embeddings.

In this work we focus specifically on the $A_4$ symmetry, further developing that line of inquiry. Compared to the analysis presented in Ref.~\cite{Palavric:2024gvu}, in this letter we offer several key improvements. 

The first improvement concerns a refined classification of possible $A_4$ flavor irreducible representations (irreps). Motivated by approaches used in the analyses involving continuous flavor symmetries, where symmetry breaking is parametrized using finite sets of flavor spurions~\cite{Greljo:2022cah,Faroughy:2020ina}, we extend the analysis to include a broader set of irreps that emerge once $A_4$ breaking effects are incorporated in the interactions between UV mediators and the SM fields. This is achieved by promoting the exact discrete $A_4$ symmetry to its modular counterpart, where symmetry breaking is intrinsically encoded in the vacuum expectation value of the complex modulus $\tau$. In this setup, Yukawa couplings are no longer treated as arbitrary parameters but are instead realized as modular forms, holomorphic functions that transform non-trivially under modular transformations and carry specific modular weights. Our framework includes UV interactions with at most one insertion of the lowest-weight modular form, capturing the minimal realization of modular symmetry breaking, which, after tree-level matching onto SMEFT, leads to effective operators involving two modular insertions~\cite{Kobayashi:2021pav}. Lastly, we note that our analysis is performed below the supersymmetry (SUSY) breaking scale~\cite{Kobayashi:2021pav, Kobayashi:2023zzc}.\footnote{This choice is well-motivated, as the modular symmetry structure, originating from the underlying superpotential and K\"ahler geometry, can imprint effective couplings that remain relevant even after SUSY breaking, provided that the SUSY-breaking scale lies above the energies of interest. Working in this regime allows us to retain the predictive power of modular flavor constructions while simplifying the field content to that of the non-supersymmetric SMEFT.}

The second key improvement over Ref.~\cite{Palavric:2024gvu} lies in the phenomenological analysis. In this work, we consider a more comprehensive set of charged lepton flavor violation (cLFV) observables and, in case of scalar and fermion mediators, we extend the analysis by computing one-loop matching contributions to the SMEFT.\footnote{Given the absence of a concrete UV completion, for vectors we employ the leading-log contributions from the RGEs.} This allows us to assess the full impact of radiative corrections and to identify flavor irreps of $A_4$ for which one-loop effects play a phenomenologically significant role.

Returning to the relevant set of UV mediators, as outlined in Ref.~\cite{deBlas:2017xtg}, we isolate 13 new physics fields whose interaction Lagrangians include couplings to lepton bilinears. These consist of four scalars ($\mathcal{S}_1$, $\mathcal{S}_2$, $\varphi$, $\Xi_1$), six fermions ($N$, $E$, $\Delta_1$, $\Delta_3$, $\Sigma$, $\Sigma_1$), and three vectors ($\mathcal{B}$, $\mathcal{W}$, $\mathcal{L}_3$).\footnote{Mediators such as $N$, $\varphi$, and $\mathcal{S}_{1,2}$ are also commonly featured in discrete flavor symmetry models of neutrino masses (see e.g. Refs.~\cite{Ferreira:2013oga, Holthausen:2012wz, Abbas:2020qzc}).} Their gauge quantum numbers, along with the relevant interaction terms are detailed in Tab.~\ref{tab:Intro_Table_Mediators_SMEFT}.

The remainder of the letter is structured as follows. In Section~\ref{sec:A4_symm_group}, we review the key properties of the $A_4$ group, with a focus on its representation theory, and briefly introduce the modular forms. Section~\ref{sec:UV_flavor_invs} outlines the construction of flavor invariants and details the irreducible representations used to define the flavor tensors. In Section~\ref{sec:phenomenology_intro}, we carry out a comprehensive analysis of the phenomenological implications associated with the various $A_4$ flavor irreps. Section~\ref{sec:conc} concludes the paper with a summary of our findings and potential avenues for future work.

\section{$A_4$ symmetry group}
\label{sec:A4_symm_group}

\subsection{Representations and decompositions}
\label{sec:A4_reps_decs}
The $A_4$ group (alternating group on four objects) is a finite, non-Abelian group comprising all even permutations of four elements. With an order of $|A_4|=4!/2=12$, it is the smallest group that admits complex, irreducible three-dimensional representations—a feature central to its utility in flavor physics. Geometrically, $A_4$ is isomorphic to the rotation symmetry group of a regular tetrahedron (excluding reflections), which underscores its role in describing discrete rotational symmetries in three dimensions.

This geometric connection motivates its adoption in particle physics: just as a tetrahedron’s rotations map its vertices onto one another, $A_4$ transformations relate fermion generations, offering a predictive framework for leptonic mixing patterns. Notably, the group’s character table includes a triplet irreducible representation ($\bm3$) alongside three singlets ($\bm1$, $\bm1'$, $\bm1''$), enabling natural embeddings of the Standard Model’s three-generation structure.

The $A_4$ group can be compactly generated by two fundamental elements, $S$ and $T$, which satisfy the presentation rules 
\begin{equation}\label{eq: A4 definition}
    S^2=T^3=(ST)^3=1,
\end{equation}
where $1$ denotes the identity element. These generators encapsulate the group’s structure and provide a convenient basis for constructing its representations. In the geometrical context of the tetrahedral symmetry, $T$ corresponds to a $2\pi/3$ rotation around an axis passing through a vertex and the center of the opposite face, while $S$ represents a $\pi$ rotation around an axis through the midpoints of opposite edges. Together, these transformations generate all 12 rotational symmetries of the tetrahedron. 

In terms of the representation theory, as previously indicated, $A_4$ proves to be the smallest finite non-Abelian group that admits an irreducible triplet representation ($\bm3$). Furthermore, in order to construct the tensor products, explicit forms of the $S$ and $T$ generators need to be specified, which turns out to be especially relevant for the nontrivial irreps.

Let us begin by examining the singlet representations of the $A_4$. The $S$ and $T$ generators take the form
\begin{equation}
    \begin{alignedat}{5}
        \bm{1}&: &\qquad S&=1,&\qquad T&=1,
        \\
        \bm{1}'&: &\qquad S&=1,&\qquad T&=\omega,
        \\
        \bm{1}''&: &\qquad S&=1,&\qquad T&=\omega^2,
    \end{alignedat}
\end{equation}
where $\omega=e^{2\pi i/3}=-1/2+i\sqrt{3}/2$ is the primitive cube root of unity. These representations explicitly satisfy the defining relations of $A_4$ given by Eq.~\eqref{eq: A4 definition}.

The tensor products between these singlet representations exhibit a cyclic structure, which plays a crucial role in constructing invariant terms in the Lagrangian:
\begin{equation}\label{eq:A4_sing_rep_dec}
    \bm1\otimes\bm1=\bm1,\quad \bm1'\otimes\bm1'=\bm1'',\quad \bm1''\otimes\bm1''=\bm1',\quad \bm1'\otimes\bm1''=\bm1.
\end{equation}
This multiplicative structure ensures that products of singlet fields preserve the group symmetry, with the phases $\omega$ and $\omega^2$ governing the interplay between $\bm1'$ and $\bm1''$.\footnote{In forming flavor invariants, we allow contractions of arbitrary $A_4$ irreps, but focus on extracting the trivial singlet component from their decomposition.} Notably, this behavior mirrors the representation theory of discrete cyclic groups, albeit embedded within the larger $A_4$ framework.

To decompose the tensor product of two $A_4$ triplet representations ($\bm3\otimes\bm3$), it is essential to work with explicit matrix forms of the generators $S$ and $T$. A particularly useful choice is the basis where $T$ is diagonal, which simplifies the analysis of representation properties. In this basis, the generators take the following $3\times3$ matrix forms
\begin{equation}\label{eq:A4_triplet_ST}
    S=\frac{1}{3}
    \begin{bmatrix}
        -1&2&2\\2&-1&2\\2&2&-1
    \end{bmatrix},
    \qquad
    T=\begin{bmatrix}
        1&0&0\\0&\omega&0\\0&0&\omega^2
    \end{bmatrix},
\end{equation}
as indicated in, among others, Refs.~\cite{Ishimori:2010au, Kobayashi:2021pav, Feruglio:2008ht, Ding:2019zxk}.
The diagonal form of $T$ reflects its role as a cyclic permutation generator weighted by phases $(1,\omega,\omega^2)$, while $S$ implements a more intricate transformation that mixes all three components of the triplet. This specific basis is computationally advantageous for two reasons: first, the diagonal $T$ matrix immediately reveals the eigenvalues $(1,\omega,\omega^2)$ associated with each component of the triplet and second, the symmetric form of $S$ ensures proper behavior under the $A_4$ group relations given by Eq.~\eqref{eq: A4 definition}.

With the generator representations established, the decomposition reads
\begin{equation}\label{eq:A4_tensor_product}
    \bm3\otimes\bm3=\bm1\oplus\bm1'\oplus\bm1''\oplus\bm{3_S}\oplus\bm{3_A},
\end{equation}
where the subscripts $S$ and $A$ denote the symmetric and antisymmetric combinations respectively. In the basis defined by Eq.~\eqref{eq:A4_triplet_ST} this decomposition becomes more transparent, as the symmetric and antisymmetric combinations of the tensor product can be systematically identified through the action of these generators.

For two generic $A_4$ triplets, $\alpha \equiv (\alpha_1, \alpha_2, \alpha_3)^\intercal \sim \bm{3}$ and $\beta \equiv (\beta_1, \beta_2, \beta_3)^\intercal \sim \bm{3}$, we provide below the explicit expressions for their decomposition into irreducible representations, as required for the construction of $A_4$-invariant flavor structures.
\\\\
\textbf{Singlet representations:}
\begin{itemize}
    \item The trivial singlet $\bm1$:
        \begin{equation}\label{eq:A4_1}
            \cO_{\bm1}\equiv[\alpha\beta]_{\bm1}=\alpha_1\beta_1+\alpha_2\beta_3+\alpha_3\beta_2.
        \end{equation}
    \item The $\omega$-charged singlet $\bm1'$:
        \begin{equation}
            \cO_{\bm1'}\equiv[\alpha\beta]_{\bm1'}=\alpha_3\beta_3+\alpha_1\beta_2+\alpha_2\beta_1.
        \end{equation}
    \item The $\omega^2$-charged singlet $\bm1''$:
        \begin{equation}
            \cO_{\bm1''}\equiv[\alpha\beta]_{\bm1''}=\alpha_2\beta_2+\alpha_1\beta_3+\alpha_3\beta_1.
        \end{equation}
\end{itemize}
\textbf{Triplet representations:}
\begin{itemize}
    \item Symmetric triplet $\bm3_S$:
        \begin{equation}
            \cO_{\bm3_S}\equiv[\alpha\beta]_{\bm3_S}=\frac{1}{3}
        \begin{bmatrix}
            2\alpha_1\beta_1-\alpha_2\beta_3-\alpha_3\beta_2
            \\
            2\alpha_3\beta_3-\alpha_1\beta_2-\alpha_2\beta_1
            \\
            2\alpha_2\beta_2-\alpha_1\beta_3-\alpha_3\beta_1
        \end{bmatrix}.
        \end{equation}
    \item Antisymmetric triplet $\bm3_A$:
        \begin{equation}\label{eq:A4_3A}
            \cO_{\bm3_A}\equiv[\alpha\beta]_{\bm3_A}=\frac{1}{2}
        \begin{bmatrix}
            \alpha_2\beta_3-\alpha_3\beta_2
            \\
            \alpha_1\beta_2-\alpha_2\beta_1
            \\
            \alpha_3\beta_1-\alpha_1\beta_3
        \end{bmatrix}.
        \end{equation}
\end{itemize}
The symmetric triplet components maintain the permutation symmetry $\alpha_i\beta_j+\alpha_j\beta_i$, while the antisymmetric components exhibit the expected sign change under index exchange. Furthermore, the particular numerical coefficients ($-2$ and $1$) in the symmetric triplet ensure orthogonality to the singlet combinations and proper normalization under group transformations.

\subsection{$A_4$ modular forms}

Building on our previous analysis of exact $A_4$ flavor symmetry~\cite{Palavric:2024gvu}, in this letter we extend this framework to account for the flavor symmetry breaking by promoting the exact $A_4$ into modular $A_4$ flavor symmetry in the lepton sector. In case of modular $A_4$ symmetry, the breaking of the original $A_4$ invariance is dynamically controlled by the vacuum expectation value (VEV) of the complex modulus field $\tau$. In contrast to the exact symmetries, modular flavor symmetries offer a compelling mechanism for generating realistic fermion mass hierarchies and mixing patterns, as the Yukawa couplings and symmetry-breaking terms become functions of $\tau$---a field that transforms non-trivially under the modular group $\overline\Gamma\cong\mathrm{PSL}(2,\mathbb Z)=\mathrm{SL}(2,\mathbb Z)/\{\pm\mathbb I\}$. This modular group is a projective special linear group of $2\times2$ matrices with integer entries and unit determinant, where the matrices $\pm M$ are identified. The group $\overline\Gamma$ is generated by two transformations $S$ and $T$, whose action on the complex modulus $\tau\in \mathbb H$ can be written as
\begin{equation}
    S:~~\tau\to -\frac{1}{\tau},\qquad T:~~ \tau\to \tau+1, 
\end{equation}
where $\mathbb H$ denotes the upper half-plane and where $S$ and $T$ satisfy the presentation rules given by Eq.~\eqref{eq: A4 definition}. Equivalently, the modular group $\overline\Gamma$ can be defined as the group of linear fractional transformations $\gamma$ acting on the modulus $\tau$ belonging to the upper half-plane as
\begin{equation}
    \tau\to\gamma\tau=\frac{a\tau+b}{c\tau+d},\qquad\im(\tau)>0.
\end{equation}
where $a,b,c,d\in\mathbb Z$ and $ad-bc=1$. The group of linear transformations defined in this way is then isomorphic to $\mathrm{PSL}(2,\mathbb Z)$. The group admits discrete congruence subgroups $\Gamma(N)\subset\overline\Gamma$, where $N\in\mathbb N$, defined as
\begin{equation}\small
    \Gamma(N)=\lzv \begin{bmatrix}a&b\\c&d\end{bmatrix}\in\mathrm{SL}(2,\mathbb Z),\quad \begin{bmatrix}a&b\\c&d\end{bmatrix}=\begin{bmatrix}1&0\\0&1\end{bmatrix}\,(\mathrm{mod}\,N)  \dzv.
\end{equation}
In the present analysis, we focus on the $N=3$ case, for which the finite modular group $\Gamma_3$, which is defined as the quotient group $\Gamma_3\equiv\overline\Gamma/\Gamma(3)$, turns out to be isomorphic to $A_4$. The modular forms $f_i(\tau)$ of weight $k$ transform as
\begin{equation}\label{eq:mod_trans}
    f_i(\tau)\to (c\tau+d)^k\rho(\gamma)_{ij}f_j(\tau),\qquad \gamma\in\overline\Gamma,
\end{equation}
where $\rho(\gamma)_{ij}$ represents a unitary matrix under $\Gamma_3$.\footnote{Fermion fields $\psi$ with assigned modular weight $k$ transform under a modular transformation $\gamma \in \bar{\Gamma}$ as $\psi \rightarrow (c\tau + d)^{-k} \rho(\gamma)_{ij} \psi_j$. For scalar and vector fields, the transformation follows Eq.~\eqref{eq:mod_trans}.} In the basis where generators $S$ and $T$ take the form given by Eq.~\eqref{eq:A4_triplet_ST}, the Yukawa couplings can be written as $A_4$ triplets of weight $2$ as~\cite{Kobayashi:2021pav,Kobayashi:2023zzc,Ding:2024fsf}
\begin{equation}
    Y_{\bm3}^{(2)}(\tau)=\begin{bmatrix}Y_1(\tau)\\Y_2(\tau)\\Y_3(\tau)\end{bmatrix},
    \qquad
    \bar Y_{\bm3}^{(2)}(\tau)=\begin{bmatrix}Y_1^*(\tau)\\Y_3^*(\tau)\\Y_2^*(\tau)\end{bmatrix},
\end{equation}
where the individual components can be expressed as
\begin{equation}
    \begin{bmatrix}Y_1(\tau) \\Y_2(\tau) \\Y_3(\tau)\end{bmatrix} =
    \begin{bmatrix}
        1 + 12q + 36q^2 + 12q^3 +\ldots
        \\
        -6q^{1/3}(1 + 7q + 8q^2+\ldots)
        \\
        -18q^{2/3}(1 + 2q + 5q^2 + \cdots)
    \end{bmatrix},\quad q=e^{2i\pi\tau}.
\end{equation}
Using these expressions, the higher-weight ($k=4,6,\ldots$) forms can be constructed accordingly~\cite{Ding:2019zxk,Kobayashi:2021pav,Ding:2024fsf}. However, in this analysis we will focus on the triplet forms of lowest weight ($k=2$) and examine their interplay in the context of UV mediators and SMEFT matching relations.\footnote{In the construction of allowed $A_4$ flavor invariants and the classification of irreducible representations at $\mathcal{O}(1)$ and $\cO(Y_{\bm3}^{(2)})$, the modular weight $k$ serves as an additional distinguishing label between different irreps.}

Moreover, under the assumption of $A_4$ (modular) flavor symmetry, we adopt the conventional assignment for the SM fields, where the $SU(2)_L$ lepton doublet transforms as an $A_4$ triplet of weight $k=2$, while the three generations of right-handed lepton singlets are assigned as three distinct $A_4$ singlets of vanishing weights:
\begin{equation}
    \ell\equiv(\ell_1,\ell_2,\ell_3)^\intercal\sim\bm3,\quad (e_1,e_2,e_3)\sim(\bm1,\bm1',\bm1'').
\end{equation}
This representation choice is widely employed in various models involving $A_4$ symmetry group, e.g. as indicated in Refs.~\cite{Kobayashi:2021pav, Feruglio:2008ht, Ding:2019zxk, Abbas:2020qzc, Pramanick:2015qga}.\footnote{Conjugated spinors follow the same arrangement, while $\bar\ell$ and $\bar e$, in case of $\cB$ and $\cW$, are taken with the arrangement outlined in Ref.~\cite{Kobayashi:2021pav}}

As previously indicated, the modulus $\tau$ ($\im(\tau)>0$) parametrizes the breaking of the modular $A_4$ symmetry and the distinct values of $\tau$ imply certain residual flavor symmetry preserved below the modular symmetry breaking scale. These residual symmetries are determined by the stabilizer subgroup of the VEV within the fundamental domain $\mathcal F$ of the modular group $\overline\Gamma$. For the $A_4$ modular group, three key cases can be distinguished:
\begin{itemize}
    \item $\langle\tau\rangle=i\infty$, which preserves $\mathbb Z_3$ subgroup generated by $T$,
    \item $\langle\tau\rangle=i$, which is invariant under $S$, yielding a $\mathbb Z_2$ remnant symmetry,
    \item $\langle\tau\rangle=\omega$, which is invariant under $ST$, generating a $\mathbb Z_3$ subgroup distinct from the $T$-preserving case.
\end{itemize}
In this work, we fix $\tau$ = $i$ as a representative VEV, corresponding to a point of enhanced symmetry in the modular group where the residual symmetry is $\mathbb{Z}_2$. This choice is particularly well-motivated: as a fixed point under the modular transformation $S$, $\tau$ = $i$ lies within the fundamental domain and leads to symmetric, non-trivial modular forms that can generate predictive and structured flavor textures~\cite{Kobayashi:2021pav,Kobayashi:2018scp,Behera:2020lpd}. Furthermore, from a bottom-up perspective, fixing $\tau$ reduces the number of free parameters and offers a minimal but non-trivial benchmark scenario for modular flavor symmetry breaking.\footnote{While we fix $\tau=i$ in our main analysis, small deviations around this point, such as $\tau=i+\epsilon$ with $|\epsilon|\ll1$, can be considered to introduce controlled breaking of the residual $\mathbb Z_2$ symmetry. This allows for additional structure in the flavor textures while retaining the predictive power of the modular framework. Such perturbations may be particularly relevant when extending the analysis beyond the leptonic sector, e.g., to accommodate the observed hierarchies and mixings in the quark sector, which typically require more pronounced breaking of flavor symmetries. Exploring the phenomenological impact of $\tau$ deviations offers a promising direction for future work~\cite{Petcov:2022fjf,Petcov:2023vws,Yao:2020qyy,Nomura:2021yjb,Okada:2019uoy}.} In the context of our EFT setup, this allows for a controlled classification of UV mediator couplings and a well-defined pattern of flavor symmetry breaking relevant for the matching to SMEFT operators.

\section{UV $A_4$ flavor invariants}
\label{sec:UV_flavor_invs}

Leveraging the theoretical framework outlined in Sec.~\ref{sec:A4_symm_group}, we now systematically analyze the flavor invariants that can be constructed for 13 UV mediators presented in Tab.~\ref{tab:Intro_Table_Mediators_SMEFT}. The flavor invariants are built from contractions of the mediator representations with the modular forms and SM fields, subject to $A_4$ covariance~\cite{Greljo:2023adz,Palavric:2024gvu}.

Using the explicit relations for the irreducible components given by Eqs.~\eqref{eq:A4_1}--\eqref{eq:A4_3A}, the flavor invariants can be written in the expanded form, which makes it possible to readily extract the UV flavor tensors. In the analysis that follows, we consider flavor invariants of $\mathcal{O}(1)$ and $\mathcal{O}(Y_{\bm{3}}^{(2)})$, which yield flavor tensors containing at most one insertion of modular forms. Once integrated out at tree level, the latter yields two insertions of modular forms at the level of SMEFT operators.\footnote{In the case of scalar and fermionic mediators, whose one-loop matching contributions are computed in the phenomenological analysis, the resulting expressions for the Wilson coefficients can involve up to four insertions of modular forms.} Below we provide some more technical details regarding the extraction of flavor invariants for scalar, fermion and vector mediators separately.

\subsubsection{Scalars}

Among the four scalar mediators included in Tab.~\ref{tab:Intro_Table_Mediators_SMEFT}, three of them, more specifically $\cS_1$, $\cS_2$ and $\Xi_1$, couple to the SM fermion  bilinears of the form $\bar f^c f$, where $f\in\{\ell,e\}$ represents the lepton fields. For these mediators, the coupling structure of this type imposes significant constraints on the allowed flavor tensor structures, which can be determined using the transformation properties under charge conjugation as well as their gauge contractions.\footnote{For these purposes, the fundamental identity of the form $\bar\psi\chi=\bar\chi^c\psi^c$ is employed, where $\psi^c$ denotes conjugated spinor.} Performing this analysis reveals that the $\cS_1$ mediator necessarily requires an antisymmetric flavor tensor, while both $\cS_2$ and $\Xi_1$ must possess symmetric ones. This structural constraint then directly determines the allowed spectrum of $A_4$ irreducible representations. In contrast to these three mediators, the remaining $\varphi$ scalar has no such constraints as it couples to the bilinear of the form $\bar\ell \,e$. Having established the symmetry properties of the scalar mediator couplings, we now systematically examine the allowed $A_4$ flavor invariants.

$\bm{\cS_1}$. Starting with $\cS_1$ mediator, at $\cO(1)$ only a single flavor invariant is permitted, necessitating that $\cS_1$ transforms as an $A_4$ triplet:
\begin{equation}
    \cI\sim y_{\cS_1}^{(1)}[\cS_1^\dag]_{\bm3}[\bar\ell\,\ell^c]_{\bm3_A},
\end{equation}
which, once expanded using the tensor product decomposition rules outlined in Sec.~\ref{sec:A4_reps_decs}, yields a flavor tensor structure of the form
\begin{equation}
    \begin{alignedat}{2}
        [y_{\cS_1}]_{rij}&=\frac{y_{\cS_1}^{(1)}}{2}\Big[ \delta_{r1}\lzm \delta_{i2}\delta_{j3}-\delta_{i3}\delta_{j2} \dzm+\delta_{r2}\lzm \delta_{i3}\delta_{j1}-\delta_{i1}\delta_{j3} \dzm
        \\&
        ~~~~~~~\,+\delta_{r3}\lzm \delta_{i1}\delta_{j2}-\delta_{i2}\delta_{j1} \dzm\Big],
    \end{alignedat}
\end{equation}
which is manifestly antisymmetric under the exchange of $i$ and $j$ indices. The same methodology in extracting the flavor tensors from the invariants applies systematically to all remaining irreps. To maintain conciseness and avoid redundancy, we suppress the explicit tensor expressions hereafter. At $\cO(Y_{\bm3}^{(2)})$, five independent flavor invariants can be formed. Three of these correspond to cases where $\cS_1$ transforms as one of the $A_4$ singlets ($\bm1$, $\bm1'$ or $\bm1''$), while the remaining two occur when $\cS_1$ transforms as a triplet. Notably, in the triplet case, the two invariants appear with distinct free parameters, resulting in a flavor tensor that depends on two independent coefficients.

$\bm{\cS_2}$. At $\cO(1)$, the $\cS_2$ mediator may transform under any of the $A_4$ singlet representations. Each of these singlet representations permits three possible flavor invariants, however, due to the symmetric nature of the flavor tensor, only two of these are linearly independent. As a concrete example, for the case where $\cS_2\sim\bm1$, the allowed $A_4$ invariants are
\begin{equation}
    \begin{alignedat}{2}
        \cI\sim  & \,y_{\cS_2}^{(1)}[\cS_2^\dag]_{\bm1}[\bar e]_{\bm1}[e^c]_{\bm1}\oplus y_{\cS_2}^{(2)}[\cS_2^\dag]_{\bm1}[\bar e]_{\bm1'}[e^c]_{\bm1''}
        \\&\oplus 
        y_{\cS_2}^{(3)}[\cS_2^\dag]_{\bm1}[\bar e]_{\bm1''}[e^c]_{\bm1'}\,,
    \end{alignedat}
\end{equation}
where the symmetry requirement for the flavor tensor enforces the relation $y_{\cS_2}^{(2)}=y_{\cS_2}^{(3)}$, ultimately yielding two independent parameters in the flavor tensor. Given that $\cS_2$ couples exclusively to right-handed electron bilinears, which transform as $A_4$ singlets (see Eq.~\ref{eq:A4_sing_rep_dec}), in order to form invariants at $\cO(Y_{\bm3}^{(2)})$, $\cS_2$ necessarily has to transform as a triplet. This construction produces nine flavor invariants, of which six remain independent after accounting for symmetry constraints.

$\bm{\varphi}$. At $\cO(1)$ the bilinear $\bar\ell\, e$ requires $\varphi$ to transform as an $A_4$ triplet. This results in three independent flavor invariants, which directly correspond to three distinct parameters in the associated flavor tensor. At $\cO(Y_{\bm3}^{(2)})$, $\varphi$ may transform under any of the $A_4$ singlet representations. For each singlet irrep, the construction admits three independent flavor invariants, corresponding to three distinct parameters in each case. Finally, the triplet representation is also permitted, yielding six linearly independent flavor invariants.

$\bm{\Xi_1}$. Owing to its symmetric flavor tensor structure, the $\Xi_1$ scalar mediator can transform under any representation of $A_4$--- both at $\cO(1)$ and $\cO(Y_{\bm3}^{(2)})$. In all scenarios except the triplet representation at $\cO(Y_{\bm3}^{(2)})$, the construction permits only a single flavor invariant leading to a one-parameter flavor tensor structure. Conversely, the triplet representation at $\cO(Y_{\bm3}^{(2)})$ accommodates five independent invariants. 

A comprehensive summary of scalar representations, accompanied by their permissible flavor invariants and their associated parameter counts, is given in Tab.~\ref{tab:scalars_a4_irreps_invs}.

\subsubsection{Fermions}

We now proceed with the analysis of the fermionic mediators. The classification of the $A_4$ irreps along with the extraction of the flavor tensors proceeds analogously to the scalar cases. In all cases presented in Tab.~\ref{tab:Intro_Table_Mediators_SMEFT}, the NP fermion couples to a single SM lepton, through the bilinear of the form $\bar F \Phi^{(\dag)} f$, where $f\in\{\ell,e\}$ and $\Phi\in\{\phi,\tilde\phi\}$, which implies that at $\cO(1)$ the NP fermions will directly inherit their flavor transformation properties from the SM leptons they couple to. At $\cO(Y_{\bm3}^{(2)})$ the space of allowed invariants expands, with several new invariant structures emerging. The flavor invariants for different new physics fermions naturally separate into two distinct classes, depending on whether the mediator couples to $\ell$ ($N$, $E$, $\Sigma$ and $\Sigma_1$) or $e$ ($\Delta_1$ and $\Delta_3$). We address each of these classes separately.

$\bm{N,E,\Sigma,\Sigma_1}$. As established previously, these four NP fermions couple to $\ell$, which implies that they necessarily transform as $A_4$ triplets at $\cO(1)$. The corresponding flavor invariants can then be written as
\begin{equation}
    \cI\sim \lambda_F^{(1)}[\bar F_R\,\ell]_{\bm1},\qquad F\in\{N,E,\Sigma,\Sigma_1\}.
\end{equation}
At $\cO(Y_{\bm3}^{(2)})$, the NP fermions can transform under any irreducible representation of $A_4$. This follows directly from the tensor product decomposition of the Yukawa modular forms and $\ell$, given by Eq.~\eqref{eq:A4_tensor_product}. Consequently, flavor invariants constructed with singlet representations depend on a single free parameter, while the triplet representation generates two independent invariants with distinct coefficients.

$\bm{\Delta_1,\Delta_3}$. Since both mediators couple exclusively to the singlet field $e$, at $\cO(1)$ they need to transform as $A_4$ singlets accordingly, resulting in flavor invariants with a single independent parameter. At $\cO(Y_{\bm3}^{(2)})$, however, these mediators necessarily adopt $A_4$ triplet representations, which permit three independent flavor invariants in each case.

Tab.~\ref{tab:fermions_a4_irreps_invs} provides a complete classification of the allowed $A_4$ representations and corresponding flavor invariants for all six new physics fermion fields.

\subsubsection{Vectors}

Finally, we examine the three vector mediators enumerated in Tab.~\ref{tab:Intro_Table_Mediators_SMEFT}. Two of them ($\cB$ and $\cW$) couple to the vector bilinears of the form $\bar f\gamma^\mu f$, where $f\in\{\ell,e\}$, resulting in a rich structure once $A_4$ symmetry is imposed, with numerous independent $A_4$ invariants permitted. In contrast, $\cL_3$ mediator couples to the $\bar e^c\gamma^\mu \ell$ bilinear, mirroring the coupling structure of the $\varphi$ scalar, and consequently generating analogous flavor invariants. We provide more details for each vector mediator in the discussion below.

$\bm{\cB}$. Among the three vector mediators considered, $\cB$ stands out as the only one for which two interaction terms are allowed in the UV. Owing to this coupling structure, the $\cB$ mediator possesses particularly rich structure, where both singlet and triplet irreducible representations of $A_4$ are allowed at both orders. At $\cO(1)$, each of the three $A_4$ singlet irreps permits the formation of four independent flavor invariants. One of these arises from the interaction with the $\bar\ell\gamma^\mu\ell$ bilinear, where a singlet component emerges through the decomposition of two triplets. The remaining three invariants originate from the coupling to the $\bar e\gamma^\mu e$ bilinear. In contrast, for the triplet case, only two independent flavor invariants can be constructed, both of which originate exclusively from the $\ell\gamma^\mu\ell$ coupling. At $\cO(Y_{\bm3}^{(2)})$, all three $A_4$ singlet irreps continue to be permitted. Each of these representations admits two independent flavor invariants, both of which originate from the interaction with the $\ell\gamma^\mu\ell$ bilinear. Lastly, the triplet irrep at $\cO(Y_{\bm3}^{(2)})$ exhibits a particularly rich structure, allowing for the construction of 16 independent flavor invariants.

$\bm{\cW}$. As illustrated in Tab.~\ref{tab:Intro_Table_Mediators_SMEFT}, this mediator interacts solely with the bilinear $\bar\ell\gamma^\mu\ell$. As a result, the flavor invariants at $\cO(1)$ and $\cO(Y_{\bm3}^{(2)})$ follow directly from those associated with the $\cB$ mediator. Due to the absence of coupling to the $\bar e\gamma^\mu e$ bilinear, the enumeration of independent invariants is modified accordingly. At $\cO(1)$, the singlet irreps allow for only one invariant, while the triplet irrep admits two, analogously to the $\cB$ scenario. At $\cO(Y_{\bm3}^{(2)})$, the only deviation arises in the counting of flavor invariants for the triplet irrep, where we now obtain seven independent structures, in contrast to the 16 present in the $\cB$ case.

$\bm{\cL_3}$. As indicated earlier, this mediator interacts with a structure analogous to that of $\varphi$, leading to an identical pattern of flavor invariants. Consequently, at $\cO(1)$, $\cL_3$ is restricted to transforming as a triplet, while at $\cO(Y_{\bm3}^{(2)})$, all $A_4$ irreducible representations are permitted. In addition, the enumeration of independent flavor invariants remains unchanged.

A complete summary of the different irreps and their flavor invariants is presented in Tab.~\ref{tab:vectors_a4_irreps_invs}.

\section{Phenomenology}
\label{sec:phenomenology_intro}
Building on the discussion of flavor irreps and invariants in Sec.~\ref{sec:UV_flavor_invs}, this section aims to employ the extracted flavor tensors to perform the SMEFT matching for various mediators and to investigate the resulting phenomenological implications. A key objective is to use the obtained SMEFT matching relations, in conjunction with various observables, to establish constraints on the characteristic scale of different mediators.

In order to extract the bounds on the scale for different $A_4$ flavor irreps of the UV mediators considered, we utilize a diverse set of observables, including the combined low-energy fit as well as $|\Delta L_{\alpha}|=1$ and $|\Delta L_{\alpha}|=2$ transitions. Furthermore, building on the phenomenological analysis presented in Ref.~\cite{Palavric:2024gvu}, we extend the study to include one-loop matching contributions to the SMEFT coefficients for scalar and fermion mediators, assessing their phenomenological implications. In contrast, for vector mediators, due to the absence of explicit UV models serving as their origin, we analyze the effects induced by renormalization group equations (RGEs) within the leading-logarithm approximation. 

This section is structured as follows: we first provide an overview of the observables relevant to the upcoming analysis, followed by a dedicated examination of each class of mediators. We highlight the most significant constraints both at tree level and when one-loop effects are taken into account.

\subsection{Overview of the observables}
As previously mentioned, the phenomenological analysis encompasses two categories of observables: lepton-flavor-conserving low-energy observables, for which we employ the combined fit, and charged lepton-flavor violating (cLFV) observables, where we explicitly differentiate between $|\Delta L_\alpha|=1$ and $|\Delta L_\alpha|=2$ processes. A brief discussion of each of these groups is presented below.

Beginning with low-energy observables, we determine constraints on the mass scales of UV mediators by utilizing the fits presented in Refs.~\cite{Falkowski:2017pss, Breso-Pla:2023tnz, Falkowski:2015krw}. Since the tree-level SMEFT matching relations involve operators with either four lepton fields (scalars and vectors) or two lepton fields (fermions), the most relevant datasets include lepton pair production in $e^+e^-$ collisions~\cite{ALEPH:2013dgf, VENUS:1997cjg}, $W/Z$ pole observables~\cite{Falkowski:2019hvp, Efrati:2015eaa}, as well as $\nu_\mu e$ scattering~\cite{ParticleDataGroup:2016lqr}. Further constraints are provided by measurements of $\tau$ polarization~\cite{VENUS:1997cjg}, low-energy parity-violating $e^+e^-$ scattering, and neutrino trident production~\cite{CHARM-II:1990dvf, CCFR:1991lpl, Altmannshofer:2014pba}.

As emphasized in Ref.~\cite{Palavric:2024gvu}, an important consequence of enforcing discrete flavor symmetries at the level of UV Lagrangians governing the interactions of NP mediators with SM fields, followed by their integration at tree level, is the generation of SMEFT operators in the matching relations that contribute to $|\Delta L_\alpha|=1$ and $|\Delta L_\alpha|=2$ processes, where $\alpha=e,\mu,\tau$. This behavior stands in contrast to scenarios involving continuous flavor symmetries, where such operators would typically arise with strong flavor suppression~\cite{Greljo:2022cah}. In the case of discrete flavor symmetries, however, these operators appear at the same order in the flavor power counting, thereby introducing a relevant phenomenological sector that provides a complementary direction for constraining the relevant NP irreps~\cite{Heeck:2024uiz, Heeck:2016xwg, Conlin:2020veq, Banerjee:2022vdd, Bigaran:2022giz, Crivellin:2013hpa, Hayasaka:2010np,Calibbi:2017uvl}.

\begin{table}[t]
\centering
\scalebox{1}{
\begin{tabular}{c@{\hspace{0.5cm}}cc@{\hspace{0.5cm}}c}
\toprule
\multirow{1}{*}{\textbf{cLFV obs.}}
&\multirow{1}{*}{\textbf{90\% CL}}
&\textbf{Exp.}
&\multirow{1}{*}{\textbf{Ref.}} 
\\
\midrule
$\mathrm{BR}(\mu\to e\gamma)$&$4.2\times10^{\eminus13}$&MEG (2016)&\cite{MEG:2016leq}
\\
$\mathrm{BR}(\tau\to e\gamma)$&$3.3\times10^{\eminus8}$&BaBar (2010)&\cite{BaBar:2009hkt}
\\
$\mathrm{BR}(\tau\to \mu\gamma)$&$4.2\times10^{\eminus8}$&Belle (2021)&\cite{Belle:2021ysv}
\\
\midrule
$\mathrm{BR}(\mu\to ee\bar e)$&$1.0\times10^{\eminus12}$&SINDRUM (1988)&\cite{SINDRUM:1987nra}
\\
$\mathrm{BR}(\tau\to ee\bar e)$&$2.7\times10^{\eminus8}$&Belle (2010)&\cite{Hayasaka:2010np}
\\
$\mathrm{BR}(\tau\to e\mu\bar\mu)$&$2.7\times10^{\eminus8}$&Belle (2010)&\cite{Hayasaka:2010np}
\\
$\mathrm{BR}(\tau\to \mu\mu\bar\mu)$&$2.1\times10^{\eminus8}$&Belle (2010)&\cite{Hayasaka:2010np}
\\
$\mathrm{BR}(\tau\to \mu e\bar e)$&$1.8\times10^{\eminus8}$&Belle (2010)&\cite{Hayasaka:2010np}
\\
\midrule
$\mathrm{CR}(\mu\to e,\mathrm{Au})$&$7.0\times10^{\eminus13}$&SINDRUM II (2006)&\cite{SINDRUMII:2006dvw}
\\
\bottomrule
\end{tabular}
}
\caption{Overview of the present upper bounds for the $|\Delta L_\alpha|=1$ cLFV transitions considered in our analysis. Further details and analytical expressions for these observables in terms of SMEFT operators are provided in Refs.~\cite{Calibbi:2021pyh,Crivellin:2013hpa}.}
\label{tab:overview_deltaL1_obs}
\end{table}

The overview of $|\Delta L_\alpha|=1$ observables is presented in Tab.~\ref{tab:overview_deltaL1_obs}. These observables fall into three main categories: radiative decays, lepton three-body decays, and muon-to-electron conversion in gold. In context of the UV mediators considered in this analysis, three-body decays play a significant role at tree level for scalar and vector irreps, as they directly generate four-fermion SMEFT operators at tree level. Similarly, for fermionic mediators, muon-to-electron conversion is particularly relevant, since the tree-level matching of fermions onto SMEFT induces operators of the type $\cO_{\phi f}$, which contribute to this process. Moreover, 
as indicated in App.~A of Ref.~\cite{Calibbi:2021pyh}, the radiative decays are induced by the SMEFT operators of the form $\cO_{e(B,W)}$, which arise at the one-loop level for scalar and fermion mediators. Consequently, the radiative decay $\mu\to e\gamma$ is particularly relevant, given its precise experimental measurement~\cite{Davidson:2020hkf,Ardu:2024bua,Crivellin:2013hpa,Pruna:2014asa,Uesaka:2024tfn}.

\begin{table}[t]
\centering
\scalebox{0.76}{
\begin{tabular}{ccc@{\hspace{0.5cm}}c}
\toprule
\multirow{1}{*}{\textbf{WC}}
&\multirow{1}{*}{\textbf{90\% CL}}
&\textbf{Exp.}
&\multirow{1}{*}{\textbf{Ref.}} 
\\
\midrule
$|[\cC_{\ell\ell}]_{2121}+[\cC_{ee}]_{2121}|$&$(3.2\,\tev)^{\eminus2}$&$\mathrm{Mu}-\mathrm{to}-\overline{\mathrm{Mu}}$&\cite{Willmann:1998gd}
\\[3pt]
$\lzu [\cC_{\ell e}]_{2121} \dzu$&$(3.8\,\tev)^{\eminus2}$&$\mathrm{Mu}-\mathrm{to}-\overline{\mathrm{Mu}}$&\cite{Willmann:1998gd}
\\
\midrule
$|[\cC_{\ell\ell}]_{2131}|$, $|[\cC_{\ell e}]_{2131}|$, $|[\cC_{\ell e}]_{3121}|$, $|[\cC_{ee}]_{2131}|$&$(10.0\,\tev)^{\eminus2}$&$\tau\to \bar\mu ee$&\cite{Hayasaka:2010np}
\\[3pt]
$|[\cC_{\ell\ell}]_{1232}|$, $|[\cC_{\ell e}]_{1232}|$, $|[\cC_{\ell e}]_{3212}|$, $|[\cC_{ee}]_{1232}|$&$(8.8\,\tev)^{\eminus2}$&$\tau\to \bar e\mu\mu $&\cite{Hayasaka:2010np}
\\
\bottomrule
\end{tabular}
}
\caption{Overview of the most stringent constraints on the relevant (combinations of) Wilson coefficients, derived from $|\Delta L_\alpha|=2$ cLFV transitions. All four-fermion SMEFT operators listed in this table follow standard definitions, as outlined in Ref.~\cite{Grzadkowski:2010es}.}
\label{tab:overview_deltaL2_obs}
\end{table}

Furthermore, in Tab.~\ref{tab:overview_deltaL2_obs} we summarize the most stringent constraints on specific Wilson coefficients and their combinations arising from $|\Delta L_\alpha|=2$ transitions. While Ref.~\cite{Heeck:2024uiz} provides a more extensive list of experimental measurements related to this type of cLFV transitions, our analysis focuses only on those that offer the most stringent bounds. This selection allows for a direct comparison across different sectors and facilitates the determination of the most restrictive upper limit on the scale of a given $A_4$ irrep, rendering weaker constraints unnecessary for our purposes. Analogous to three-body lepton decay modes, the $|\Delta L_\alpha|=2$ transitions summarized in Tab.~\ref{tab:overview_deltaL2_obs} primarily constrain scalar and vector irreps, as they generate four-fermion SMEFT operators at tree level. However, the presence of more stringent $|\Delta L_\alpha|=1$ constraints significantly limits the impact of $|\Delta L_\alpha|=2$ transitions in the one-loop analysis, rendering them non-competitive in comparison.

\subsection{Phenomenology of UV mediators}
The subsections that follow provide a detailed examination of each class of UV mediators, focusing on the constraints imposed by the observables discussed above. We adopt a systematic strategy, first addressing tree-level constraints for all $A_4$ irreps. In cases where an irrep supports only a single flavor invariant, we can directly extract a bound on the mass-to-coupling ratio. In cases where multiple independent parameters exist within the flavor tensor, we proceed under the assumption that all couplings are $\sim1$. This methodology extends to our analysis of loop-generated constraints accordingly.\footnote{The one-loop matching relations were obtained using \texttt{Matchete}~\cite{Fuentes-Martin:2022jrf}, and independently cross-checked against the results of Ref.~\cite{Gargalionis:2024jaw}.} Finally, for irreps that permit two independent flavor invariants, we conduct a more detailed examination, enabling a comparative two-dimensional study of different observables.

\subsubsection{Scalars}
\label{sec:scalars_pheno}
\begin{table}[t]
\centering
\scalebox{0.65}{
\begin{tabular}{cccccccc}
\toprule
\textbf{Field}
&\textbf{$\bm{(A_4,k)}$}
&\textbf{Order} 
&\textbf{LE}
&$\bm{|\Delta L_\alpha|=1}$ 
&\textbf{Obs.}
&$\bm{|\Delta L_\alpha|=2}$
&\textbf{Obs.}
\\
\midrule
\multirow{5}{*}{\vspace{-0.3cm}$\cS_1\sim(\bm1,\bm1)_1$}
&$(\bm3,4)$
&$\cO(1)$
&{\color{customcolor}{5.20}}
&-
&-&-&-
\\
&$(\bm1,2)$&$\cO(\bar Y_{\bm3}^{(2)})$&3.93&{\color{customcolor}{5.46}}&$\tau\to e\mu\bar\mu $&-&-
\\
&$(\bm1',2)$&$\cO(\bar Y_{\bm3}^{(2)})$&1.74&{\color{customcolor}{3.66}}&$\tau\to\mu e\bar e$&-&-
\\
&$(\bm1'',2)$&$\cO(\bar Y^{(2)}_{\bm3})$&5.28&{\color{customcolor}{6.04}}&$\tau\to\mu e\bar e$&-&-
\\
&$(\bm3,2)$&$\cO(\bar Y_{\bm3}^{(2)})$&3.43&{\color{customcolor}{4.36}}&$\tau\to e\mu\bar\mu $&-&-
\\
\midrule
\multirow{4}{*}{$\cS_2\sim(\bm1,\bm1)_2$}
&$(\bm1,0)$
&$\cO(1)$
&1.78
&{\color{customcolor}{12.16}}
&$\tau\to ee\bar\mu$
&10.00
&$\tau \to\bar\mu ee$
\\
&$(\bm1',0)$
&$\cO(1)$
&3.56
&{\color{customcolor}{11.79}}
&$\tau\to\mu\mu\bar e$
&8.80
&$\tau\to \bar e\mu\mu$
\\
&$(\bm1'',0)$
&$\cO(1)$
&5.21
&-
&-
&-
&-
\\
&$(\bm3,-2)$
&$\cO(\bar Y_{\bm3}^{(2)})$
&6.63
&{\color{customcolor}{33.79}}
&$\mu\to ee \bar e$
&12.97
&$\tau\to \bar\mu e e$
\\
\midrule
\multirow{5}{*}{\vspace{-0.3cm}$\varphi\sim(\bm1,\bm2)_{1/2}$}
&$(\bm3,-2)$
&$\cO(1)$
&{\color{customcolor}{4.43}}
&-
&-
&2.69
&$\mathrm{Mu}-\mathrm{to}-\overline{\mathrm{Mu}}$
\\
&$(\bm1,0)$
&$\cO(\bar Y_{\bm3}^{(2)})$
&4.58
&{\color{customcolor}{22.80}}
&$\mu\to ee\bar e$
&7.70
&$\tau\to \bar e \mu\mu $
\\
&$(\bm1',0)$
&$\cO(\bar Y_{\bm3}^{(2)})$
&3.18
&{\color{customcolor}{11.80}}
&$\mu\to ee\bar e$
&5.29
&$\tau\to \bar\mu e e$
\\
&$(\bm1'',0)$
&$\cO(\bar Y_{\bm3}^{(2)})$
&2.42
&{\color{customcolor}{12.91}}
&$\mu\to ee\bar e$
&10.23
&$\tau\to \bar\mu e e$
\\
&$(\bm3,0)$
&$\cO(\bar Y_{\bm3}^{(2)})$
&3.31
&5.29
&$\tau\to ee\bar e$
&{\color{customcolor}{6.98}}
&$\tau\to \bar\mu e e$
\\
\midrule
\multirow{8}{*}{\vspace{-0.3cm}$\Xi_1\sim(\bm1,\bm3)_1$}
&$(\bm1,4)$
&$\cO(1)$
&[2.15,\,4.95]
&{\color{customcolor}{17.19}}
&$\tau\to ee\bar\mu $
&14.14
&$\tau\to \bar\mu ee$
\\
&$(\bm1',4)$
&$\cO(1)$
&5.84
&{\color{customcolor}{16.67}}
&$\tau\to \mu\mu \bar e$
&12.45
&$\tau\to \bar e\mu\mu$
\\
&$(\bm1'',4)$
&$\cO(1)$
&{\color{customcolor}{14.15}}
&-
&-
&-
&-
\\
&$(\bm3,4)$
&$\cO(1)$
&3.95
&{\color{customcolor}{8.11}}
&$\tau\to e e\bar\mu$
&6.67
&$\tau\to \bar\mu e e$
\\
&$(\bm1,2)$
&$\cO(\bar Y^{(2)}_{\bm3})$
&2.69
&{\color{customcolor}{21.50}}
&$\mu\to ee\bar e$
&6.82
&$\tau\to \bar\mu ee$
\\
&$(\bm1',2)$
&$\cO(\bar Y^{(2)}_{\bm3})$
&1.69
&{\color{customcolor}{11.13}}
&$\mu\to ee\bar e$
&6.00
&$\tau\to \bar e\mu\mu$
\\
&$(\bm1'',2)$
&$\cO(\bar Y^{(2)}_{\bm3})$
&4.77
&{\color{customcolor}{13.01}}
&$\mu\to ee\bar e$
&4.39
&$\tau\to \bar e \mu\mu $
\\
&$(\bm3,2)$
&$\cO(\bar Y^{(2)}_{\bm3})$
&18.24
&{\color{customcolor}{60.63}}
&$\mu \to ee\bar e$
&19.00
&$\tau\to\bar\mu ee$
\\
\bottomrule
\end{tabular}
}
\caption{Tree-level bounds on the various $A_4$ scalar flavor irreps. The first column lists the UV scalar fields along with their charges under the SM gauge group while the second column specifies their corresponding $A_4$ irreps.  The third column indicates whether the flavor tensor includes an insertion of modular forms. The fourth column compiles the bounds obtained from the low-energy (LE) combined fit of lepton-flavor conserving observables. The fifth and sixth columns report constraints from $|\Delta L_\alpha|=1$ transitions and the associated observables, respectively. Finally, the last two columns summarize the bounds derived from $|\Delta L_\alpha|=2$ processes. The bounds in {\color{customcolor}{purple}} denote the best tree-level bound for each irrep. All bounds are expressed in TeV.}
\label{tab:Scalars_treelevel_bounds}
\end{table}

\begin{figure*}[t]
\centering
\begin{tabular}{ccc}
\includegraphics[width=60mm]{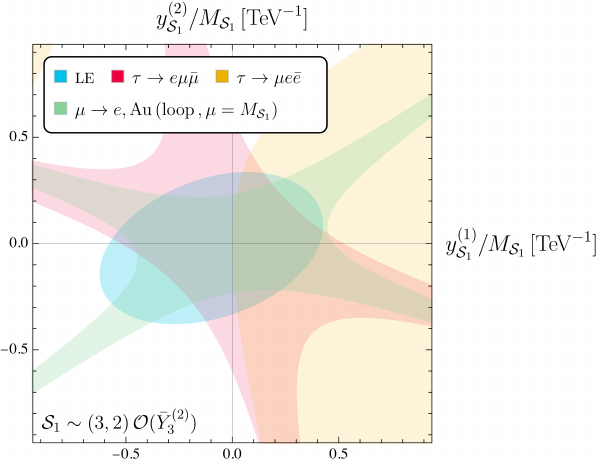} 
&   
\includegraphics[width=60mm]{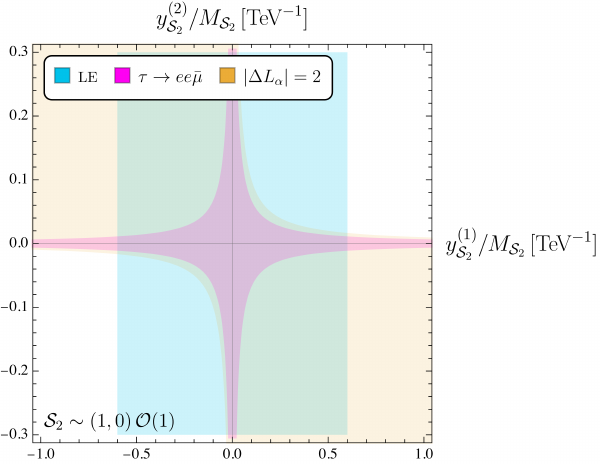}  
&
\includegraphics[width=60mm]{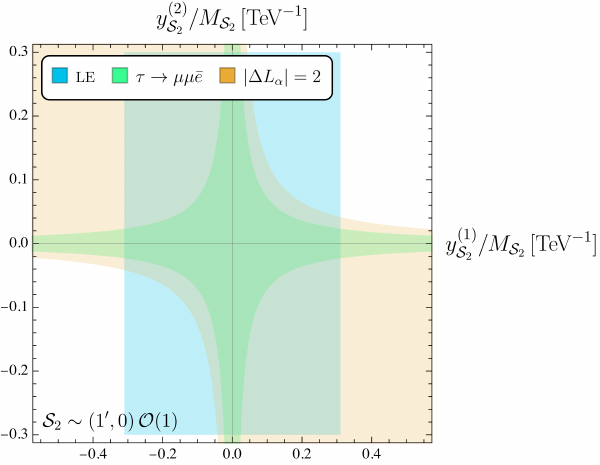}
\end{tabular}
\caption{Complementary overview of the constraints on three scalar irreps leading to two independent parameters in the corresponding flavor tensors. Each plot shows the parameter space regions constrained by different observables, including the low-energy combined fit, $|\Delta L_\alpha|=1$, $|\Delta L_\alpha|=2$, as well as one-loop matching effects. See Sec.~\ref{sec:scalars_pheno} for more details.}
\label{fig:scalars_2D_bounds}
\end{figure*}

\begin{table}[t]
\centering
\scalebox{0.61}{
\begin{tabular}{cccccccc}
\toprule
\multirow{2}{*}{\textbf{Field}}
&\multirow{2}{*}{\textbf{$\bm{(A_4,k)}$}}
&\multirow{2}{*}{\textbf{Order}} 
&\multirow{2}{*}{$\bm{\mu\to e\gamma} $}
&$\bm{\mathrm{CR}(\mu\to e,\mathrm{Au})}$ 
&$\bm{\mathrm{CR}(\mu\to e,\mathrm{Au})}$ 
&$\bm{\mu\to ee\bar e}$ 
&$\bm{\mu\to ee\bar e}$ 
\\
&
&
&
&$\bm{\bar\mu=M_{\sscript{NP}}}$
&$\bm{\bar\mu=M_{Z}}$
&$\bm{\bar\mu=M_{\sscript{NP}}}$
&$\bm{\bar\mu=M_{Z}}$
\\
\midrule
\multirow{4}{*}{\vspace{-0.25cm}$\cS_1\sim(\bm1,\bm1)_1$}
&$(\bm1,2)$
&$\cO(\bar Y_{\bm3}^{(2)})$
&1.96
&{\color{customcolor}{2.72}}
&{\color{black}{15.65}}
&0.67
&0.68
\\
&$(\bm1',2)$
&$\cO(\bar Y_{\bm3}^{(2)})$
&3.96
&{\color{customcolor}{4.50}}
&{\color{black}{27.38}}
&1.42
&1.43
\\
&$(\bm1'',2)$
&$\cO(\bar Y^{(2)}_{\bm3})$
&{\color{black}{1.68}}
&{\color{customcolor}{2.33}}
&{\color{black}{13.13}}
&0.87
&0.87
\\
&$(\bm3,2)$
&$\cO(\bar Y_{\bm3}^{(2)})$
&0.74
&{\color{customcolor}{1.01}}
&{\color{black}{0.08}}
&0.36
&0.36
\\
\midrule
\multirow{1}{*}{$\cS_2\sim(\bm1,\bm1)_2$}
&$(\bm3,-2)$
&$\cO(\bar Y_{\bm3}^{(2)})$
&23.40
&14.11
&90.25
&{\color{customcolor}{24.48}}
&{\color{black}{58.21}}
\\
\midrule
\multirow{5}{*}{\vspace{-0.45cm}$\varphi\sim(\bm1,\bm2)_{1/2}$}
&$(\bm3,-2)$
&$\cO(1)$
&{\color{customcolor}{61.90}}
&57.54
&57.54
&-&-
\\
&$(\bm1,0)$
&$\cO(\bar Y_{\bm3}^{(2)})$
&{\color{customcolor}{47.09}}
&43.23
&{\color{black}{49.41}}
&22.40
&25.26
\\
&$(\bm1',0)$
&$\cO(\bar Y_{\bm3}^{(2)})$
&{\color{customcolor}{15.76}}
&12.99
&{\color{black}{27.05}}
&11.97
&13.37
\\
&$(\bm1'',0)$
&$\cO(\bar Y_{\bm3}^{(2)})$
&{\color{customcolor}{31.39}}
&29.82
&32.84
&12.69
&14.15
\\
&$(\bm3,0)$
&$\cO(\bar Y_{\bm3}^{(2)})$
&{\color{customcolor}{38.64}}
&{\color{black}{36.82}}
&25.94
&6.97
&7.05
\\
\midrule
\multirow{4}{*}{\vspace{-0.3cm}$\Xi_1\sim(\bm1,\bm3)_1$}
&$(\bm1,2)$
&$\cO(\bar Y^{(2)}_{\bm3})$
&6.72
&7.29
&{\color{black}{30.43}}
&{\color{customcolor}{21.24}}
&25.90
\\
&$(\bm1',2)$
&$\cO(\bar Y^{(2)}_{\bm3})$
&5.75
&7.09
&{\color{black}{24.75}}
&{\color{customcolor}{10.90}}
&13.02
\\
&$(\bm1'',2)$
&$\cO(\bar Y^{(2)}_{\bm3})$
&11.10
&{\color{black}{12.05}}
&{\color{black}{50.52}}
&{\color{customcolor}{12.55}}
&15.06
\\
&$(\bm3,2)$
&$\cO(\bar Y^{(2)}_{\bm3})$
&38.88
&{\color{customcolor}{51.59}}
&{\color{black}{196.98}}
&21.02
&0.08
\\
\bottomrule
\end{tabular}
}
\caption{Overview of the bounds obtained after including one-loop matching for the various scalar irreps is presented. For the description of the first three columns, we refer the reader to Tab.~\ref{tab:Scalars_treelevel_bounds}. This table focuses on the three most precisely measured $|\Delta L_\alpha|=1$ observables. In case of $\mu\to e$ and $\mu\to ee\bar e$, the effect of the logarithmic dependence in the matching relations is demonstrated by evaluating the contributions at two distinct matching scales, namely $\bar\mu=M_{\sscript{NP}}$ and $\bar\mu=M_Z$. For the purposes of the discussion below, we perform all comparisons using the matching results evaluated at the NP scale. With this in mind, the best bound for each irrep is denoted in {\color{customcolor}{purple}}. All bounds are expressed in TeV.}
\label{tab:Scalars_looplevel_bounds}
\end{table}

Our analysis starts with the UV scalar mediators, whose corresponding tree-level and one-loop constraints are summarized in Tabs.~\ref{tab:Scalars_treelevel_bounds} and~\ref{tab:Scalars_looplevel_bounds}, respectively. In the following discussion, we analyze each scalar mediator individually. As previously noted, for those $A_4$ irreps permitting multiple independent flavor invariants, for the corresponding couplings we assume $y_\cS^{(i)}\sim1$.

$\bm{\cS_1}$. At $\cO(1)$, the triplet irrep is subject to a bound of approximately 5 TeV. For this particular case, the flavor tensor does not lead to any flavor-violating terms in the matching, which implies that no constraints arise from $|\Delta L_\alpha|=1$ and $|\Delta L_\alpha|=2$ processes. However, at $\cO(Y_{\bm3}^{(2)})$, loop-level contributions become relevant and can compete with the tree-level bounds, which can be verified for $\bm1$ and $\bm1'$ irreps. Finally, among the contributions at $\cO(Y_{\bm3}^{(2)})$, the $A_4$ triplet is mostly constrained from $|\Delta L_\alpha|=1$ transitions.

$\bm{\cS_2}$. At $\cO(1)$, the singlet irreps are dominantly constrained by tree-level cLFV transitions. For the $\bm1$ and $\bm1'$ representations, the dominant constraints arise from both $|\Delta L_\alpha|=1$ and $|\Delta L_\alpha|=2$ transitions, whereas for $\bm1''$, only the low-energy lepton-flavor conserving observables contribute significantly. At $\cO(Y_{\bm3}^{(2)})$, the triplet irrep receives a tree-level bound from the $\mu\to ee\bar e$ transition. At one-loop level, the constraints from $\mu\to e\gamma$ and $\mu\to ee\bar e$ are comparable, with the bound from $\mu\to ee\bar e$ being reduced by approximately 25\% compared to its tree-level value.

$\bm{\varphi}$. For all irreps, whether at $\cO(1)$ or $\cO(Y_{\bm3}^{(2)})$ the strongest bounds arise from one-loop contributions. In particular, the most constraining observables are $\mu\to e\gamma$ and $\mu\to e$ conversion.

$\bm{\Xi_1}$. The irreducible representations appearing at $\cO(1)$ are predominantly constrained by tree-level contributions. For the $\bm1$, $\bm1'$ and $\bm3$ irreps, the constraints from $|\Delta L_\alpha|=1$ and $|\Delta L_\alpha|=2$ transitions are of comparable strength. The $\bm1''$ irrep, in contrast, is constrained solely via the low-energy combined fit. At $\cO(Y_{\bm3}^{(2)})$, the $\mu\to ee\bar e$ process provides the dominant bounds for the $\bm1$ and $\bm1'$ irreps, while for $\bm1''$, $\mu\to e$ conversion becomes equally significant. This behavior is mirrored by the triplet irrep as well.

As the final part of the phenomenological analysis of scalar fields, we focus on the $A_4$ irreps that give rise to two independent flavor invariants. From Tab.~\ref{tab:scalars_a4_irreps_invs}, we identify the relevant cases: the $\cS_1$ triplet irrep at $\cO(Y_{\bm3}^{(2)})$ and three $\cS_2$ singlet irreps at $\cO(1)$. Fig.~\ref{fig:scalars_2D_bounds} displays the constraints obtained from various observables in cases where the flavor tensors contain two independent parameters. As indicated in Tabs.~\ref{tab:Scalars_treelevel_bounds} and \ref{tab:Scalars_looplevel_bounds}, the $\bm1''$ $\cS_2$ irrep is solely constrained by low-energy observables and does not induce any other processes, which could serve as complementary bounds. For this reason, we do not include its bounds in the figure.

As a final remark, we note that a similar analysis can, in principle, be carried out for flavor irreps that give rise to more than two independent flavor invariants. In the absence of specific assumptions, one can select a pair of couplings and construct a profiled likelihood over the remaining parameters. Several illustrative examples of such approach are provided in the Supplemental Material.

\subsubsection{Fermions}
\label{sec:fermions_pheno}
The next part of our phenomenological analysis is dedicated to fermionic UV mediators. Tab.~\ref{tab:Fermions_treelevel_bounds} summarizes the tree-level constraints for the various fermionic $A_4$ irreps. As indicated in Tab.~\ref{tab:Fermions_treelevel_bounds}, all fermionic irreps are subject to bounds from the low-energy combined fit. The low-energy constraints lead to bounds ranging from about 3 TeV to 10 TeV, with the exact value depending on the particular fermionic irrep under consideration. Conversely, fermionic $A_4$ irreps, whose flavor tensors permit flavor-violating effects already at tree level, are strongly constrained by $\mu\to e$ conversion, which, in all such cases, yields the dominant bound. While tree-level constraints are already dominant for these mediators, we have explicitly verified that the inclusion of one-loop contributions does not qualitatively alter the phenomenological conclusions. For instance, in the case of $\mu \to e$ conversion, one-loop effects lead to $\mathcal{O}(10\%)$ shifts in the bounds.

Having established the above, we now revisit the irreps that exhibit no contributions to $|\Delta L_\alpha|=1$ processes at tree level. Beginning with the $\cO(1)$ $A_4$ singlets corresponding to $\Delta_1$ and $\Delta_3$ fermions, we note that the lack of $|\Delta L_\alpha|=1$ constraints at tree level arises from the particularly simple form of the associated flavor invariants and flavor tensors. Tab.~\ref{tab:fermions_a4_irreps_invs} illustrates that the flavor tensor corresponding to these irreps can be expressed as
\begin{equation}
    \cI\sim 
    \lambda_{\Delta_n}^{(1)}[\bar\Delta_{n_L}e]_{\bm1}
    \implies[\lambda_{\Delta_n}]_i=\lambda_{\Delta_n}^{(1)}\delta_{ik},
\end{equation}
where $n=1,3$ and $k=1,2,3$ for $\bm1,\bm1'',\bm1'$ irreps, respectively. As seen from this expression, the flavor tensors corresponding to the singlet irreps are non-vanishing only for a single value of the flavor index. This is precisely why flavor-violating effects are absent, both at tree level as well as at one-loop level.

\begin{figure*}[t]
\centering
\begin{tabular}{cccc}
\includegraphics[width=43.5mm]{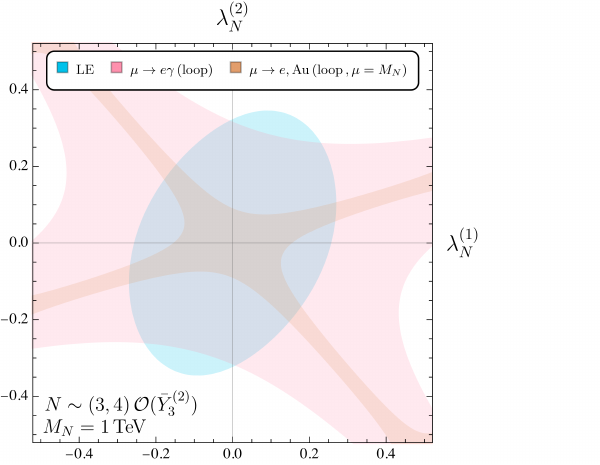}
&
\includegraphics[width=43.5mm]{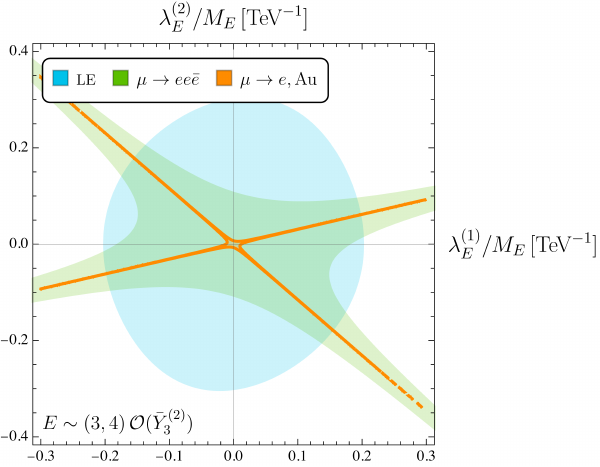} 
&   
\includegraphics[width=43.5mm]{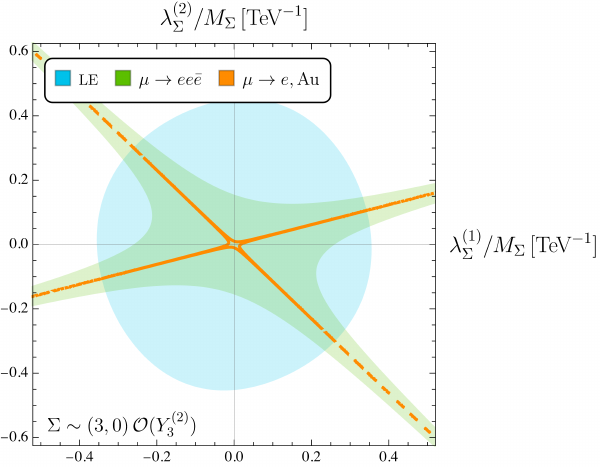}  
&
\includegraphics[width=43.5mm]{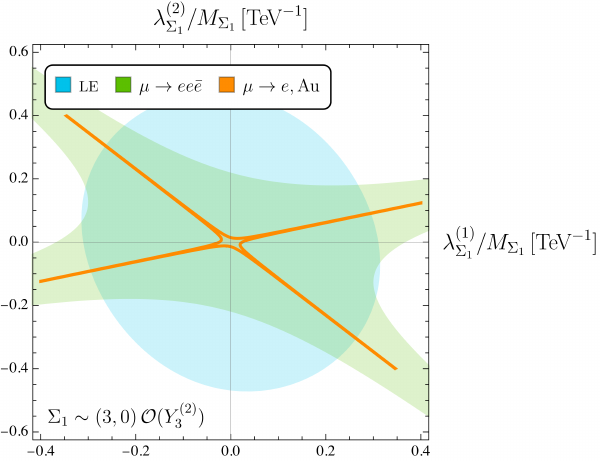}
\end{tabular}
\caption{Overview of the complementary constraints on four $A_4$ flavor triplet irreps corresponding to the $N$, $E$, $\Sigma$ and $\Sigma_1$ fermionic mediators, each featuring two independent flavor invariants. For the $N$ mediator, we fix the scale to $M_N=1\,\tev$, while for the remaining cases, constraints are shown in terms of the coupling-to-mass ratios. For further details, see Sec.~\ref{sec:fermions_pheno}.}
\label{fig:fermions_treeloop_bounds}
\end{figure*}

\begin{table}[t]
\centering
\scalebox{0.8}{
\begin{tabular}{cccccc}
\toprule
\textbf{Field}&\textbf{$\bm{(A_4,k)}$}&\textbf{Order} & \textbf{LE}& \textbf{$\bm{|\Delta L_\alpha|=1}$} &\textbf{Obs.}
\\
\midrule
\multirow{5}{*}{\vspace{-0.43cm}$N\sim(\bm1,\bm1)_{0}$}
&$(\bm3,2)$&$\cO(1)$&{\color{customcolor}{6.58}}&-&-
\\
&$(\bm1,4)$&$\cO(\bar Y_{\bm3}^{(2)})$&{\color{customcolor}{4.06}}&-&-
\\
&$(\bm1',4)$&$\cO(\bar Y_{\bm3}^{(2)})$&{\color{customcolor}{4.67}}&-&-
\\
&$(\bm1'',4)$&$\cO(\bar Y_{\bm3}^{(2)})$&{\color{customcolor}{6.51}}&-&-
\\
&$(\bm3,4)$&$\cO(\bar Y_{\bm3}^{(2)})$&{\color{customcolor}{4.50}}&-&-
\\
\midrule
\multirow{5}{*}{\vspace{-0.5cm}$E\sim(\bm1,\bm1)_{-1}$}
&$(\bm3,2)$&$\cO(1)$&7.92&-&-
\\
&$(\bm1,4)$&$\cO(\bar Y_{\bm3}^{(2)})$&7.28&{\color{customcolor}{167.54}}&$\mathrm{CR}(\mu\to e,\mathrm{Au})$
\\
&$(\bm1',4)$&$\cO(\bar Y_{\bm3}^{(2)})$&4.84&{\color{customcolor}{143.35}}&$\mathrm{CR}(\mu\to e,\mathrm{Au})$
\\
&$(\bm1'',4)$&$\cO(\bar Y_{\bm3}^{(2)})$&7.15&{\color{customcolor}{276.93}}&$\mathrm{CR}(\mu\to e,\mathrm{Au})$
\\
&$(\bm3,4)$&$\cO(\bar Y_{\bm3}^{(2)})$&6.66&{\color{customcolor}{168.34}}&$\mathrm{CR}(\mu\to e,\mathrm{Au})$
\\
\midrule
\multirow{4}{*}{\vspace{-0.1cm}$\Delta_1\sim(\bm1,\bm2)_{-1/2}$}
&$(\bm1,0)$
&$\cO(1)$
&7.91
&-
&-
\\
&$(\bm1',0)$
&$\cO(1)$
&3.86
&-
&-
\\
&
$(\bm1'',0)$
&$\cO(1)$
&3.33
&-
&-
\\
&$(\bm3,2)$
&$\cO(\bar Y_{\bm3}^{(2)})$
&9.56
&{\color{customcolor}{290.19}}
&$\mathrm{CR}(\mu\to e,\mathrm{Au})$
\\
\midrule
\multirow{4}{*}{\vspace{-0.2cm}$\Delta_3\sim(\bm1,\bm2)_{-3/2}$}
&$(\bm1,0)$
&$\cO(1)$
&5.54
&-
&-
\\
&$(\bm1',0)$
&$\cO(1)$
&3.87
&-
&-
\\
&
$(\bm1'',0)$
&$\cO(1)$
&5.51
&-
&-
\\
&$(\bm3,2)$
&$\cO(\bar Y_{\bm3}^{(2)})$
&8.37
&{\color{customcolor}{290.19}}
&$\mathrm{CR}(\mu\to e,\mathrm{Au})$
\\
\midrule
\multirow{5}{*}{\vspace{-0.3cm}$\Sigma\sim(\bm1,\bm3)_{0}$}
&$(\bm3,2)$
&$\cO(1)$
&4.82
&-
&-
\\
&$(\bm1,0)$
&$\cO(Y_{\bm3}^{(2)})$
&4.25
&{\color{customcolor}{118.47}}
&$\mathrm{CR}(\mu\to e,\mathrm{Au})$
\\
&$(\bm1',0)$
&$\cO(Y_{\bm3}^{(2)})$
&3.65
&{\color{customcolor}{101.36}}
&$\mathrm{CR}(\mu\to e,\mathrm{Au})$
\\
&$(\bm1'',0)$
&$\cO(Y_{\bm3}^{(2)})$
&4.11
&{\color{customcolor}{195.81}}
&$\mathrm{CR}(\mu\to e,\mathrm{Au})$
\\
&$(\bm3,0)$
&$\cO(Y_{\bm3}^{(2)})$
&4.05
&{\color{customcolor}{119.43}}
&$\mathrm{CR}(\mu\to e,\mathrm{Au})$
\\
\midrule
\multirow{5}{*}{\vspace{-0.35cm}$\Sigma_1\sim(\bm1,\bm3)_{-1}$}
&$(\bm3,2)$
&$\cO(1)$
&5.46
&-
&-
\\
&$(\bm1,0)$
&$\cO(Y_{\bm3}^{(2)})$
&5.34
&{\color{customcolor}{83.77}}
&$\mathrm{CR}(\mu\to e,\mathrm{Au})$
\\
&$(\bm1',0)$
&$\cO(Y_{\bm3}^{(2)})$
&2.70
&{\color{customcolor}{71.67}}
&$\mathrm{CR}(\mu\to e,\mathrm{Au})$
\\
&$(\bm1'',0)$
&$\cO(Y_{\bm3}^{(2)})$
&4.28
&{\color{customcolor}{138.46}}
&$\mathrm{CR}(\mu\to e,\mathrm{Au})$
\\
&$(\bm3,0)$
&$\cO(Y_{\bm3}^{(2)})$
&4.90
&{\color{customcolor}{84.45}}
&$\mathrm{CR}(\mu\to e,\mathrm{Au})$
\\
\bottomrule
\end{tabular}
}
\caption{Overview of the tree-level bounds on various $A_4$ flavor irreps for fermionic mediators. The best bound for each irrep is denoted in {\color{customcolor}{purple}}. For a description of the first four columns, we refer the reader to Tab.~\ref{tab:Scalars_treelevel_bounds}. All bounds are expressed in TeV.}
\label{tab:Fermions_treelevel_bounds}
\end{table}

Compared to $\Delta_{1,3}$, $N$ mediator features a more elaborate structure, leading to richer phenomenological implications. Tab.~\ref{tab:Fermions_treelevel_bounds} indicates that $|\Delta L_\alpha|=1$ constraints are not present at tree level. The absence of these constraints can be traced back to the fact that the tree-level matching generates dimension-6 $\cC_{\phi\ell}^{(1)}$ and $\cC_{\phi\ell}^{(3)}$ Wilson coefficients with opposite signs:
\begin{equation}\label{eq:N_treelevel_WCs}
    [\cC_{\phi\ell}^{(1)}]_{ij}=-[\cC_{\phi\ell}^{(3)}]_{ij}=\frac{1}{4M_N^2}[\lambda_N]^*_{ri}[\lambda_N]_{rj}.
\end{equation}
Below the electroweak scale, this SMEFT configuration induces corrections to the couplings of neutrinos with the Z boson, modifying their neutral current interactions, while leaving the couplings of left-handed electrons unchanged to leading order~\cite{Jenkins:2017jig}:
\begin{equation}
    \begin{alignedat}{2}
        [\delta Z_{\nu_L}]_{ij}&=-\frac{v^2}{2}[\cC_{\phi\ell}^{(1)}-\cC_{\phi\ell}^{(3)}]_{ij}=-\frac{v^2}{4M_N^2}[\lambda_N]^*_{ri}[\lambda_N]_{rj},
        \\
        [\delta Z_{e_L}]_{ij}&=-\frac{v^2}{2}[\cC_{\phi\ell}^{(1)}+\cC_{\phi\ell}^{(3)}]_{ij}
        =0,
    \end{alignedat}
\end{equation}
where $v\approx246\,\gev$ denotes the VEV of the Higgs field. Once one-loop matching effects are taken into account, the tree-level relation between the Wilson coefficients given by Eq.~\eqref{eq:N_treelevel_WCs} is no longer preserved. Consequently, the left-handed lepton coupling to the $Z$ boson evaluated at $\mu=M_N$ contains the terms
\begin{equation}
    \begin{alignedat}{2}
        [\delta Z_{e_L}]_{ij}&\supset
        \frac{v^2}{32\pi^2}\frac{7}{8M_N^2}[\lambda_N]^*_{rk}[\lambda_N]_{rk}[\lambda_N]^*_{r'i}[\lambda_N]_{r'j}
        \\&
        -\frac{v^2}{32\pi^2}\frac{11g_1^2+187g_2^2}{144M_N^2}[\lambda_N]^*_{ri}[\lambda_N]_{rj},
    \end{alignedat}
\end{equation}
which contribute to $|\Delta L_\alpha|=1$ transitions.\footnote{This serves as a representative one-loop contribution. As outlined in Ref.~\cite{Gargalionis:2024jaw}, additional SMEFT operators are induced, contributing to lepton flavor-violating processes.} Tab.~\ref{tab:N_looplevel_bounds} provides a summary of the one-loop constraints on the relevant $A_4$ irreps for $N$ mediator.

\begin{table}[t]
\centering
\scalebox{0.625}{
\begin{tabular}{cccccccc}
\toprule
\multirow{2}{*}{\textbf{Field}}
&\multirow{2}{*}{\textbf{$\bm{(A_4,k)}$}}
&\multirow{2}{*}{\textbf{Order}} 
&\multirow{2}{*}{$\bm{\mu\to e\gamma}$}
&$\bm{\mathrm{CR}(\mu\to e,\mathrm{Au})}$ 
&$\bm{\mathrm{CR}(\mu\to e,\mathrm{Au})}$
&$\bm{\mu\to ee\bar e}$
&$\bm{\mu\to ee\bar e}$ 
\\
&
&
&
&$\bm{\bar\mu=M_{N}}$
&$\bm{\bar\mu=M_{Z}}$
&$\bm{\bar\mu=M_{N}}$
&$\bm{\bar\mu=M_{Z}}$
\\
\midrule
\multirow{4}{*}{\vspace{-0.3cm}$N\sim(\bm1,\bm1)_{0}$}
&$(\bm1,4)$&$\cO(\bar Y_{\bm3}^{(2)})$
&3.21
&{\color{customcolor}{10.88}}
&16.22
&1.42
&2.39
\\
&$(\bm1',4)$&$\cO(\bar Y_{\bm3}^{(2)})$
&3.28
&{\color{customcolor}{9.31}}
&13.24
&1.52
&3.18
\\
&$(\bm1'',4)$&$\cO(\bar Y_{\bm3}^{(2)})$
&6.33
&{\color{customcolor}{17.98}}
&30.31
&0.95
&0.05
\\
&$(\bm3,4)$&$\cO(\bar Y_{\bm3}^{(2)})$
&{\color{customcolor}{3.85}}
&{\color{black}{2.50}}
&22.24
&1.16
&2.09
\\
\bottomrule
\end{tabular}
}
\caption{Summary of one-loop constraints for different $A_4$ irreps associated with the $N$ mediator. The reader is referred to Tab.~\ref{tab:Scalars_looplevel_bounds} for details on the first three columns. We denote the best bounds in {\color{customcolor}{purple}}. All bounds are expressed in TeV.}
\label{tab:N_looplevel_bounds}
\end{table}

As shown by the one-loop bounds, $\mu\to e$ conversion, although generated at one loop, yields stronger constraints than the tree-level flavor-conserving bounds across all irreps at $\cO(Y_{\bm3}^{(2)})$. The $\cO(1)$ triplet irrep, due to its flavor tensor structure, is the only case constrained solely by low-energy observables at tree level.

Following the approach used for the scalar irreps, here we also present the flavor irreps that lead to two independent flavor invariants. Specifically, this includes the $\cO(Y_{\bm3}^{(2)})$ triplet irreps for the $N$, $E$, $\Sigma$ and $\Sigma_1$ mediators. Fig.~\ref{fig:fermions_treeloop_bounds} provides a summary of the complementary constraints for these cases.

\subsubsection{Vectors}
\label{sec:vectors_pheno}

\begin{table}[t]
\centering
\scalebox{0.7}{
\begin{tabular}{cccccccc}
\toprule
\textbf{Field}&\textbf{$\bm{(A_4,k)}$}&\textbf{Order} & \textbf{LE}& \textbf{$\bm{|\Delta L_\alpha|=1}$} &\textbf{Obs.}&\textbf{$\bm{|\Delta L_\alpha|=2}$}&\textbf{Obs.}
\\
\midrule
\multirow{8}{*}{\vspace{-0.5cm}$\cB\sim(\bm1,\bm1)_0$}
&$(\bm1,0)$&$\cO(1)$&{\color{customcolor}{8.00}}&-&-&-&-
\\
&$(\bm1',0)$&$\cO(1)$&7.38&{\color{customcolor}{16.01}}&$\tau\to ee\bar\mu$&14.14&$\tau\to\bar\mu ee$
\\
&$(\bm1'',0)$&$\cO(1)$&7.38&{\color{customcolor}{16.01}}&$\tau\to ee\bar\mu$&14.14&$\tau\to\bar\mu ee$
\\
&$(\bm3,0)$&$\cO(1)$&9.10&{\color{customcolor}{12.42}}&$\tau\to \mu\mu\bar e$&9.28&$\tau\to \bar e\mu\mu $
\\
&$(\bm1,2)$&$\cO(\bar Y_{\bm3}^{(2)})$&7.53&{\color{customcolor}{17.77}}&$\mu\to ee\bar e$&3.81&$\tau\to \bar\mu ee$
\\
&$(\bm1',2)$&$\cO(\bar Y_{\bm3}^{(2)})$&5.51&{\color{customcolor}{13.01}}&$\mu\to ee\bar e$&2.79&$\tau\to\bar\mu ee$
\\
&$(\bm1'',2)$&$\cO(\bar Y_{\bm3}^{(2)})$&2.02&{\color{customcolor}{4.76}}&$\mu\to ee\bar e$&1.02&$\tau\to\bar\mu ee$
\\
&$(\bm3,2)$&$\cO(\bar Y_{\bm3}^{(2)})$&18.79&{\color{customcolor}{84.24}}&$\mu\to ee\bar e$&24.06&$\tau\to\bar\mu ee$
\\
\midrule
\multirow{8}{*}{\vspace{-0.4cm}$\cW\sim(\bm1,\bm3)_0$}
&$(\bm1,0)$&$\cO(1)$&{\color{customcolor}{5.24}}&-&-&-&-
\\
&$(\bm1',0)$&$\cO(1)$&4.65&{\color{customcolor}{6.91}}&$\tau\to \mu e\bar e$&5.00&$\tau\to\bar \mu ee$
\\
&$(\bm1'',0)$&$\cO(1)$&4.65&{\color{customcolor}{6.91}}&$\tau\to \mu e\bar e$&5.00&$\tau\to\bar \mu ee$
\\
&$(\bm3,0)$&$\cO(1)$&4.38&{\color{customcolor}{6.58}}&$\tau\to e\mu\bar\mu$&4.64&$\tau\to\bar e\mu\mu$
\\
&$(\bm1,2)$&$\cO(\bar Y_{\bm3}^{(2)})$&{\color{customcolor}{10.47}}&8.88&$\mu\to ee\bar e$&1.91&$\tau\to\bar \mu ee$
\\
&$(\bm1',2)$&$\cO(\bar Y_{\bm3}^{(2)})$&{\color{customcolor}{7.66}}&6.50&$\mu\to ee\bar e$&1.39&$\tau\to \bar\mu ee$
\\
&$(\bm1'',2)$&$\cO(\bar Y_{\bm3}^{(2)})$&2.80&{\color{customcolor}{4.19}}&$\tau\to e\mu\bar\mu $&0.51&$\tau\to \bar\mu ee$
\\
&$(\bm3,2)$&$\cO(\bar Y_{\bm3}^{(2)})$&11.53&{\color{customcolor}{35.68}}&$\mu\to ee\bar e$&7.81&$\tau\to\bar\mu ee$
\\
\midrule
\multirow{5}{*}{\vspace{-0.4cm}$\cL_3\sim(\bm1,\bm2)_{-3/2}$}
&$(\bm3,-2)$
&$\cO(1)$
&4.53
&12.16
&$\tau\to ee\bar\mu$&{\color{customcolor}{14.14}}&$\tau\to\bar\mu ee$
\\
&$(\bm1,0)$
&$\cO(Y_{\bm3}^{(2)})$
&3.92
&{\color{customcolor}{27.98}}
&$\mu\to ee\bar e$&12.37&$\tau\to \bar\mu ee$
\\
&$(\bm1',0)$
&$\cO(Y_{\bm3}^{(2)})$
&4.03
&{\color{customcolor}{16.69}}&$\mu\to ee\bar e$&12.73&$\tau\to\bar e\mu\mu $
\\
&$(\bm1'',0)$
&$\cO(Y_{\bm3}^{(2)})$
&3.66
&{\color{customcolor}{32.25}}
&$\mu\to ee\bar e$&10.59&$\tau\to \bar\mu ee$
\\
&$(\bm3,0)$
&$\cO(Y_{\bm3}^{(2)})$
&3.50
&8.25
&$\tau\to \mu\mu\bar e$&{\color{customcolor}{10.29}}&$\tau\to \bar\mu ee$
\\
\bottomrule
\end{tabular}
}
\caption{Overview of the tree-level bounds on various $A_4$ flavor irreps of the vector mediators. For a detailed description of the columns, see Tab.~\ref{tab:Scalars_treelevel_bounds}. The best tree-level bound for each irrep is denoted in {\color{customcolor}{purple}}. All bounds are expressed in TeV.}
\label{tab:vectors_treelevel_bounds}
\end{table}

In the final stage of our analysis, we focus on the phenomenological implications of vector mediators. As outlined at the beginning of Sec.~\ref{sec:phenomenology_intro}, in line with the bottom-up approach, our EFT framework remains agnostic to the details of UV completions and the mechanisms of spontaneous symmetry breaking associated with vector mediators. Therefore, instead of pursuing a full one-loop matching, which would necessitate a complete model, we make use of one-loop RGE effects to perform a more comprehensive numerical analysis. Tree-level bounds are shown in Tab.~\ref{tab:vectors_treelevel_bounds}, whereas Tab.~\ref{tab:vectors_looplevel_bounds} summarizes the results after incorporating one-loop RGE effects~\cite{Alonso:2013hga}. In what follows, we briefly comment on each mediator individually.

\begin{table}[t]
\centering
\scalebox{0.615}{
\begin{tabular}{ccccccc}
\toprule
\multirow{2}{*}{\textbf{Field}}
&\multirow{2}{*}{$\bm{(A_4,k)}$}
&\multirow{2}{*}{\textbf{Order}} 
&\textbf{$\bm{\mu\to ee\bar e}$}
&\textbf{$\bm{\mu\to ee\bar e}$}
&$\bm{\mathrm{CR}(\mu\to e,\mathrm{Au})}$
&$\bm{\mathrm{CR}(\mu\to e,\mathrm{Au})}$
\\
&&
&$\bm{\bar\mu_i=M_{\sscript{NP}}}$
&$\bm{\bar\mu_i=3\,\mathrm{TeV}}$
&$\bm{\bar\mu_i=M_{\sscript{NP}}}$
&$\bm{\bar\mu_i=3\,\mathrm{TeV}}$
\\
\midrule
\multirow{7}{*}{\vspace{-0.4cm}$\cB\sim(\bm1,\bm1)_0$}
&$(\bm1',0)$&$\cO(1)$
&4.73
&4.45
&{\color{customcolor}{33.70}}
&25.91
\\
&$(\bm1'',0)$&$\cO(1)$
&4.64
&4.38
&{\color{customcolor}{134.80}}
&93.27
\\
&$(\bm3,0)$&$\cO(1)$
&1.93
&2.06
&{\color{customcolor}{59.91}}
&43.97
\\
&$(\bm1,2)$&$\cO(\bar Y_{\bm3}^{(2)})$
&32.83
&{\color{customcolor}{33.27}}
&20.24
&16.28
\\
&$(\bm1',2)$&$\cO(\bar Y_{\bm3}^{(2)})$
&13.75
&13.71
&{\color{customcolor}{101.46}}
&71.61
\\
&$(\bm1'',2)$&$\cO(\bar Y_{\bm3}^{(2)})$
&19.59
&19.89
&{\color{customcolor}{83.89}}
&60.03
\\
&$(\bm3,2)$&$\cO(\bar Y_{\bm3}^{(2)})$
&78.19
&79.17
&{\color{customcolor}{173.34}}
&117.93
\\
\midrule
\multirow{7}{*}{$\cW\sim(\bm1,\bm3)_0$}
&$(\bm1'',0)$&$\cO(1)$
&3.23
&3.20
&{\color{customcolor}{40.14}}
&30.41
\\
&$(\bm3,0)$&$\cO(1)$
&0.85
&1.07
&{\color{customcolor}{11.98}}
&10.14
\\
&$(\bm1,2)$&$\cO(\bar Y_{\bm3}^{(2)})$
&16.48
&{\color{customcolor}{16.63}}
&3.89
&3.75
\\
&$(\bm1',2)$&$\cO(\bar Y_{\bm3}^{(2)})$
&6.88
&6.87
&{\color{customcolor}{20.56}}
&16.51
\\
&$(\bm1'',2)$&$\cO(\bar Y_{\bm3}^{(2)})$
&9.84
&9.94
&{\color{customcolor}{16.92}}
&13.84
\\
&$(\bm3,2)$&$\cO(\bar Y_{\bm3}^{(2)})$
&30.90
&31.40
&{\color{customcolor}{122.15}}
&85.10
\\
\midrule
\multirow{4}{*}{\vspace{-0.3cm}$\cL_3\sim(\bm1,\bm2)_{-3/2}$}
&$(\bm1,0)$
&$\cO(Y_{\bm3}^{(2)})$
&28.29
&28.17
&{\color{customcolor}{35.37}}
&27.08
\\
&$(\bm1',0)$
&$\cO(Y_{\bm3}^{(2)})$
&17.06
&16.94
&{\color{customcolor}{35.37}}
&27.08
\\
&$(\bm1'',0)$
&$\cO(Y_{\bm3}^{(2)})$
&32.62
&32.47
&{\color{customcolor}{35.37}}
&27.08
\\
&$(\bm3,0)$
&$\cO(Y_{\bm3}^{(2)})$
&7.39
&7.40
&{\color{customcolor}{13.60}}
&11.36
\\
\bottomrule
\end{tabular}
}
\caption{Bounds on various $A_4$ irreps for vector mediators after including RGE effects. The table shows the constraints derived from the most precise observables, i.e. $\mu\to 3e$ and $\mu\to e$ conversion. Bounds are computed for two initial matching scales, $\bar\mu_i=M_{\sscript{NP}}$ and $\bar\mu_i=3\,\tev$, with the final running scale taken to $\mu_f\sim M_Z$. The best bounds are denoted in {\color{customcolor}{purple}}. All bounds are expressed in TeV.}
\label{tab:vectors_looplevel_bounds}
\end{table}

\begin{figure*}[t]
\centering
\begin{tabular}{cccc}
\includegraphics[width=43.5mm]{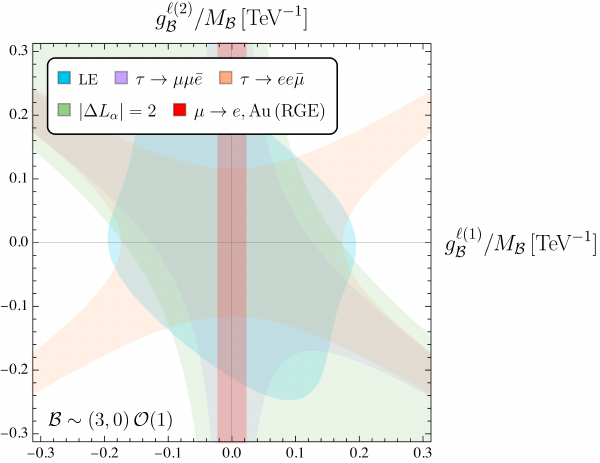} 
&   
\includegraphics[width=43.5mm]{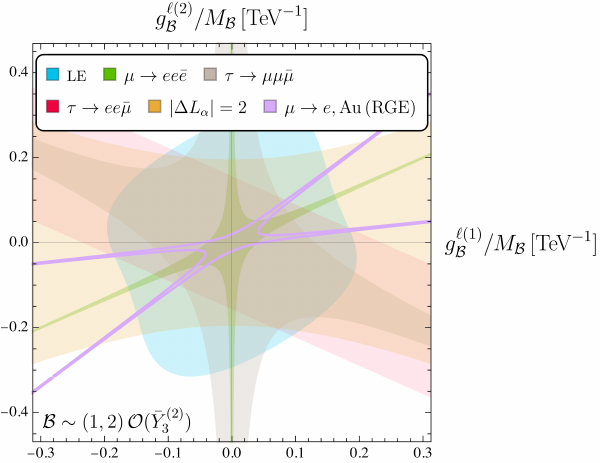}  
&
\includegraphics[width=43.5mm]{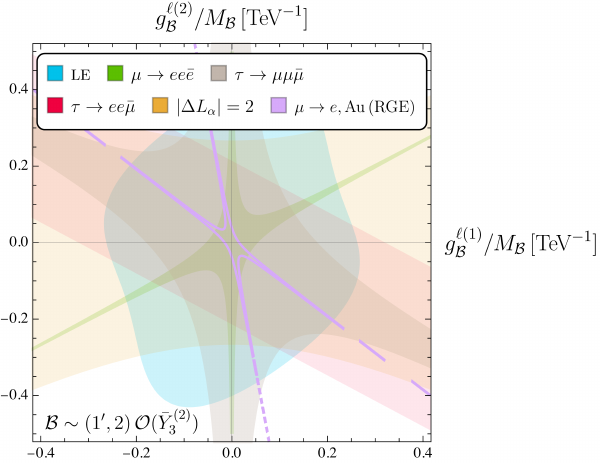}
&
\includegraphics[width=43.5mm]{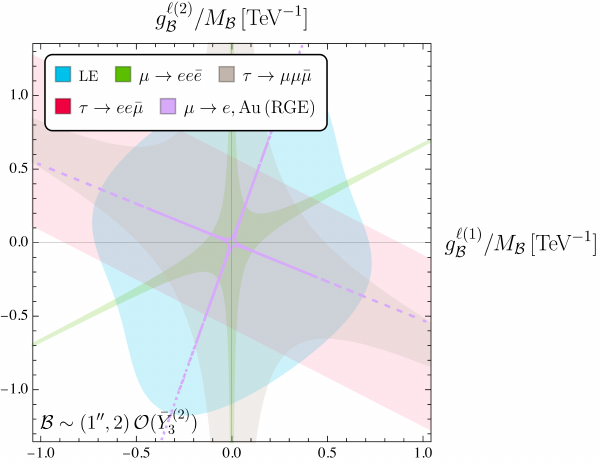}
\\
\includegraphics[width=43.5mm]{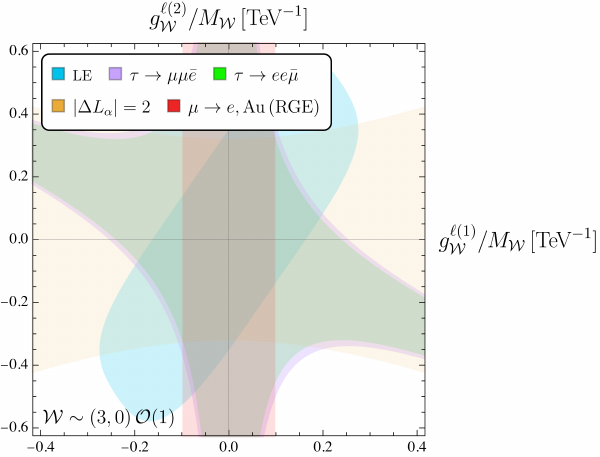}
&
\includegraphics[width=43.5mm]{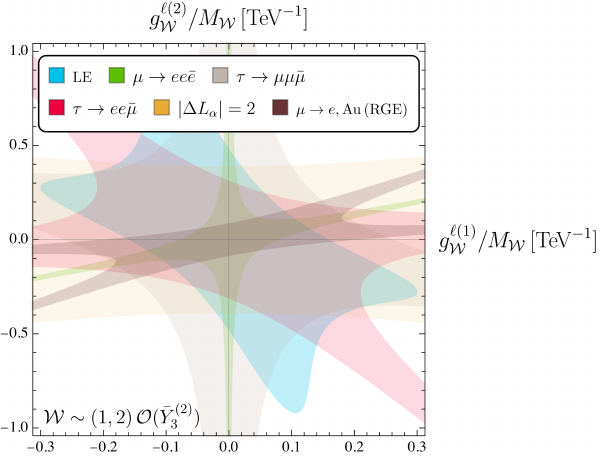}
&
\includegraphics[width=43.5mm]{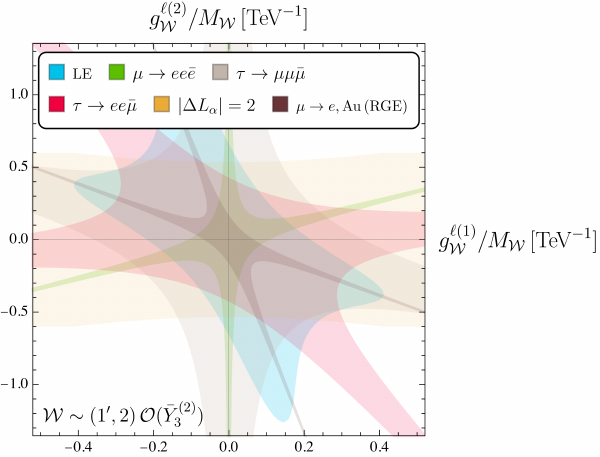}
&
\includegraphics[width=43.5mm]{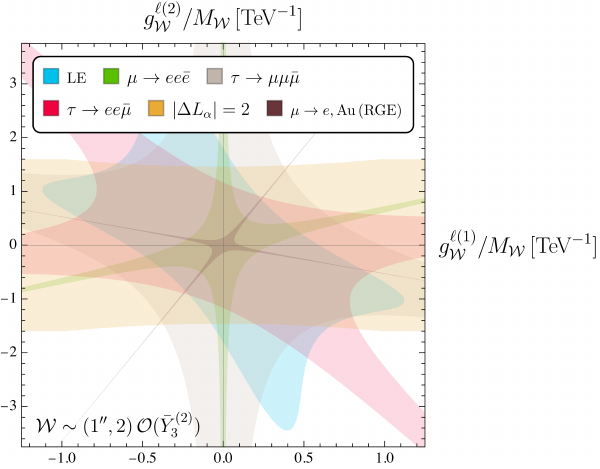}
\end{tabular}
\caption{Overview of the combined constraints on the vector irreps whose flavor tensors contain two independent parameters. The upper four plots display the complementary bounds on the four relevant flavor irreps of the $\mathcal{B}$ mediator, while the lower four correspond to the $\mathcal{W}$ mediator.}
\label{fig:vectors_compl_bounds}
\end{figure*}

$\bm{\cB}$.
For the $\cO(1)$ flavor irreps, the $\bm{1}$ irrep yields a flavor tensor that does not induce lepton flavor-violating transitions, and is thus constrained only by the low-energy combined fit.  In contrast, the $\bm{1}'$ and $\bm{1}''$ irreps are subject to stronger tree-level bounds arising from both $|\Delta L_\alpha| = 1$ and $|\Delta L_\alpha| = 2$ transitions.

At $\mathcal{O}(Y_{\bm{3}}^{(2)})$, the leading constraint at tree level arises from $\mu \to ee\bar e$, while RGE effects shift the dominant bounds to $\mu \to e$ conversion for all irreps, except for $\cO(Y_{\bm3}^{(2)})$ trivial singlet irrep, which remains most strongly constrained by $\mu \to ee\bar e$.

$\bm{\cW}$.
Starting with $\cO(1)$, the trivial singlet does not contribute to lepton flavor-violating processes at tree level and is therefore constrained solely through the low-energy combined fit. In contrast, the non-trivial singlets $\bm{1}'$ and $\bm{1}''$ receive bounds from both $|\Delta L_\alpha| = 1$ and $|\Delta L_\alpha| = 2$ transitions, with processes such as $\mu \to e e \bar{e}$ and $\tau \to \mu e \bar e$ playing a dominant role from each sector. The $\cO(1)$ triplet also receives bounds from $|\Delta L_\alpha| = 1$ and $|\Delta L_\alpha| = 2$ transitions, which are comparable to those obtained from the combined fit. $\cO(Y_{\bm3}^{(2)})$ singlets are dominantly constrained from low-energy and $|\Delta L_\alpha| = 1$ transitions, whereas for the triplet irrep, bounds become substantially stronger. In particular, the triplet irrep is constrained up to 35 TeV from $\mu \to e e \bar{e}$ transition.

Once RGE effects are included, the dominant constraints shift to $\mu \to e$ conversion, which provides the most stringent bounds across all irreps apart from the trivial singlet at $\cO(Y_{\bm3}^{(2)})$, which receives the strongest constraint from $\mu\to ee\bar e$. The $\cO(Y_{\bm3}^{(2)})$ triplet irrep receives the strongest constraint, exceeding 120 TeV, for $\mu_i=M_{\sscript{NP}}$.

$\bm{\cL_3}$.
At $\cO(1)$, the $A_4$ triplet irrep receives constraints from both $|\Delta L_\alpha| = 1$ and $|\Delta L_\alpha| = 2$ transitions, with $\tau \to e e \bar\mu$ and $\tau \to \mu e \bar e$ providing the leading sensitivity. Furthermore, the $\cO(Y_{\bm3}^{(2)})$ singlets exhibit stronger bounds, especially from $\mu \to e e \bar{e}$ and $|\Delta L_\alpha| = 2$ processes. Lastly, the $\cO(Y_{\bm3}^{(2)})$ triplet receives weaker bounds, dominated by $|\Delta L_\alpha| = 2$ transitions.

Once RGE effects are taken into account, $\mu \to e$ conversion provides the most stringent bounds across all irreps, surpassing the corresponding tree-level constraints for every irrep.

Analogous to the scalar and fermion mediators, Tab.~\ref{tab:vectors_a4_irreps_invs} reveals eight irreps for which the flavor tensors admit two independent parameters. For these cases, the complementary constraints on the independent coupling-to-mass ratios are presented in Fig.~\ref{fig:vectors_compl_bounds}.

\section{Conclusion}
\label{sec:conc}
The intricate structure of the lepton sector motivates a detailed exploration of various flavor symmetry groups as frameworks for explaining the observed mass hierarchies and mixing patterns. Expanding upon earlier efforts that connected flavor symmetries with UV mediator structures~\cite{Greljo:2022cah, Greljo:2023adz, Greljo:2023bdy, Palavric:2024gvu}, this work incorporates modular $A_4$ symmetry into the UV-SMEFT matching framework, where the breaking of flavor symmetry is dynamically controlled by the modulus $\tau$. By classifying mediators according to their modular weights and constructing the corresponding flavor invariants, we obtain a refined understanding of how modular symmetry shapes the effective flavor structure. This approach enables a symmetry-guided interpretation of low-energy observables and provides a concrete framework to assess both tree-level and loop-level effects in charged lepton flavor transitions.

Compared to the framework developed in Ref.~\cite{Palavric:2024gvu}, this work presents a notable advancement in the classification of UV mediators by incorporating modular $A_4$ flavor symmetry and the structure of modular forms into the construction of flavor invariants. The classification is performed by systematically organizing the allowed flavor contractions involving SM fields, mediators, and up to one insertion of a lowest-weight modular form in the UV interaction terms. These modular insertions enrich the flavor structure while preserving symmetry consistency, and upon tree-level matching, they result in effective SMEFT operators containing up to two modular form insertions in the Wilson coefficients. The modular weight thus serves as a distinguishing label across different irreps, allowing for a more refined and structured mapping of mediator couplings beyond what is accessible in the exact $A_4$ limit. A summary of the classification results, including the allowed irreps and associated flavor invariants for scalar, fermionic, and vector mediators, is provided in Tabs.~\ref{tab:scalars_a4_irreps_invs}, \ref{tab:fermions_a4_irreps_invs}, and \ref{tab:vectors_a4_irreps_invs}, respectively.

The phenomenological analysis presented in this work provides a systematic study of the low-energy implications of modular $A_4$-invariant mediator couplings across all relevant irreps. For mediators whose flavor invariants predict a single independent parameter, we directly extract bounds on the ratio of coupling to mass using combined fit of the low-energy observables, which conserve lepton flavor, as well as the cLFV $|\Delta L_\alpha|=1$ and $|\Delta L_\alpha|=2$ observables. Moreover, in cases where two parameters are present, we perform a dedicated two-dimensional analysis, scanning over the parameter plane to determine the allowed regions. In addition to tree-level matching relations~\cite{deBlas:2017xtg}, we incorporate one-loop corrections for scalars and fermions, and leading-log RGE effects in the case of vector mediators. While some observables first arise at one-loop order, we find that in a significant number of cases, these loop-induced contributions actually provide the leading constraint. This highlights the importance of incorporating radiative effects even in symmetry-constrained frameworks. Tabs.~\ref{tab:Scalars_treelevel_bounds}--\ref{tab:vectors_looplevel_bounds} summarize the full set of results for all $\cO(1)$ and $\cO(Y_{\bm3}^{(2)})$ irreps. Additionally, Figs.~\ref{fig:scalars_2D_bounds}, \ref{fig:fermions_treeloop_bounds} and \ref{fig:vectors_compl_bounds} show complementary 2D constraints for various scalar, fermion, and vector irreps with two-parameter flavor tensors. For the cases where the flavor tensor involves more than two independent parameters, we include a multidimensional profiling analysis with selected examples in the Supplemental Material.

While the framework and results presented in this work offer useful guidance for exploring modular flavor symmetries in effective field theory, we emphasize that this study serves primarily as a phenomenological benchmark. The bounds we derive are obtained under simplified assumptions, considering one mediator at a time and without specifying the details of the underlying mechanism for spontaneous symmetry breaking. In realistic model-building scenarios, where multiple states and dynamical symmetry breaking are typically involved, a more detailed and comprehensive analysis would be required. Nevertheless, our results provide a reference point and a practical starting guide for constructing and testing explicit $A_4$ modular models, helping to identify viable directions and relevant experimental constraints.

To conclude, this work suggests several avenues for further investigation that build naturally on the framework developed here.

A natural extension of this work lies in applying modular flavor symmetries to the quark sector. While the lepton sector can often be accommodated by low-weight modular forms and simple group assignments, the quark mass hierarchies and CKM structure may require more intricate modular weights or higher-weight modular forms. Exploring how modular $A_4$, or potentially larger groups, can reproduce realistic quark textures remains an open and compelling challenge. This direction could also illuminate whether a unified modular framework is viable across both lepton and quark sectors, and in doing so, provide the foundation for analogous SMEFT analyses in the quark sector, similar in spirit to the present study.

From the EFT perspective, incorporating higher-order modular form insertions would provide a more complete picture of symmetry-breaking effects in flavor structures. Such terms may capture subleading contributions that become relevant in precision observables or in scenarios where leading-order effects are suppressed. Additionally, studying how modular weights accumulate in higher insertions could reveal new symmetry-driven patterns of suppression or enhancement in the effective couplings.

Altogether, as modular symmetries continue to mature into a predictive framework, the interplay between them, UV dynamics, and effective field theory continues to offer a rich and promising arena for uncovering the underlying structure of flavor.

\vspace{+0.3cm}
\begin{acknowledgments}
\noindent 
We thank Miguel Levy for useful discussions and feedback on the manuscript. We are also grateful to Tatsuo Kobayashi, Hajime Otsuka, Morimitsu Tanimoto, and Kei Yamamoto for carefully reading the manuscript. The work of AMS is supported by Ministerio de Ciencia, Innovaci\'on y Universidades under  grant FPU23/01639,  by MICIU/AEI/10.13039/501100011033 and FEDER/UE under grant PID2022-139466NB-C21, by Junta de Andaluc\'ia under grants P21\_00199 and FQM10 and by Consejer\'ia de Universidad, Investigaci\'on e Innovaci\'on, Gobierno de Espa\~{n}a and Uni\'on Europea – NextGenerationEU under grant AST22 6.5. 
The work of AP has received funding from the Swiss National Science Foundation (SNF) through the Eccellenza Professorial Fellowship
“Flavor Physics at the High Energy Frontier” project number 186866.
\end{acknowledgments}

\begin{table*}[t]
    \centering
\scalebox{0.9}{
\begin{tabular}{ccc|@{\hspace{0.5cm}}ccccc}
\toprule
\multirow{1}{*}{\textbf{UV Field}}
&\multirow{1}{*}{\textbf{$\bm{(A_4,k)}$}}
&\multirow{1}{*}{\textbf{Order}} 
&\multirow{1}{*}{\textbf{Invariants}} 
&\multirow{1}{*}{\textbf{\# of parameters}}
\\
\midrule
\multirow{5}{*}{\vspace{-0.5cm}$\cS_1\sim(\bm1,\bm1)_1$}
&$(\bm3,4)$
&$\cO(1)$
&\vspace{+0.05cm}$y_{\cS_1}^{(1)}[\cS_1^\dag]_{\bm3}[\bar\ell\,\ell^c]_{\bm3_A}$
&1
\\
&$(\bm1,2)$
&$\cO(\bar Y_{\bm3}^{(2)})$
&\vspace{+0.05cm}$y_{\cS_1}^{(1)}[\cS_1^\dag]_{\bm1}[[\bar\ell\,\ell^c]_{\bm3_A}\,\bar Y_{\bm3}^{(2)}(\tau)]_{\bm1}$
&1
\\
&$(\bm1',2)$
&$\cO(\bar Y_{\bm3}^{(2)})$
&\vspace{+0.05cm}$y_{\cS_1}^{(1)}[\cS_1^\dag]_{\bm1'}[[\bar\ell\,\ell^c]_{\bm3_A}\,\bar Y_{\bm3}^{(2)}(\tau)]_{\bm1''}$
&1
\\
&$(\bm1'',2)$
&$\cO(\bar Y^{(2)}_{\bm3})$
&\vspace{+0.05cm}$y_{\cS_1}^{(1)}[\cS_1^\dag]_{\bm1''}[[\bar\ell\,\ell^c]_{\bm3_A}\,\bar Y_{\bm3}^{(2)}(\tau)]_{\bm1'}$
&1
\\
&$(\bm3,2)$
&$\cO(\bar Y_{\bm3}^{(2)})$
&\vspace{+0.05cm}$y_{\cS_1}^{(1)}[\cS_1^\dag \bar Y_{\bm3}^{(2)}]_{\bm3_S}[\bar\ell\,\ell^c]_{\bm3_A}
			\oplus y_{\cS_1}^{(2)}[\cS_1^\dag \bar Y_{\bm3}^{(2)}]_{\bm3_A}[\bar\ell\,\ell^c]_{\bm3_A}$
&2
\\
\midrule
\multirow{6}{*}{\vspace{-0.6cm}$\cS_2\sim(\bm1,\bm1)_2$}
&$(\bm1,0)$
&$\cO(1)$
&
\vspace{+0.05cm}$y_{\cS_2}^{(1)}[\cS_2^\dag]_{\bm1}[\bar e]_{\bm1}[e^c]_{\bm1}
\oplus 
y_{\cS_2}^{(2)}[\cS_2^\dag]_{\bm1}[\bar e]_{\bm1'}[e^c]_{\bm1''}
\oplus
y_{\cS_2}^{(2)}[\cS_2^\dag]_{\bm1}[\bar e]_{\bm1''}[e^c]_{\bm1'}
$
&2
\\
&$(\bm1',0)$
&$\cO(1)$
&
\vspace{+0.05cm}$y_{\cS_2}^{(2)}[\cS_2^\dag]_{\bm1'}[\bar e]_{\bm1}[e^c]_{\bm1''}
\oplus 
y_{\cS_2}^{(2)}[\cS_2^\dag]_{\bm1'}[\bar e]_{\bm1''}[e^c]_{\bm1'}
\oplus
y_{\cS_2}^{(1)}[\cS_2^\dag]_{\bm1'}[\bar e]_{\bm1'}[e^c]_{\bm1'}
$
&2
\\
&$(\bm1'',0)$
&$\cO(1)$
&
\vspace{+0.05cm}$y_{\cS_2}^{(2)}[\cS_2^\dag]_{\bm1''}[\bar e]_{\bm1}[e^c]_{\bm1'}
\oplus 
y_{\cS_2}^{(2)}[\cS_2^\dag]_{\bm1''}[\bar e]_{\bm1'}[e^c]_{\bm1}
\oplus
y_{\cS_2}^{(1)}[\cS_2^\dag]_{\bm1''}[\bar e]_{\bm1''}[e^c]_{\bm1''}
$
&2
\\
\cmidrule{2-5}
&\multirow{3}{*}{$(\bm3,-2)$}
&\multirow{3}{*}{$\cO(\bar Y_{\bm3}^{(2)})$}
&
\vspace{+0.05cm}$
y_{\cS_2}^{(1)}[\cS_2^\dag \bar Y_{\bm3}^{(2)}]_{\bm1}[\bar e]_{\bm1}[e^c]_{\bm1}
\oplus
y_{\cS_2}^{(2)}[\cS_2^\dag \bar Y_{\bm3}^{(2)}]_{\bm1}[\bar e]_{\bm1'}[e^c]_{\bm1''}
\oplus 
y_{\cS_2}^{(2)}[\cS_2^\dag \bar Y_{\bm3}^{(2)}]_{\bm1}[\bar e]_{\bm1''}[e^c]_{\bm1'}
$
\\
&&&
\vspace{+0.05cm}$
\oplus y_{\cS_2}^{(3)}[\cS_2^\dag \bar Y_{\bm3}^{(2)}]_{\bm1'}[\bar e]_{\bm1}[e^c]_{\bm1''}
\oplus
y_{\cS_2}^{(3)}[\cS_2^\dag \bar Y_{\bm3}^{(2)}]_{\bm1'}[\bar e]_{\bm1''}[e^c]_{\bm1}
\oplus 
y_{\cS_2}^{(4)}[\cS_2^\dag \bar Y_{\bm3}^{(2)}]_{\bm1'}[\bar e]_{\bm1'}[e^c]_{\bm1'}
\,$
&6
\\
&&&
\vspace{+0.05cm}$
\oplus y_{\cS_2}^{(5)}[\cS_2^\dag \bar Y_{\bm3}^{(2)}]_{\bm1''}[\bar e]_{\bm1}[e^c]_{\bm1'}
\oplus
y_{\cS_2}^{(5)}[\cS_2^\dag \bar Y_{\bm3}^{(2)}]_{\bm1''}[\bar e]_{\bm1'}[e^c]_{\bm1}
\oplus 
y_{\cS_2}^{(6)}[\cS_2^\dag \bar Y_{\bm3}^{(2)}]_{\bm1''}[\bar e]_{\bm1''}[e^c]_{\bm1''}
$
\\
\midrule
\multirow{6}{*}{\vspace{-0.5cm}$\varphi\sim(\bm1,\bm2)_{1/2}$}
&$(\bm3,-2)$
&$\cO(1)$
\vspace{+0.05cm}&
$y_\varphi^{(1)}[\varphi\bar\ell]_{\bm1}[e]_{\bm1}
\oplus
y_\varphi^{(2)}[\varphi\bar\ell]_{\bm1'}[e]_{\bm1''}
\oplus 
y_\varphi^{(3)}[\varphi\bar\ell]_{\bm1''}[e]_{\bm1'} 
$
&3
\\
&$(\bm1,0)$
&$\cO(\bar Y_{\bm3}^{(2)})$
&
\vspace{+0.05cm}$
y_\varphi^{(1)}[\varphi]_{\bm1}[\bar\ell \bar Y^{(2)}_{\bm3}]_{\bm1}[e]_{\bm1}
\oplus
y_\varphi^{(2)}[\varphi]_{\bm1}[\bar\ell \bar Y^{(2)}_{\bm3}]_{\bm1'}[e]_{\bm1''}
\oplus
y_\varphi^{(3)}[\varphi]_{\bm1}[\bar\ell \bar Y^{(2)}_{\bm3}]_{\bm1''}[e]_{\bm1'} 
$
&3
\\
&$(\bm1',0)$
&$\cO(\bar Y_{\bm3}^{(2)})$
&
\vspace{+0.05cm}$
y_\varphi^{(1)}[\varphi]_{\bm1'}[\bar\ell \bar Y^{(2)}_{\bm3}]_{\bm1}[e]_{\bm1''}
\oplus
y_\varphi^{(2)}[\varphi]_{\bm1'}[\bar\ell \bar Y^{(2)}_{\bm3}]_{\bm1'}[e]_{\bm1'}
\oplus
y_\varphi^{(3)}[\varphi]_{\bm1'}[\bar\ell \bar Y^{(2)}_{\bm3}]_{\bm1''}[e]_{\bm1} 
$
&3
\\
&$(\bm1'',0)$
&$\cO(\bar Y_{\bm3}^{(2)})$
&
\vspace{+0.05cm}$
y_\varphi^{(1)}[\varphi]_{\bm1''}[\bar\ell \bar Y^{(2)}_{\bm3}]_{\bm1}[e]_{\bm1'}
\oplus
y_\varphi^{(2)}[\varphi]_{\bm1''}[\bar\ell \bar Y^{(2)}_{\bm3}]_{\bm1'}[e]_{\bm1}
\oplus
y_\varphi^{(3)}[\varphi]_{\bm1''}[\bar\ell \bar Y^{(2)}_{\bm3}]_{\bm1''}[e]_{\bm1''} 
$
&3
\\
\cmidrule{2-5}
&\multirow{2}{*}{$(\bm3,0)$}
&\multirow{2}{*}{$\cO(\bar Y_{\bm3}^{(2)})$}
&
\vspace{+0.05cm}$
~~y_\varphi^{(1)}[[\varphi\bar\ell]_{\bm3_S} \bar Y_{\bm3}^{(2)}]_{\bm1}[e]_{\bm1}
\oplus
y_\varphi^{(2)}[[\varphi\bar\ell]_{\bm3_S} \bar Y_{\bm3}^{(2)}]_{\bm1''}[e]_{\bm1'}
\oplus 
y_\varphi^{(3)}[[\varphi\bar\ell]_{\bm3_S} \bar Y_{\bm3}^{(2)}]_{\bm1'}[e]_{\bm1''}
$
&\multirow{2}{*}{6}
\\
&&&
\vspace{+0.05cm}$
\oplus y_\varphi^{(4)}[[\varphi\bar\ell]_{\bm3_A} \bar Y_{\bm3}^{(2)}]_{\bm1}[e]_{\bm1}
\oplus
y_\varphi^{(5)}[[\varphi\bar\ell]_{\bm3_A}\bar Y_{\bm3}^{(2)}]_{\bm1''}[e]_{\bm1'}
\oplus 
y_\varphi^{(6)}[[\varphi\bar\ell]_{\bm3_A} \bar Y_{\bm3}^{(2)}]_{\bm1'}[e]_{\bm1''}
$
\\
\midrule
\multirow{9}{*}{\vspace{-0.8cm}$\Xi_1\sim(\bm1,\bm3)_1$}
&$(\bm1,4)$
&$\cO(1)$
&
\vspace{+0.05cm}$y_\Xi^{(1)}[\Xi_1^\dag]_{\bm1}[\bar\ell\ell^c]_{\bm1}$
&1
\\
&$(\bm1',4)$
&$\cO(1)$
&\vspace{+0.05cm}$y_\Xi^{(1)}[\Xi_1^\dag]_{\bm1'}[\bar\ell\ell^c]_{\bm1''}$
&1
\\
&$(\bm1'',4)$
&$\cO(1)$
&\vspace{+0.05cm}$y_\Xi^{(1)}[\Xi_1^\dag]_{\bm1''}[\bar\ell\ell^c]_{\bm1'}$
&1
\\
&$(\bm3,4)$
&$\cO(1)$
&\vspace{+0.05cm}$y_\Xi^{(1)}[\Xi_1^\dag]_{\bm3}[\bar\ell \ell^c]_{\bm3_S}$
&1
\\
&$(\bm1,2)$
&$\cO(\bar Y^{(2)}_{\bm3})$
&\vspace{+0.05cm}$y_\Xi^{(1)}[\Xi_1^\dag]_{\bm1}[[\bar\ell\ell^c]_{\bm3_S} \bar Y^{(2)}_{\bm3}]_{\bm1}$
&1
\\
&$(\bm1',2)$
&$\cO(\bar Y^{(2)}_{\bm3})$
&\vspace{+0.05cm}$y_\Xi^{(1)}[\Xi_1^\dag]_{\bm1'}[[\bar\ell\ell^c]_{\bm3_S} \bar Y^{(2)}_{\bm3}]_{\bm1''}$
&1
\\
&$(\bm1'',2)$
&$\cO(\bar Y^{(2)}_{\bm3})$
&\vspace{+0.05cm}$y_\Xi^{(1)}[\Xi_1^\dag]_{\bm1''}[[\bar\ell\ell^c]_{\bm3_S} \bar Y^{(2)}_{\bm3}]_{\bm1'}$
&1
\\
\cmidrule{2-5}
&\multirow{2}{*}{$(\bm3,2)$}
&\multirow{2}{*}{$\cO(\bar Y^{(2)}_{\bm3})$}
&
\vspace{+0.05cm}$y_\Xi^{(1)}[\bar\ell\ell^c]_{\bm1}[\Xi_1^\dag \bar Y^{(2)}_{\bm3}]_{\bm1}
\oplus
y_\Xi^{(2)}[\bar\ell\ell^c]_{\bm1'}[\Xi_1^\dag \bar Y^{(2)}_{\bm3}]_{\bm1''} 
\oplus
y_\Xi^{(3)}[\bar\ell\ell^c]_{\bm1''}[\Xi_1^\dag \bar Y^{(2)}_{\bm3}]_{\bm1'}
$
&\multirow{2}{*}{5}
\\
&&&
\vspace{+0.05cm}$
\oplus
y_\Xi^{(4)}[\bar\ell\ell^c]_{\bm3_S}[\Xi_1^\dag \bar Y^{(2)}_{\bm3}]_{\bm3_S}
\oplus
y_\Xi^{(5)}[\bar\ell\ell^c]_{\bm3_S}[\Xi_1^\dag \bar Y^{(2)}_{\bm3}]_{\bm3_A} 
$
\\
\bottomrule
\end{tabular}
}
    \caption{Overview of the possible $A_4$ flavor irreducible representations at $\cO(1)$ and $\cO(\bar Y_{\bm3}^{(2)})$ for UV scalar fields. In the first column we collect the scalar fields along with their SM gauge representations. In the second and third column we denote the possible $A_4$ flavor irreps and the order in the Yukawa expansion at which they are extracted. In the third column we collect the possible flavor invariants and in the last column we note how many independent parameters are implied by each set of invariants. For more details see discussion in Sec.~\ref{sec:UV_flavor_invs}.}
    \label{tab:scalars_a4_irreps_invs}
\end{table*}

\begin{table*}[t]
    \centering
\scalebox{0.85}
{
\begin{tabular}{ccc|@{\hspace{0.5cm}}cc}
\toprule
\multirow{1}{*}{\textbf{UV Field}}
&\multirow{1}{*}{\textbf{$\bm{(A_4,k)}$}}
&\multirow{1}{*}{\textbf{Order}} 
&\multirow{1}{*}{\textbf{Invariants}} 
&\multirow{1}{*}{\textbf{\# of parameters}}
\\
\midrule
\multirow{5}{*}{\vspace{-0.5cm}$N\sim(\bm1,\bm1)_{0}$}
&$(\bm3,2)$&$\cO(1)$
&\vspace{+0.05cm}$\lambda_N^{(1)}[\bar N_R\ell ]_{\bm1}$
&1
\\
&$(\bm1,4)$&$\cO(\bar Y_{\bm3}^{(2)})$
&\vspace{+0.05cm}$\lambda_N^{(1)}[\bar N_R]_{\bm1}[\ell \bar Y_{\bm3}^{(2)}]_{\bm1}$
&1
\\
&$(\bm1',4)$&$\cO(\bar Y_{\bm3}^{(2)})$
&\vspace{+0.05cm}$\lambda_N^{(1)}[\bar N_R]_{\bm1'}[\ell \bar Y_{\bm3}^{(2)}]_{\bm1''}$
&1
\\
&$(\bm1'',4)$&$\cO(\bar Y_{\bm3}^{(2)})$
&\vspace{+0.05cm}$\lambda_N^{(1)}[\bar N_R]_{\bm1''}[\ell \bar Y_{\bm3}^{(2)}]_{\bm1'}$
&1
\\
&$(\bm3,4)$&$\cO(\bar Y_{\bm3}^{(2)})$
&\vspace{+0.05cm}$
\lambda_N^{(1)}[[\bar N_R\ell]_{\bm3_S}\bar Y_{\bm3}^{(2)}]_{\bm1}
\oplus 
\lambda_N^{(2)}[[\bar N_R\ell]_{\bm3_A}\bar Y_{\bm3}^{(2)}]_{\bm1}
$
&2
\\
\midrule
\multirow{5}{*}{\vspace{-0.5cm}$E\sim(\bm1,\bm1)_{-1}$}
&$(\bm3,2)$&$\cO(1)$
&\vspace{+0.05cm}$\lambda_E^{(1)}[\bar E_R\ell ]_{\bm1}$
&1
\\
&$(\bm1,4)$&$\cO(\bar Y_{\bm3}^{(2)})$
&\vspace{+0.05cm}$\lambda_E^{(1)}[\bar E_R]_{\bm1}[\ell \bar Y_{\bm3}^{(2)}]_{\bm1}$
&1
\\
&$(\bm1',4)$&$\cO(\bar Y_{\bm3}^{(2)})$
&\vspace{+0.05cm}$\lambda_E^{(1)}[\bar E_R]_{\bm1'}[\ell \bar Y_{\bm3}^{(2)}]_{\bm1''}$
&1
\\
&$(\bm1'',4)$&$\cO(\bar Y_{\bm3}^{(2)})$
&\vspace{+0.05cm}$\lambda_E^{(1)}[\bar E_R]_{\bm1''}[\ell \bar Y_{\bm3}^{(2)}]_{\bm1'}$
&1
\\
&$(\bm3,4)$&$\cO(\bar Y_{\bm3}^{(2)})$
&
\vspace{+0.05cm}$
\lambda_E^{(1)}[[\bar E_R\ell]_{\bm3_S}\bar Y_{\bm3}^{(2)}]_{\bm1}
\oplus 
\lambda_E^{(2)}[[\bar E_R\ell]_{\bm3_A}\bar Y_{\bm3}^{(2)}]_{\bm1}
$
&2
\\
\midrule
\multirow{4}{*}{\vspace{-0.4cm}$\Delta_1\sim(\bm1,\bm2)_{-1/2}$}
&$(\bm1,0)$
&$\cO(1)$
&\vspace{+0.05cm}$\lambda_{\Delta_1}^{(1)}[\bar\Delta_{1L}]_{\bm1}[e]_{\bm1}$
&1
\\
&$(\bm1',0)$
&$\cO(1)$
&\vspace{+0.05cm}$\lambda_{\Delta_1}^{(1)}[\bar\Delta_{1L}]_{\bm1'}[e]_{\bm1''}$
&1
\\
&
$(\bm1'',0)$
&$\cO(1)$
&\vspace{+0.05cm}$\lambda_{\Delta_1}^{(1)}[\bar\Delta_{1L}]_{\bm1''}[e]_{\bm1'}$
&1
\\
&$(\bm3,2)$
&$\cO(\bar Y_{\bm3}^{(2)})$
&
\vspace{+0.05cm}$
~~~\lambda_{\Delta_1}^{(1)}[\bar\Delta_{1L}\bar Y_{\bm3}^{(2)}]_{\bm1}[e]_{\bm1}
\oplus 
\lambda_{\Delta_1}^{(2)}[\bar\Delta_{1L}\bar Y_{\bm3}^{(2)}]_{\bm1'}[e]_{\bm1''}
\oplus
\lambda_{\Delta_1}^{(3)}[\bar\Delta_{1L}\bar Y_{\bm3}^{(2)}]_{\bm1''}[e]_{\bm1'}
$
&3
\\
\midrule
\multirow{4}{*}{\vspace{-0.5cm}$\Delta_3\sim(\bm1,\bm2)_{-3/2}$}
&$(\bm1,0)$
&$\cO(1)$
&\vspace{+0.05cm}$\lambda_{\Delta_3}^{(1)}[\bar\Delta_{3L}]_{\bm1}[e]_{\bm1}$
&1
\\
&$(\bm1',0)$
&$\cO(1)$
&\vspace{+0.05cm}$\lambda_{\Delta_3}^{(1)}[\bar\Delta_{3L}]_{\bm1'}[e]_{\bm1''}$
&1
\\
&
$(\bm1'',0)$
&$\cO(1)$
&\vspace{+0.05cm}$\lambda_{\Delta_3}^{(1)}[\bar\Delta_{3L}]_{\bm1''}[e]_{\bm1'}$
&1
\\
&$(\bm3,2)$
&$\cO(\bar Y_{\bm3}^{(2)})$
&
\vspace{+0.05cm}$
~~~\lambda_{\Delta_3}^{(1)}[\bar\Delta_{3L}\bar Y_{\bm3}^{(2)}]_{\bm1}[e]_{\bm1}
\oplus 
\lambda_{\Delta_3}^{(2)}[\bar\Delta_{3L}\bar Y_{\bm3}^{(2)}]_{\bm1'}[e]_{\bm1''}
\oplus
\lambda_{\Delta_3}^{(3)}[\bar\Delta_{3L}\bar Y_{\bm3}^{(2)}]_{\bm1''}[e]_{\bm1'}
$
&3
\\
\midrule
\multirow{5}{*}{\vspace{-0.4cm}$\Sigma\sim(\bm1,\bm3)_{0}$}
&$(\bm3,2)$
&$\cO(1)$
&\vspace{+0.05cm}$\lambda_{\Sigma}^{(1)}[\bar \Sigma_R\ell ]_{\bm1}$
&1
\\
&$(\bm1,0)$
&$\cO(Y_{\bm3}^{(2)})$
&\vspace{+0.05cm}$\lambda_{\Sigma}^{(1)}[\bar \Sigma_R]_{\bm1}[\ell  Y_{\bm3}^{(2)}]_{\bm1}$
&1
\\
&$(\bm1',0)$
&$\cO(Y_{\bm3}^{(2)})$
&\vspace{+0.05cm}$\lambda_{\Sigma}^{(1)}[\bar \Sigma_R]_{\bm1'}[\ell  Y_{\bm3}^{(2)}]_{\bm1''}$
&1
\\
&$(\bm1'',0)$
&$\cO(Y_{\bm3}^{(2)})$
&\vspace{+0.05cm}$\lambda_{\Sigma}^{(1)}[\bar \Sigma_R]_{\bm1''}[\ell  Y_{\bm3}^{(2)}]_{\bm1'}$
&1
\\
&$(\bm3,0)$
&$\cO(Y_{\bm3}^{(2)})$
&\vspace{+0.05cm}$
\lambda_{\Sigma}^{(1)}[[\bar \Sigma_R\ell]_{\bm3_S} Y_{\bm3}^{(2)}]_{\bm1}
\oplus 
\lambda_{\Sigma}^{(2)}[[\bar \Sigma_R\ell]_{\bm3_A} Y_{\bm3}^{(2)}]_{\bm1}
$
&2
\\
\midrule
\multirow{5}{*}{\vspace{-0.5cm}$\Sigma_1\sim(\bm1,\bm3)_{-1}$}
&$(\bm3,2)$
&$\cO(1)$
&\vspace{+0.05cm}$\lambda_{\Sigma_1}^{(1)}[\bar \Sigma_{1R}\ell ]_{\bm1}$
&1
\\
&$(\bm1,0)$
&$\cO(Y_{\bm3}^{(2)})$
&\vspace{+0.05cm}$\lambda_{\Sigma_1}^{(1)}[\bar \Sigma_{1R}]_{\bm1}[\ell  Y_{\bm3}^{(2)}]_{\bm1}$
&1
\\
&$(\bm1',0)$
&$\cO(Y_{\bm3}^{(2)})$
&\vspace{+0.05cm}$\lambda_{\Sigma_1}^{(1)}[\bar \Sigma_{1R}]_{\bm1'}[\ell  Y_{\bm3}^{(2)}]_{\bm1''}$
&1
\\
&$(\bm1'',0)$
&$\cO(Y_{\bm3}^{(2)})$
&\vspace{+0.05cm}$\lambda_{\Sigma_1}^{(1)}[\bar \Sigma_{1R}]_{\bm1''}[\ell  Y_{\bm3}^{(2)}]_{\bm1'}$
&1
\\
&$(\bm3,0)$
&$\cO(Y_{\bm3}^{(2)})$
&\vspace{+0.05cm}$
\lambda_{\Sigma_1}^{(1)}[[\bar \Sigma_{1R}\ell]_{\bm3_S} Y_{\bm3}^{(2)}]_{\bm1}
\oplus 
\lambda_{\Sigma_1}^{(2)}[[\bar \Sigma_{1R}\ell]_{\bm3_A} Y_{\bm3}^{(2)}]_{\bm1}
$
&2
\\
\bottomrule
\end{tabular}
}
    \caption{Overview of the possible $A_4$ flavor irreducible representations at $\cO(1)$ and $\cO( Y_{\bm3}^{(2)},\bar Y_{\bm3}^{(2)})$ for UV fermion fields. In the first column we collect the fermion fields along with their SM gauge representations. In the second and third column we denote the possible $A_4$ flavor irreps and the order in the Yukawa expansion at which they are extracted. In the third column we collect the possible flavor invariants and in the last column we note how many independent parameters are implied by each set of invariants. For more details see discussion in Sec.~\ref{sec:UV_flavor_invs}.}
    \label{tab:fermions_a4_irreps_invs}
\end{table*}

\begin{table*}[t]
    \centering
\scalebox{0.8}{
\begin{tabular}{ccc|@{\hspace{0.5cm}}cc}
\toprule
\multirow{1}{*}{\textbf{UV Field}}
&\multirow{1}{*}{\textbf{$\bm{(A_4,k)}$}}
&\multirow{1}{*}{\textbf{Order}} 
&\multirow{1}{*}{\textbf{Invariants}} 
&\multirow{1}{*}{\textbf{\# of parameters}}
\\
\midrule
\multirow{12}{*}{\vspace{-1cm}$\cB\sim(\bm1,\bm1)_0$}
&$(\bm1,0)$
&$\cO(1)$
&
\vspace{+0.05cm}$
g_\cB^{\ell(1)}[\cB_\mu]_{\bm1}[\bar\ell\ell]_{\bm1}
\oplus 
g_\cB^{e(1)}[\cB_\mu]_{\bm1}[\bar e]_{\bm1}[e]_{\bm1}
\oplus g_\cB^{e(2)}[\cB_\mu]_{\bm1}[\bar e]_{\bm1'}[e]_{\bm1''}
\oplus g_\cB^{e(3)}[\cB_\mu]_{\bm1}[\bar e]_{\bm1''}[e]_{\bm1'}
$
&4
\\
&$(\bm1',0)$&
$\cO(1)$
&
\vspace{+0.05cm}$
g_\cB^{\ell(1)}[\cB_\mu]_{\bm1'}[\bar\ell\ell]_{\bm1''}
\oplus g_\cB^{e(1)}[\cB_\mu]_{\bm1'}[\bar e]_{\bm1'}[e]_{\bm1'}
\oplus g_\cB^{e(2)}[\cB_\mu]_{\bm1'}[\bar e]_{\bm1}[e]_{\bm1''}
\oplus g_\cB^{e(3)}[\cB_\mu]_{\bm1'}[\bar e]_{\bm1''}[e]_{\bm1}
$
&4
\\
&$(\bm1'',0)$
&$\cO(1)$
&
\vspace{+0.05cm}$
g_\cB^{\ell(1)}[\cB_\mu]_{\bm1''}[\bar\ell\ell]_{\bm1'}
\oplus g_\cB^{e(1)}[\cB_\mu]_{\bm1''}[\bar e]_{\bm1}[e]_{\bm1'}
\oplus g_\cB^{e(2)}[\cB_\mu]_{\bm1''}[\bar e]_{\bm1'}[e]_{\bm1'}
\oplus g_\cB^{e(3)}[\cB_\mu]_{\bm1''}[\bar e]_{\bm1''}[e]_{\bm1''}
$
&4
\\
&$(\bm3,0)$
&$\cO(1)$
&
\vspace{+0.05cm}$
g_\cB^{\ell(1)} [\cB_\mu]_{\bm3}[\bar\ell\ell]_{\bm3_S}
\oplus
g_\cB^{\ell(2)}[\cB_\mu]_{\bm3}[\bar\ell\ell]_{\bm3_A}
$
&2
\\
&$(\bm1,2)$
&$\cO(\bar Y_{\bm3}^{(2)})$
&\vspace{+0.05cm}$
g_\cB^{\ell(1)}[\cB_\mu]_{\bm1}[[\bar\ell\ell]_{\bm3_S}Y_{\bm3}^{(2)}]_{\bm1}\oplus 
g_\cB^{\ell(2)}[\cB_\mu]_{\bm1}[[\bar\ell\ell]_{\bm3_A}Y_{\bm3}^{(2)}]_{\bm1}$
&2
\\
&$(\bm1',2)$
&$\cO(\bar Y_{\bm3}^{(2)})$
&\vspace{+0.05cm}$
g_\cB^{\ell(1)}[\cB_\mu]_{\bm1'}[[\bar\ell\ell]_{\bm3_S}Y_{\bm3}^{(2)}]_{\bm1''}\oplus 
g_\cB^{\ell(2)}[\cB_\mu]_{\bm1'}[[\bar\ell\ell]_{\bm3_A}Y_{\bm3}^{(2)}]_{\bm1''}$
&2
\\
&$(\bm1'',2)$
&$\cO(\bar Y_{\bm3}^{(2)})$
&\vspace{+0.05cm}$
g_\cB^{\ell(1)}[\cB_\mu]_{\bm1''}[[\bar\ell\ell]_{\bm3_S}Y_{\bm3}^{(2)}]_{\bm1'}\oplus 
g_\cB^{\ell(2)}[\cB_\mu]_{\bm1''}[[\bar\ell\ell]_{\bm3_A}Y_{\bm3}^{(2)}]_{\bm1'}$
&2
\\
\cmidrule{2-5}
&\multirow{5}{*}{\vspace{-0.45cm}$(\bm3,2)$}
&\multirow{5}{*}{\vspace{-0.45cm}$\cO(\bar Y_{\bm3}^{(2)})$}
&
\vspace{+0.05cm}$
g_\cB^{\ell(1)}[\cB_\mu Y_{\bm3}^{(2)}]_{\bm1}[\bar\ell\ell]_{\bm1}\oplus 
g_\cB^{\ell(2)}[\cB_\mu Y_{\bm3}^{(2)}]_{\bm1'}[\bar\ell\ell]_{\bm1''}\oplus
g_\cB^{\ell(3)}[\cB_\mu Y_{\bm3}^{(2)}]_{\bm1''}[\bar\ell\ell]_{\bm1'}\oplus g_\cB^{\ell(4)}[\cB_\mu Y_{\bm3}^{(2)}]_{\bm3_S}[\bar\ell\ell]_{\bm3_S}
$
\\
&&&
\vspace{+0.05cm}$
\oplus g_\cB^{\ell(5)}[\cB_\mu Y_{\bm3}^{(2)}]_{\bm3_A}[\bar\ell\ell]_{\bm3_S}
\oplus g_\cB^{\ell(6)}[\cB_\mu Y_{\bm3}^{(2)}]_{\bm3_S}[\bar\ell\ell]_{\bm3_A}
\oplus g_\cB^{\ell(7)}[\cB_\mu Y_{\bm3}^{(2)}]_{\bm3_A}[\bar\ell\ell]_{\bm3_A}~~~~~~~~~~~~~~
$
\\
&&&
\vspace{+0.05cm}$
\oplus g_\cB^{e(1)}[\cB_\mu Y_{\bm3}^{(2)}]_{\bm1}[\bar e]_{\bm1}[e]_{\bm1} 
\oplus g_\cB^{e(2)}[\cB_\mu Y_{\bm3}^{(2)}]_{\bm1}[\bar e]_{\bm1'}[e]_{\bm1''}
\oplus g_\cB^{e(3)}[\cB_\mu Y_{\bm3}^{(2)}]_{\bm1}[\bar e]_{\bm1''}[e]_{\bm1'}~~~~~~~~~
$
&
16
\\
&&&
\vspace{+0.05cm}$
\oplus g_\cB^{e(4)}[\cB_\mu Y_{\bm3}^{(2)}]_{\bm1'}[\bar e]_{\bm1}[e]_{\bm1''}
\oplus g_\cB^{e(5)}[\cB_\mu Y_{\bm3}^{(2)}]_{\bm1'}[\bar e]_{\bm1''}[e]_{\bm1}
\oplus g_\cB^{e(6)}[\cB_\mu Y_{\bm3}^{(2)}]_{\bm1'}[\bar e]_{\bm1'}[e]_{\bm1'}~~~~~~~    
$
\\
&&&
\vspace{+0.05cm}$
\oplus g_\cB^{e(7)}[\cB_\mu Y_{\bm3}^{(2)}]_{\bm1''}[\bar e]_{\bm1}[e]_{\bm1'}
\oplus g_\cB^{e(8)}[\cB_\mu Y_{\bm3}^{(2)}]_{\bm1''}[\bar e]_{\bm1'}[e]_{\bm1}
\oplus g_\cB^{e(9)}[\cB_\mu Y_{\bm3}^{(2)}]_{\bm1''}[\bar e]_{\bm1''}[e]_{\bm1''}~~~~~
$
\\
\midrule
\multirow{10}{*}{\vspace{-0.8cm}$\cW\sim(\bm1,\bm3)_0$}
&$(\bm1,0)$
&$\cO(1)$
&\vspace{+0.05cm}$g_\cW^{\ell(1)}[\cW_\mu]_{\bm1}[\bar\ell\ell]_{\bm1}$
&1
\\
&$(\bm1',0)$
&$\cO(1)$
&\vspace{+0.05cm}$g_\cW^{\ell(1)}[\cW_\mu]_{\bm1'}[\bar\ell\ell]_{\bm1''}$
&1
\\
&$(\bm1'',0)$
&$\cO(1)$
&\vspace{+0.05cm}$g_\cW^{\ell(1)}[\cW_\mu]_{\bm1''}[\bar\ell\ell]_{\bm1'}$
&1
\\
&$(\bm3,0)$
&$\cO(1)$
&\vspace{+0.05cm}$
g_\cW^{\ell(1)}[\cW_\mu]_{\bm3}[\bar\ell\ell]_{\bm3_S}
\oplus
g_\cW^{\ell(2)}[\cW_\mu]_{\bm3}[\bar\ell\ell]_{\bm3_A}
$
&2
\\
&$(\bm1,2)$
&$\cO(\bar Y_{\bm3}^{(2)})$
&\vspace{+0.05cm}$
g_\cW^{\ell(1)}[\cW_\mu]_{\bm1}[[\bar\ell\ell]_{\bm3_S}Y_{\bm3}^{(2)}]_{\bm1}\oplus 
g_\cW^{\ell(2)}[\cW_\mu]_{\bm1}[[\bar\ell\ell]_{\bm3_A}Y_{\bm3}^{(2)}]_{\bm1}$
&2
\\
&$(\bm1',2)$
&$\cO(\bar Y_{\bm3}^{(2)})$
&\vspace{+0.05cm}$
g_\cW^{\ell(1)}[\cW_\mu]_{\bm1'}[[\bar\ell\ell]_{\bm3_S}Y_{\bm3}^{(2)}]_{\bm1''}\oplus 
g_\cW^{\ell(2)}[\cW_\mu]_{\bm1'}[[\bar\ell\ell]_{\bm3_A}Y_{\bm3}^{(2)}]_{\bm1''}$
&2
\\
&$(\bm1'',2)$
&$\cO(\bar Y_{\bm3}^{(2)})$
&\vspace{+0.05cm}$
g_\cW^{\ell(1)}[\cW_\mu]_{\bm1''}[[\bar\ell\ell]_{\bm3_S}Y_{\bm3}^{(2)}]_{\bm1'}\oplus 
g_\cW^{\ell(2)}[\cW_\mu]_{\bm1''}[[\bar\ell\ell]_{\bm3_A}Y_{\bm3}^{(2)}]_{\bm1'}$
&2
\\
\cmidrule{2-5}
&\multirow{2}{*}{$(\bm3,2)$}
&\multirow{2}{*}{$\cO(\bar Y_{\bm3}^{(2)})$}
&
\vspace{+0.05cm}$
g_\cW^{\ell(1)}[\cW_\mu Y_{\bm3}^{(2)}]_{\bm1}[\bar\ell\ell]_{\bm1}
\oplus 
g_\cW^{\ell(2)}[\cW_\mu Y_{\bm3}^{(2)}]_{\bm1'}[\bar\ell\ell]_{\bm1''}\oplus
g_\cW^{\ell(3)}[\cW_\mu Y_{\bm3}^{(2)}]_{\bm1''}[\bar\ell\ell]_{\bm1'}\oplus
g_\cW^{\ell(4)}[\cW_\mu Y_{\bm3}^{(2)}]_{\bm3_S}[\bar\ell\ell]_{\bm3_S}$
&\multirow{2}{*}{7}
\\
&&&
\vspace{+0.05cm}$ 
\oplus g_\cW^{\ell(5)}[\cW_\mu Y_{\bm3}^{(2)}]_{\bm3_A}[\bar\ell\ell]_{\bm3_S}
\oplus g_\cW^{\ell(6)}[\cW_\mu Y_{\bm3}^{(2)}]_{\bm3_S}[\bar\ell\ell]_{\bm3_A}
\oplus g_\cW^{\ell(7)}[\cW_\mu Y_{\bm3}^{(2)}]_{\bm3_A}[\bar\ell\ell]_{\bm3_A}
~~~~~~~~~~~$
\\
\midrule
\multirow{7}{*}{\vspace{-0.45cm}$\cL_3\sim(\bm1,\bm2)_{-3/2}$}
&$(\bm3,-2)$
&$\cO(1)$
&
\vspace{+0.05cm}$
g_{\cL_3}^{(1)}[\bar e^c]_{\bm1}[[\cL_3^\dag]_{\bm3}[\ell]_{\bm3}]_{\bm1}
\oplus
g_{\cL_3}^{(2)}[\bar e^c]_{\bm1''}[[\cL_3^\dag]_{\bm3}[\ell]_{\bm3}]_{\bm1'}
\oplus
g_{\cL_3}^{(3)}[\bar e^c]_{\bm1'}[[\cL_3^\dag]_{\bm3}[\ell]_{\bm3}]_{\bm1''}
$
&3
\\
&$(\bm1,0)$
&$\cO(Y_{\bm3}^{(2)})$
&
\vspace{+0.05cm}$
g_{\cL_3}^{(1)}[\bar e^c]_{\bm1}[\cL_3^\dag]_{\bm1}[Y_{\bm3}^{(2)}\ell]_{\bm1}\oplus
g_{\cL_3}^{(2)}[\bar e^c]_{\bm1'}[\cL_3^\dag]_{\bm1}[Y_{\bm3}^{(2)}\ell]_{\bm1''}\oplus
g_{\cL_3}^{(3)}[\bar e^c]_{\bm1''}[\cL_3^\dag]_{\bm1}[Y_{\bm3}^{(2)}\ell]_{\bm1'}
$
&3
\\
&$(\bm1',0)$
&$\cO(Y_{\bm3}^{(2)})$
&
\vspace{+0.05cm}$
g_{\cL_3}^{(1)}[\bar e^c]_{\bm1}[\cL_3^\dag]_{\bm1'}[Y_{\bm3}^{(2)}\ell]_{\bm1''}\oplus
g_{\cL_3}^{(2)}[\bar e^c]_{\bm1''}[\cL_3^\dag]_{\bm1'}[Y_{\bm3}^{(2)}\ell]_{\bm1}\oplus
g_{\cL_3}^{(3)}[\bar e^c]_{\bm1'}[\cL_3^\dag]_{\bm1'}[Y_{\bm3}^{(2)}\ell]_{\bm1'}
$
&3
\\
&$(\bm1'',0)$
&$\cO(Y_{\bm3}^{(2)})$
&
\vspace{+0.05cm}$
g_{\cL_3}^{(1)}[\bar e^c]_{\bm1}[\cL_3^\dag]_{\bm1''}[Y_{\bm3}^{(2)}\ell]_{\bm1'}\oplus
g_{\cL_3}^{(2)}[\bar e^c]_{\bm1'}[\cL_3^\dag]_{\bm1''}[Y_{\bm3}^{(2)}\ell]_{\bm1}\oplus
g_{\cL_3}^{(3)}[\bar e^c]_{\bm1''}[\cL_3^\dag]_{\bm1''}[Y_{\bm3}^{(2)}\ell]_{\bm1''}
$
&3
\\
\cmidrule{2-5}
&\multirow{2}{*}{$(\bm3,0)$}
&\multirow{2}{*}{$\cO(Y_{\bm3}^{(2)})$}
&
\vspace{+0.05cm}$
g_{\cL_3}^{(1)}[\bar e^c]_{\bm1}[\cL_3^\dag [\ell Y_{\bm3}^{(2)}]_{\bm3_S}]_{\bm1}\oplus
g_{\cL_3}^{(2)}[\bar e^c]_{\bm1'}[\cL_3^\dag [\ell Y_{\bm3}^{(2)}]_{\bm3_S}]_{\bm1''}\oplus
g_{\cL_3}^{(3)}[\bar e^c]_{\bm1''}[\cL_3^\dag [\ell Y_{\bm3}^{(2)}]_{\bm3_S}]_{\bm1'}
$
&\multirow{2}{*}{6}
\\
&&&
\vspace{+0.05cm}$
\oplus
g_{\cL_3}^{(4)}[\bar e^c]_{\bm1}[\cL_3^\dag [\ell Y_{\bm3}^{(2)}]_{\bm3_A}]_{\bm1}\oplus
g_{\cL_3}^{(5)}[\bar e^c]_{\bm1'}[\cL_3^\dag [\ell Y_{\bm3}^{(2)}]_{\bm3_A}]_{\bm1''}\oplus
g_{\cL_3}^{(6)}[\bar e^c]_{\bm1''}[\cL_3^\dag [\ell Y_{\bm3}^{(2)}]_{\bm3_A}]_{\bm1'}
$
\\
\bottomrule
\end{tabular}
}
    \caption{Overview of the possible $A_4$ flavor irreducible representations at $\cO(1)$ and $\cO( Y_{\bm3}^{(2)},\bar Y_{\bm3}^{(2)})$ for UV vector fields. In the first column we collect the vector fields along with their SM gauge representations. In the second and third column we denote the possible $A_4$ flavor irreps and the order in the Yukawa expansion at which they are extracted. In the third column we collect the possible flavor invariants and in the last column we note how many independent parameters are implied by each set of invariants. For more details see discussion in Sec.~\ref{sec:UV_flavor_invs}.}
    \label{tab:vectors_a4_irreps_invs}
\end{table*}

\bibliography{letter_main}
\newpage

\onecolumngrid
\clearpage
\appendix
\section*{Supplemental Material}
\renewcommand{\theequation}{S\arabic{equation}}
\setcounter{equation}{0}

\subsection*{Analysis of Mediators with Multi-Parameter Flavor Tensors} 
\label{supl_sec:profiled_bounds}
\noindent
We focus here on mediators whose flavor tensors—derived from the allowed $A_4$ invariants—contain more than two independent parameters (see Tabs.~\ref{tab:scalars_a4_irreps_invs}–\ref{tab:vectors_a4_irreps_invs} for an overview). These scenarios lead to multidimensional parameter spaces, where correlations among flavor structures can significantly impact the resulting constraints. As noted in Sec.~\ref{sec:phenomenology_intro}, they require a more refined phenomenological treatment. To that end, we perform a combined fit to low-energy observables~\cite{Falkowski:2017pss,Breso-Pla:2023tnz}, with the number of fit parameters determined by the structure of the flavor tensors at $\mathcal{O}(1)$ or $\mathcal{O}(Y_{\bm{3}}^{(2)},\bar Y_{\bm{3}}^{(2)})$, depending on the mediator’s modular weight. We then profile over all but two of these parameters to obtain projected two-dimensional likelihoods, from which we extract the allowed regions. Below, we show several representative examples of profiled likelihood contours for different mediators.

\begin{figure*}[h]
\centering
\begin{tabular}{ccccc}
\includegraphics[width=32.50mm]{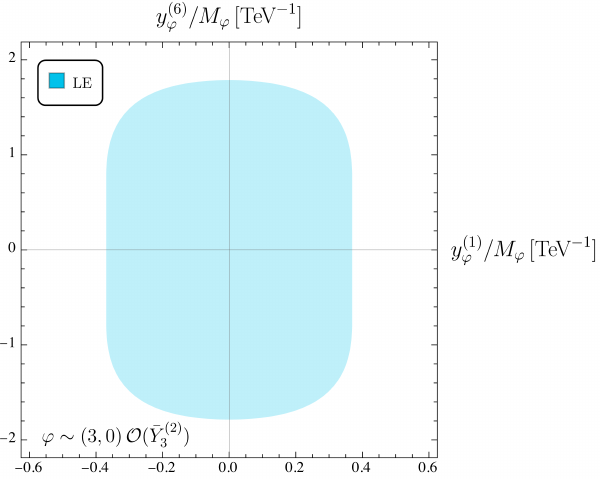} 
&   
\includegraphics[width=32.50mm]{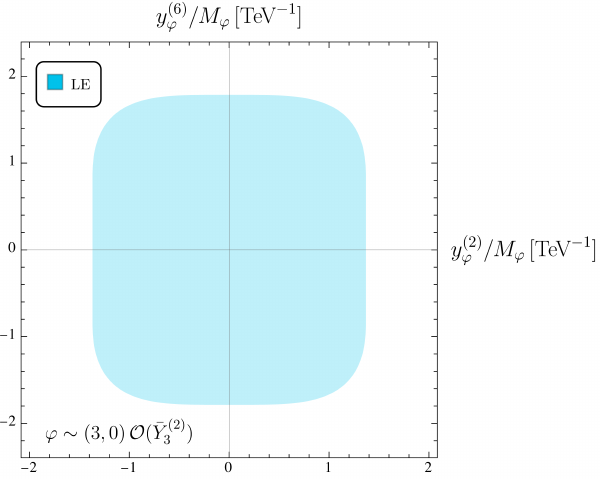}  
&
\includegraphics[width=32.50mm]{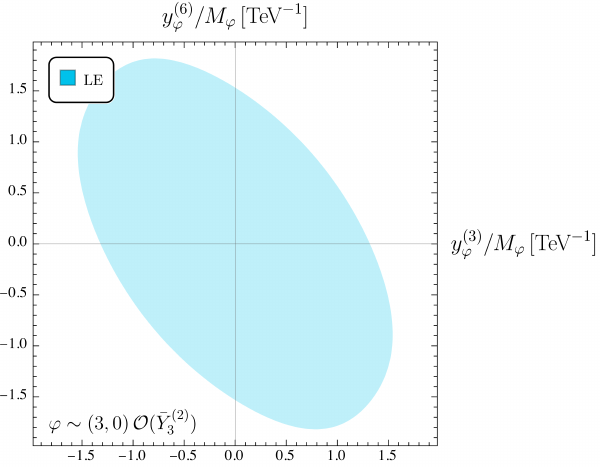}
&
\includegraphics[width=32.50mm]{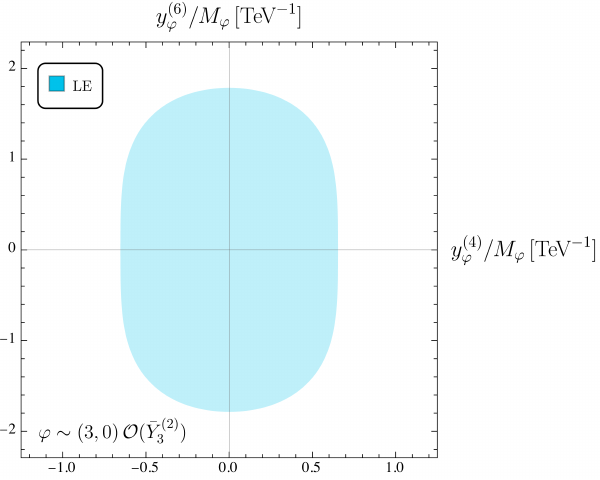} 
&   
\includegraphics[width=32.50mm]{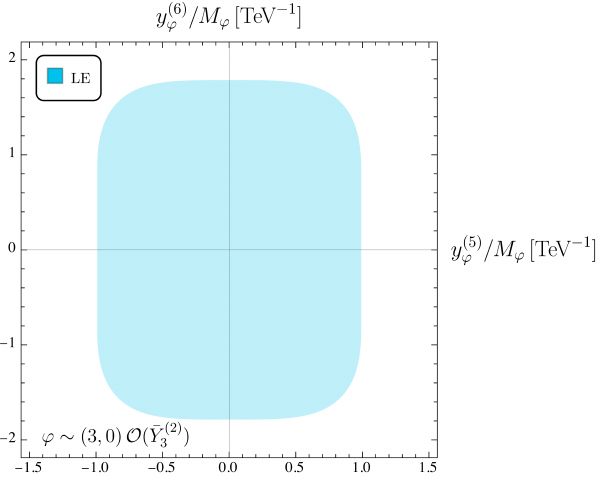}  
\\
\includegraphics[width=32.50mm]{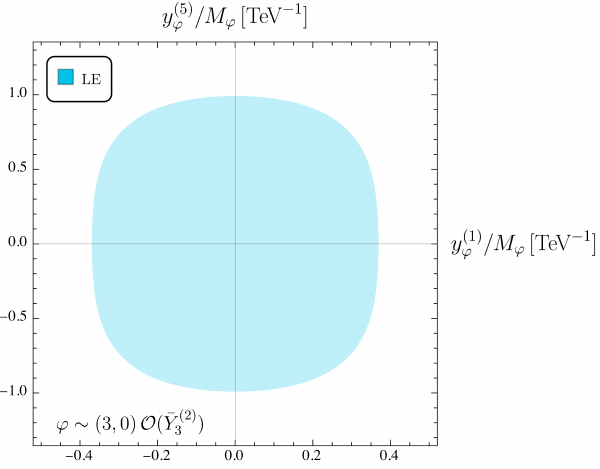}
&
\includegraphics[width=32.50mm]{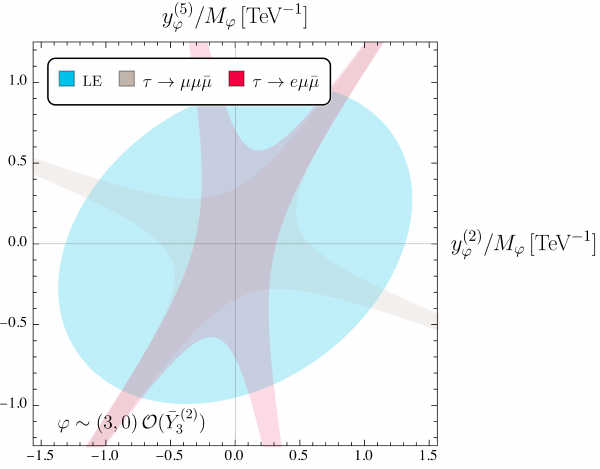} 
&   
\includegraphics[width=32.50mm]{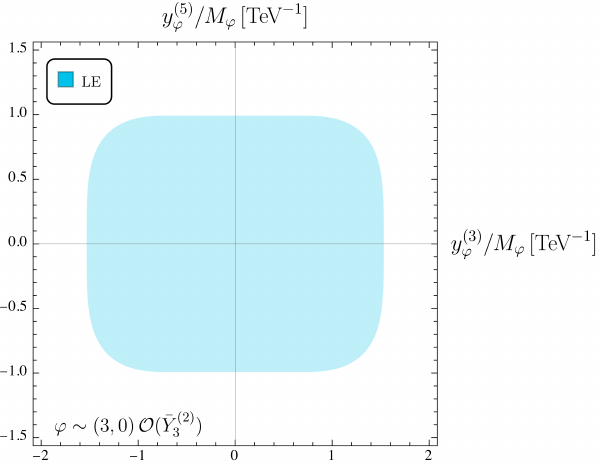}  
&
\includegraphics[width=32.50mm]{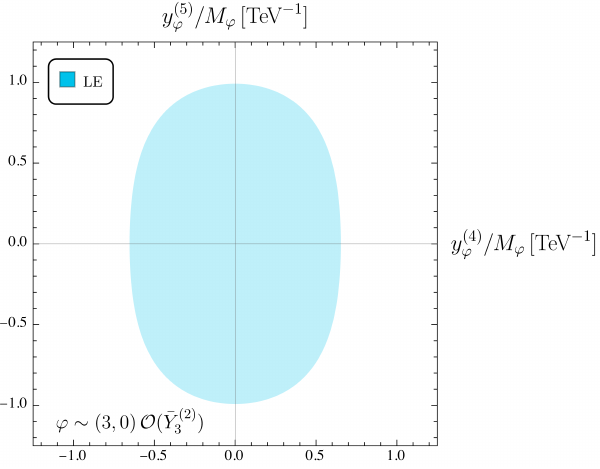}
&
\\
\includegraphics[width=32.50mm]{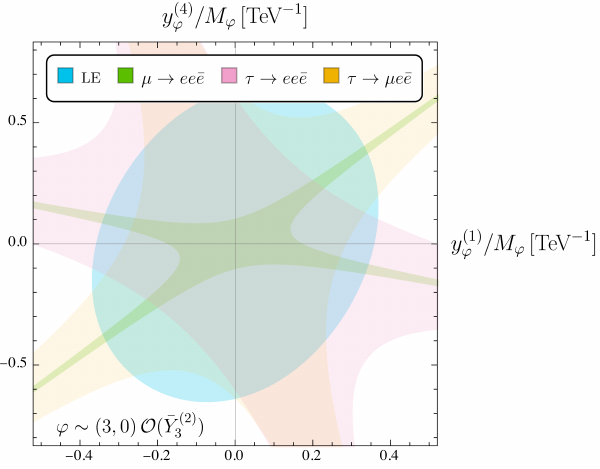}
&
\includegraphics[width=32.50mm]{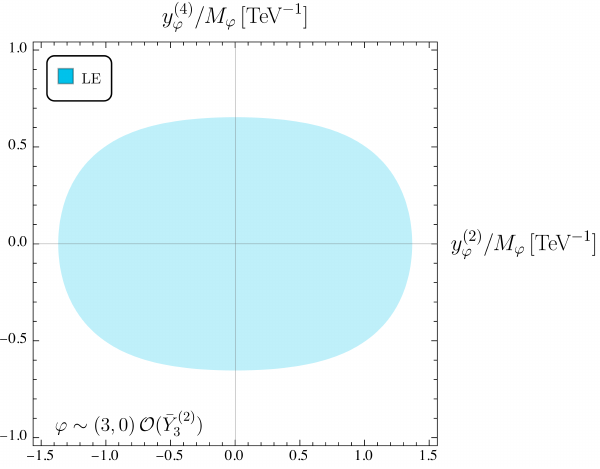}
&
\includegraphics[width=32.50mm]{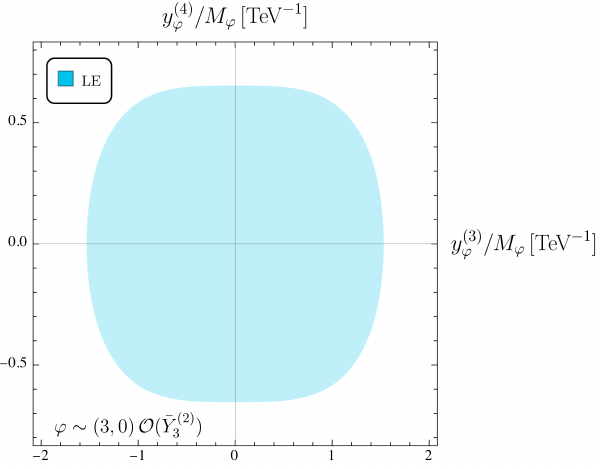}
&
&
\\
\includegraphics[width=32.50mm]{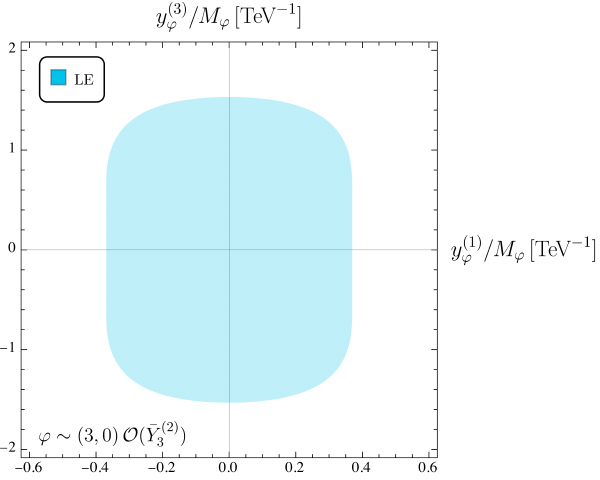}
&\includegraphics[width=32.50mm]{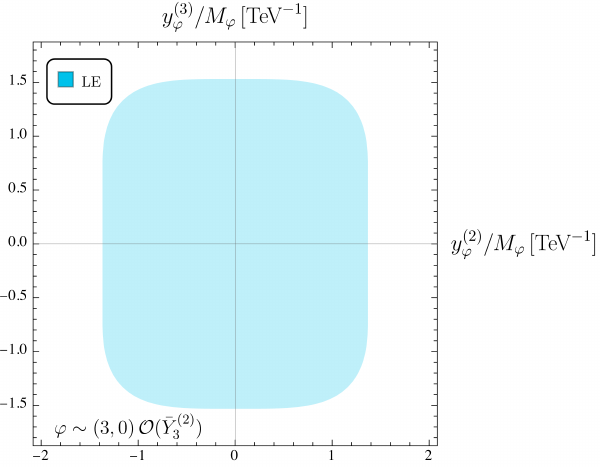}
&&&
\\
\includegraphics[width=32.50mm]{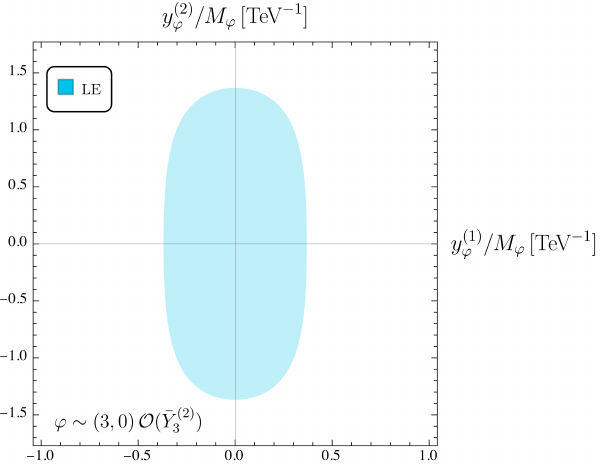}
&&&&
\end{tabular}
\caption{Two-dimensional profiled likelihood contours for the $A_4$ triplet irrep at $\cO(\bar Y_{\bm3}^{(2)})$, where the flavor structure admits six independent parameters (see Tab.~\ref{tab:scalars_a4_irreps_invs}). 15 configurations correspond to all distinct pairwise projections obtained by profiling over the remaining parameters. The $(y_\varphi^{(1)}, y_\varphi^{(4)})$ and $(y_\varphi^{(2)}, y_\varphi^{(5)})$ projections include additional allowed regions from the complementary cLFV constraints.}
\label{fig_supl:varphi_triplet}
\end{figure*}

\begin{figure*}[h]
\centering
\begin{tabular}{ccc}
\includegraphics[width=45.0mm]{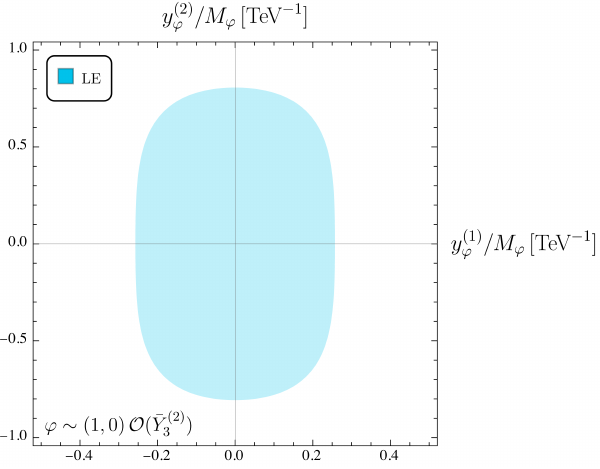} 
&   
\includegraphics[width=45.0mm]{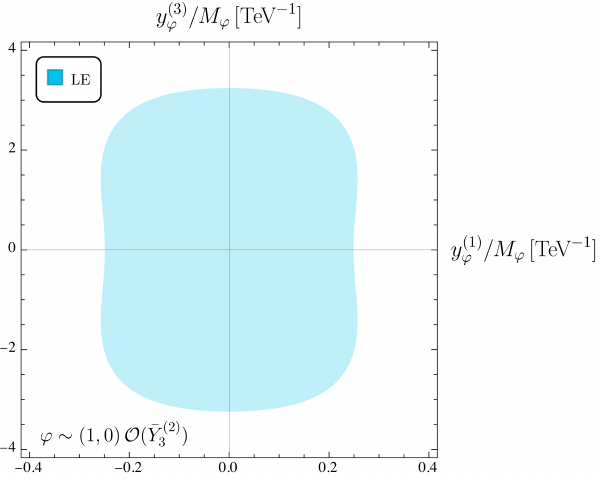}  
&
\includegraphics[width=45.0mm]{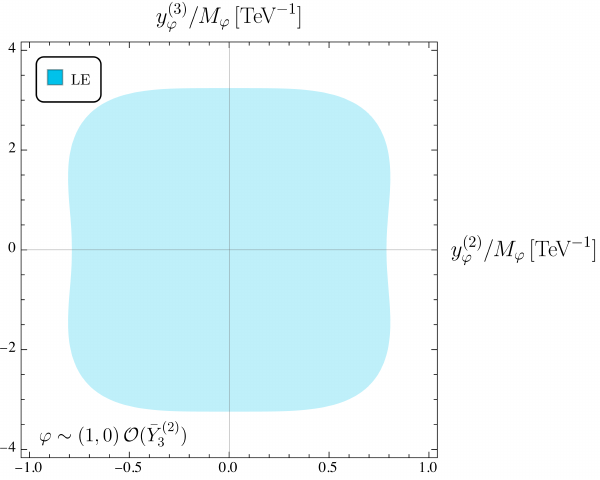}
\\
\includegraphics[width=45.0mm]{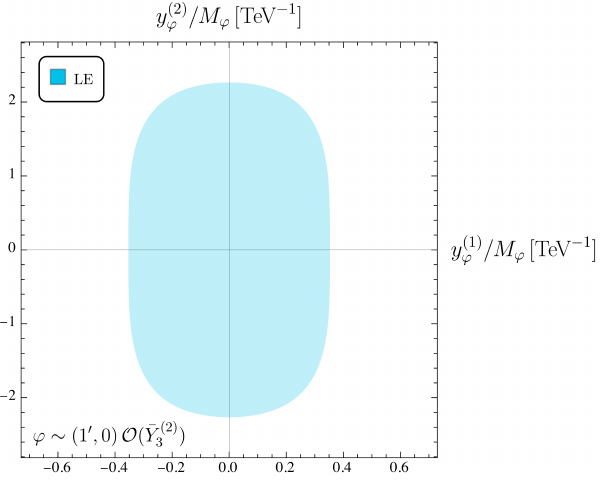} 
&   
\includegraphics[width=45.0mm]{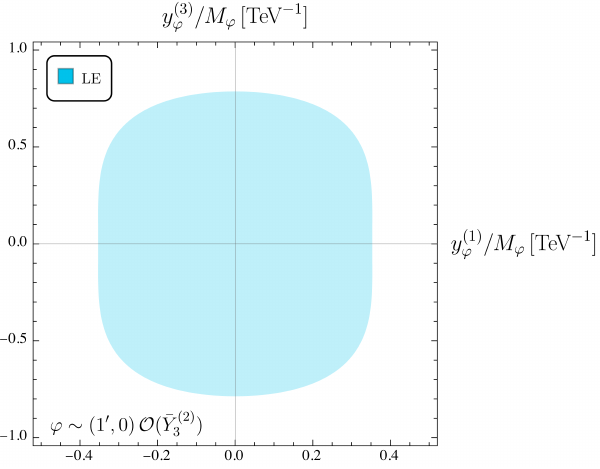}  
&
\includegraphics[width=45.0mm]{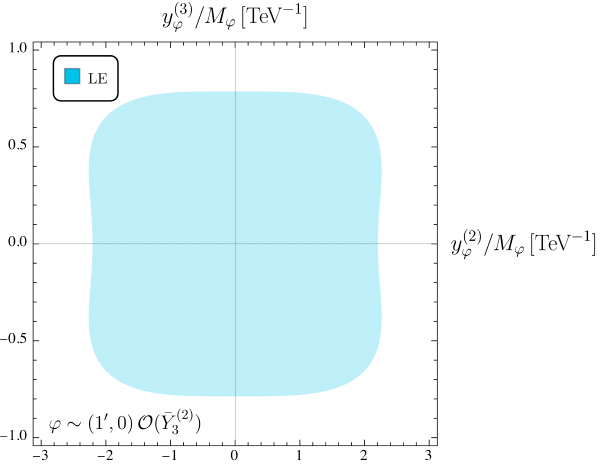}
\\
\includegraphics[width=45.0mm]{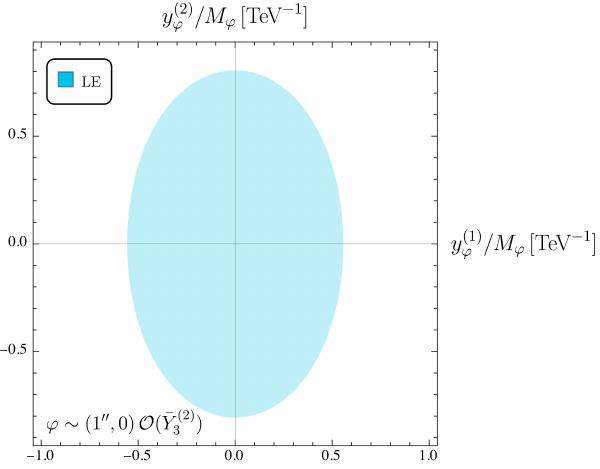} 
&   
\includegraphics[width=45.0mm]{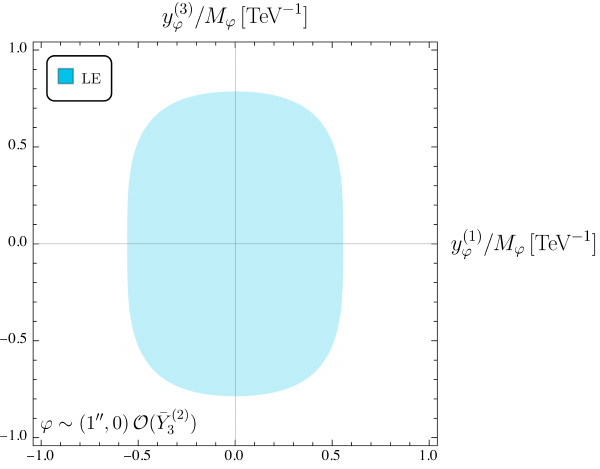}  
&
\includegraphics[width=45.0mm]{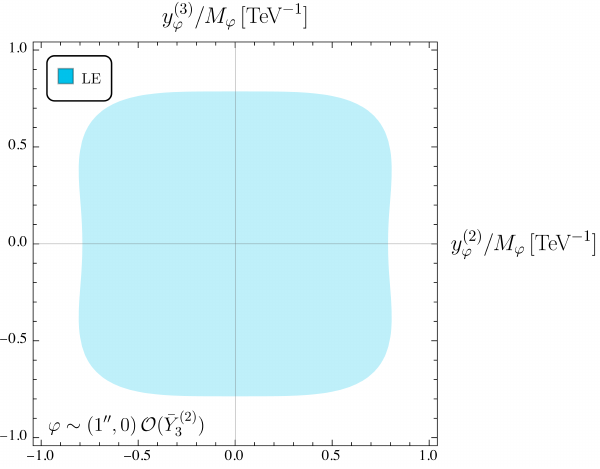}
\end{tabular}
\caption{Profiled two-dimensional likelihoods for all $A_4$ singlet irreducible representations at $\cO(\bar Y_{\bm3}^{(2)})$ of the scalar mediator $\varphi$. The upper, middle, and lower panels display the profiled constraints for the $\bm{1}$, $\bm{1}'$, and $\bm{1}''$ irreps, respectively.}
\label{fig_supl:varphi_singlets}
\end{figure*}

\vspace{0.5cm}
\noindent
$\bm{\varphi}$. For this particular mediator, all allowed $A_4$ irreducible representations lead to flavor tensors containing more than two independent parameters, as shown in Tab.~\ref{tab:scalars_a4_irreps_invs}. We present the profiled likelihoods for $\cO(\bar Y_{\bm3}^{(2)})$ triplet in Fig.~\ref{fig_supl:varphi_triplet}, while $\cO(\bar Y_{\bm3}^{(2)})$ singlets are depicted in Fig.~\ref{fig_supl:varphi_singlets}.

\clearpage

\begin{figure*}[h!]
\centering
\begin{tabular}{ccc}
\includegraphics[width=45.0mm]{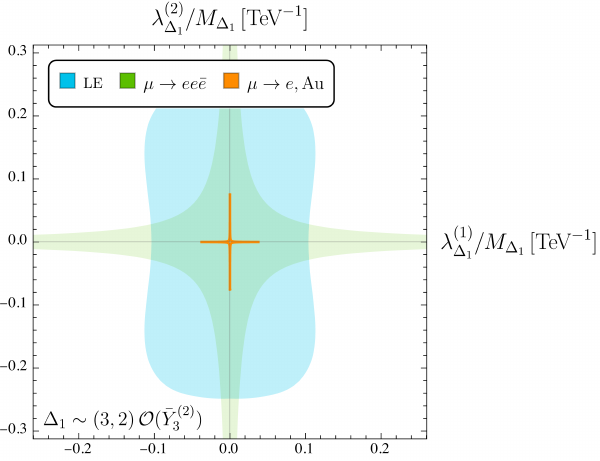} 
&   
\includegraphics[width=45.0mm]{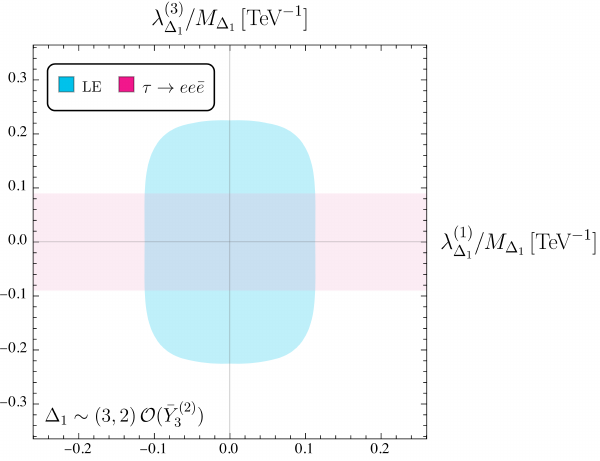}
&
\includegraphics[width=45.0mm]{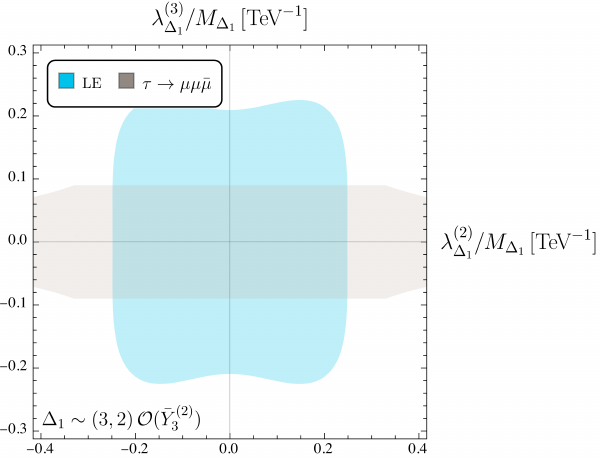} 
\\
\includegraphics[width=45.0mm]{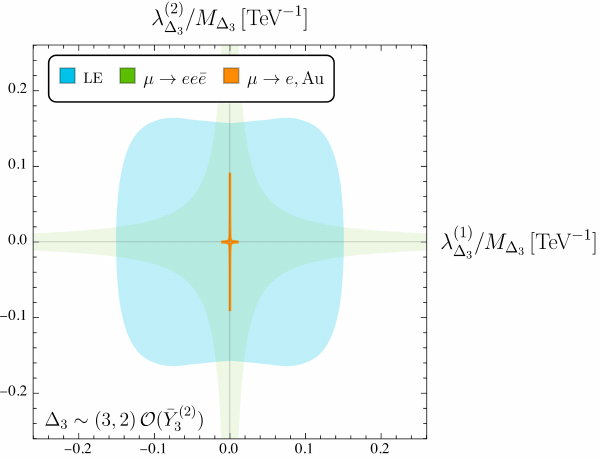} 
&   
\includegraphics[width=45.0mm]{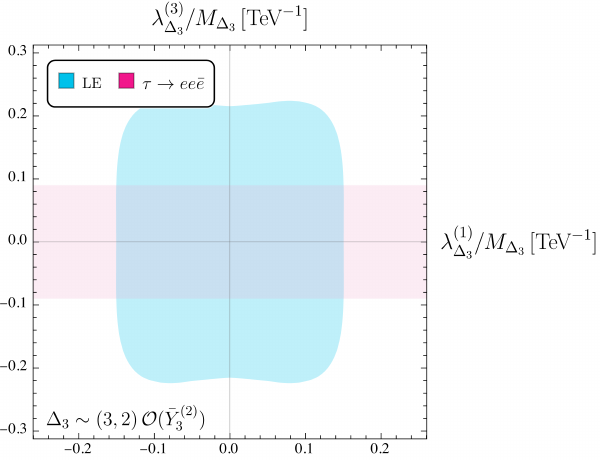}
&
\includegraphics[width=45.0mm]{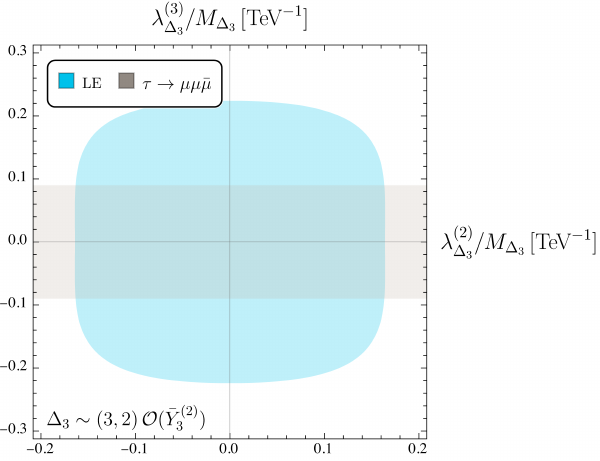}
\end{tabular}
\caption{Two-dimensional profiled likelihood contours for the $A_4$ triplet irreps at $\cO(\bar Y_{\bm3}^{(2)})$ for $\Delta_1$ (upper panel) and $\Delta_3$ (lower panel) fermionic mediators, where the flavor structure admits three independent parameters (see Tab.~\ref{tab:fermions_a4_irreps_invs}). Constraints emerging from the low-energy observables are complemented by the relevant cLFV transitions.}
\label{fig_supl:Delta13}
\end{figure*}

\noindent
$\bm{\Delta_1}$, $\bm{\Delta_3}$. As indicated in Tab.~\ref{tab:fermions_a4_irreps_invs}, the triplet irreps at $\cO(\bar Y_{\bm3}^{(2)})$ lead to flavor tensors with 3 independent parameters in case of both mediators. Constraints after profiling are presented in Fig.~\ref{fig_supl:Delta13}.

\begin{figure*}[h!]
\centering
\begin{tabular}{ccc}
\includegraphics[width=35.50mm]{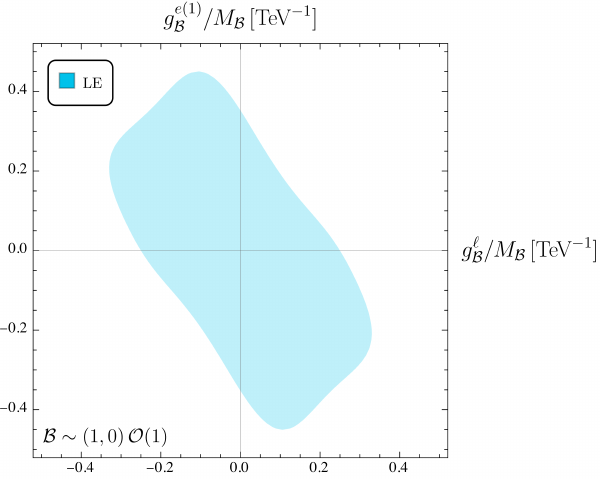} 
&   
\includegraphics[width=35.50mm]{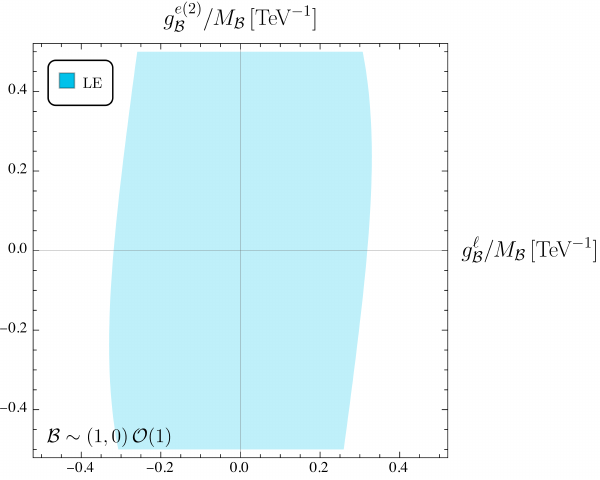} 
&
\includegraphics[width=35.50mm]{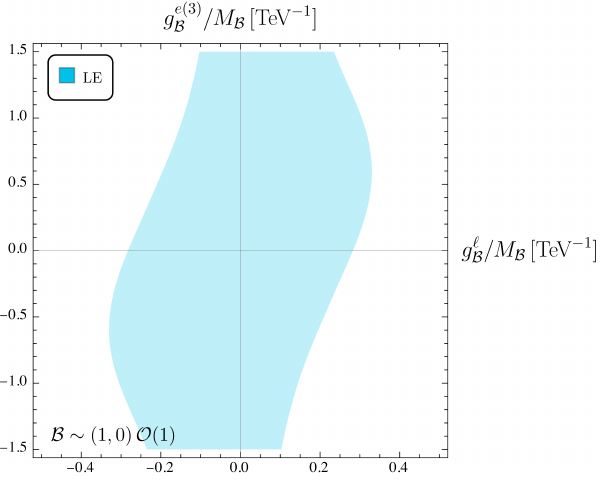}
\\
\includegraphics[width=35.50mm]{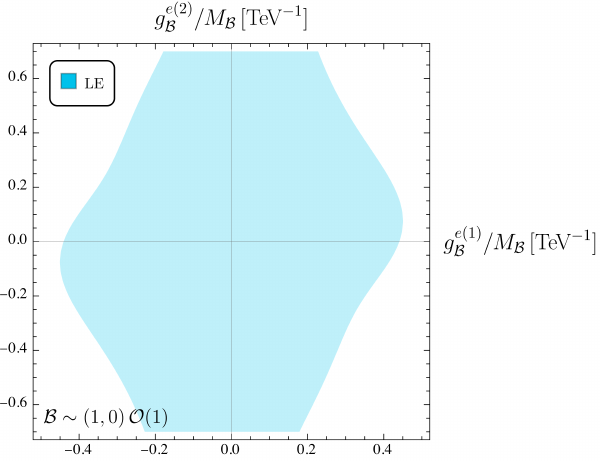}
&
\includegraphics[width=35.50mm]{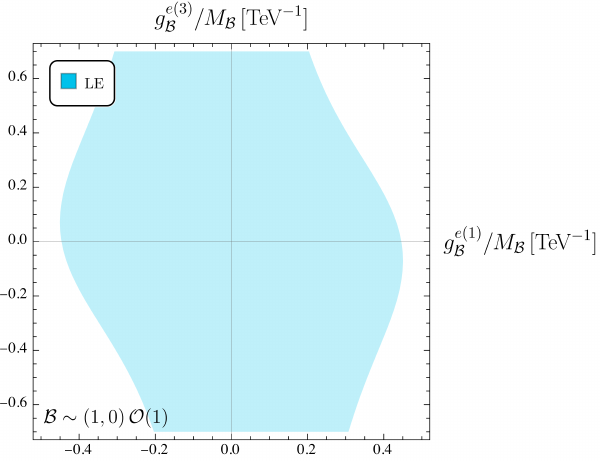}
&
\includegraphics[width=35.50mm]{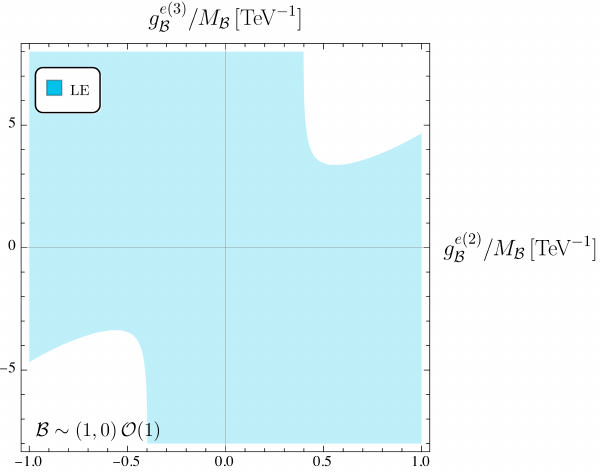}
\end{tabular}
\caption{Two-dimensional profiled likelihood contours in case of $\bm1$ irrep of $\cB$ mediator.}
\label{fig:B_1}
\end{figure*}

\begin{figure*}[h!]
\centering
\begin{tabular}{ccc}
\includegraphics[width=35.50mm]{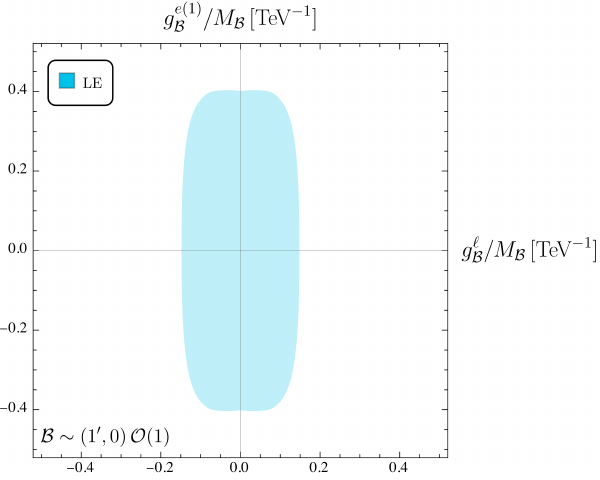} 
&   
\includegraphics[width=35.50mm]{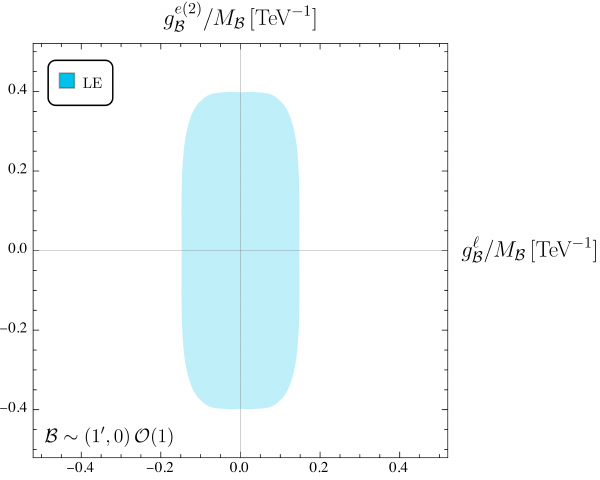} 
&
\includegraphics[width=35.50mm]{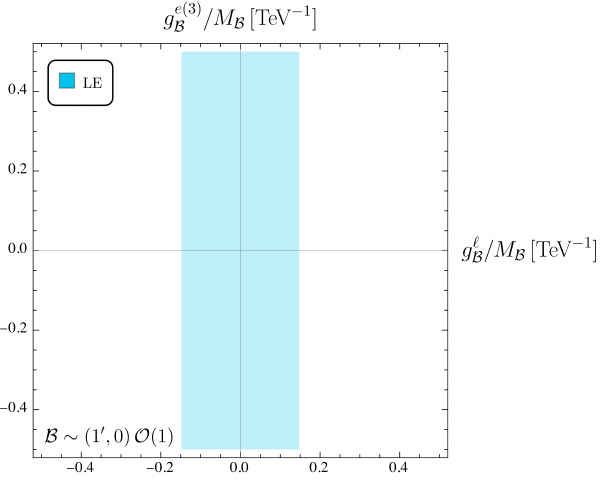}
\\
\includegraphics[width=35.50mm]{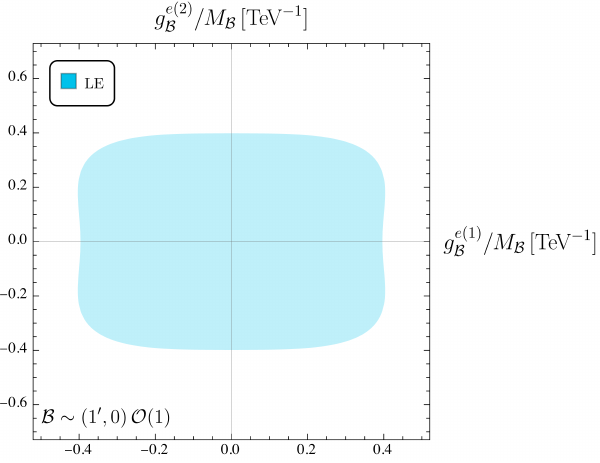}
&
\includegraphics[width=35.50mm]{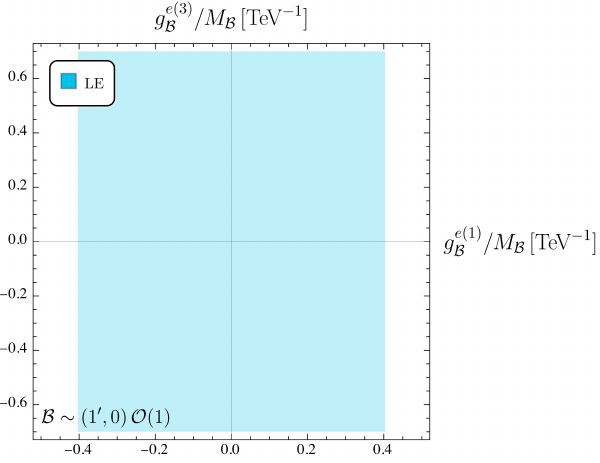}
&
\includegraphics[width=35.50mm]{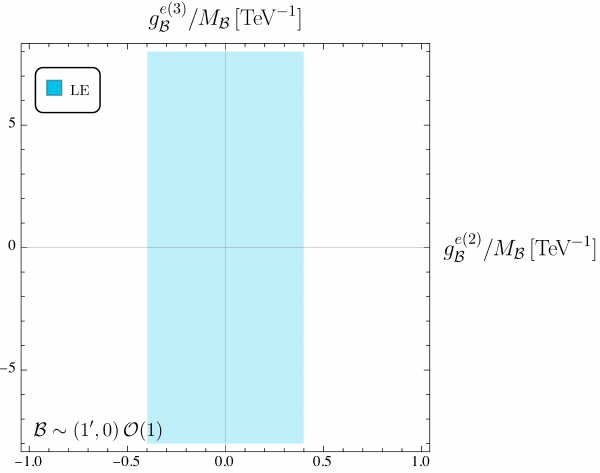}
\end{tabular}
\caption{Two-dimensional profiled likelihood contours in case of $\bm1'$ irrep of $\cB$ mediator.}
\label{fig:B_1'}
\end{figure*}

\begin{figure*}[h!]
\centering
\begin{tabular}{ccc}
\includegraphics[width=35.50mm]{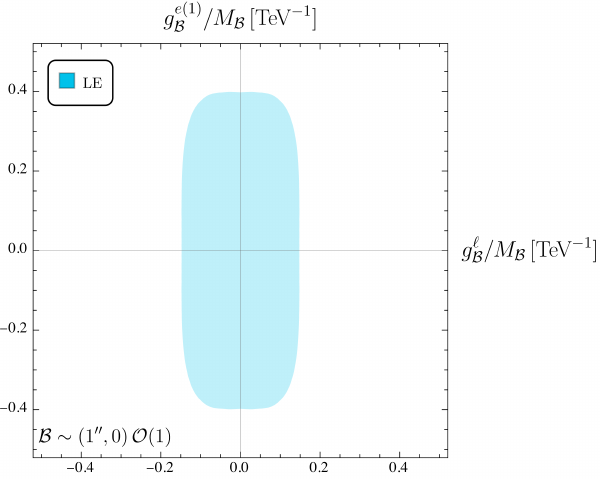} 
&   
\includegraphics[width=35.50mm]{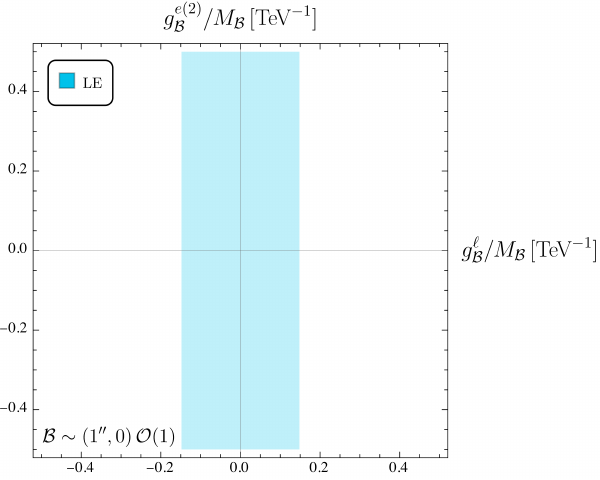} 
&
\includegraphics[width=35.50mm]{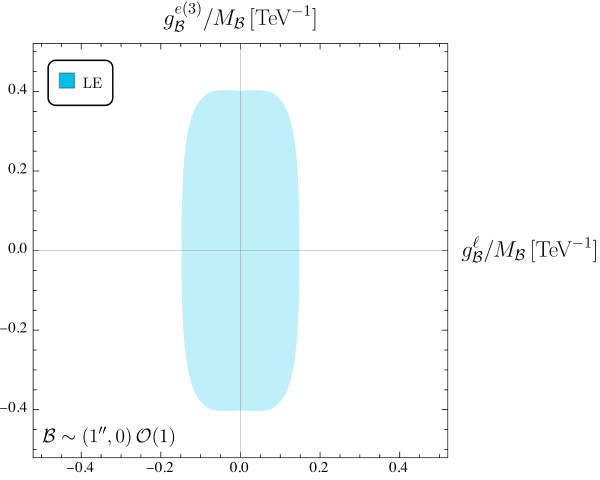}
\\
\includegraphics[width=35.50mm]{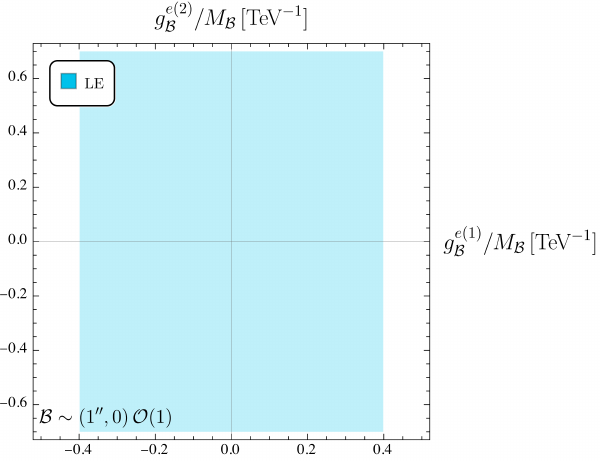}
&
\includegraphics[width=35.50mm]{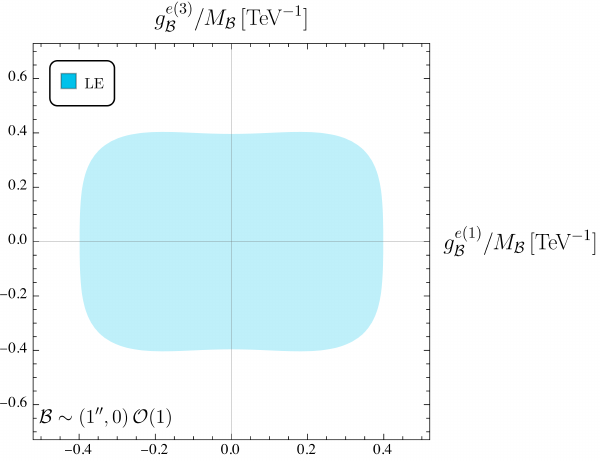}
&
\includegraphics[width=35.50mm]{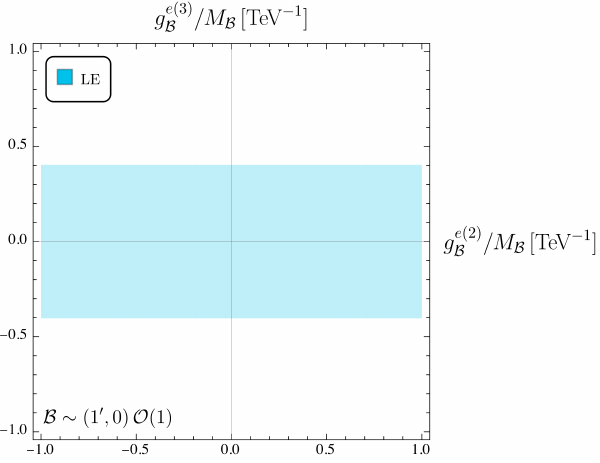}
\end{tabular}
\caption{Two-dimensional profiled likelihood contours in case of $\bm1''$ irrep of $\cB$ mediator.}
\label{fig:B_1''}
\end{figure*}

\clearpage
\noindent
$\bm{\cB}$. As our last example, we take the $\cO(1)$ singlet irreps, where each singlet irrep leads to the flavor tensor containing 4 independent parameters. Constraints obtained after profiling for $\bm1$, $\bm1'$ and $\bm1''$ are denoted in Figs.~\ref{fig:B_1}, \ref{fig:B_1'} and \ref{fig:B_1''}, respectively.

\end{document}